\documentclass[a4paper,11pt]{article}

\usepackage{amssymb,amsmath,amsfonts}
\usepackage{float}
\usepackage{graphicx}
\usepackage{tikz}
\usetikzlibrary{shapes, arrows, shadows}

\def\be{\begin{equation}}
\def\ee{\end{equation}}
\def\bea{\begin{eqnarray}}
\def\eea{\end{eqnarray}}

\begin{document}

\begin{titlepage}
\date{27 October, 2016}       \hfill

\hfill {\it To the memory of}

\hfill { O.I. Zavialov}

\begin{center}

\vskip .5in
{\LARGE \bf   RG boundaries and interfaces in Ising field theory}\\
\vspace{5mm}

\today
 
\vskip .250in

\vskip .5in
{\large Anatoly Konechny}

\vskip 0.5cm
{\it Department of Mathematics,  Heriot-Watt University\\
Edinburgh EH14 4AS, United Kingdom\\[10pt]
and \\[10pt]
Maxwell Institute for Mathematical Sciences\\
Edinburgh, United Kingdom\\[10pt]
}
E-mail: A.Konechny@hw.ac.uk
\end{center}

\vskip .5in
\begin{abstract} \large
Perturbing a CFT by a relevant operator on a half space and letting the perturbation flow to the far infrared we obtain an RG interface 
between the UV and IR CFTs. If the IR CFT is trivial we obtain an RG boundary condition. The space of massive 
perturbations thus breaks up into regions labelled by conformal boundary conditions of the UV fixed point.
For the 2D critical Ising model 
perturbed by a generic relevant operator we find the assignment of RG boundary conditions to all flows. We use 
some analytic results but mostly rely on  TCSA and TFFSA numerical techniques.
We investigate real as well as imaginary values of the magnetic field and, in particular, the RG trajectory 
that ends at the Yang-Lee CFT. We argue that the RG interface  in the latter case does not approach a single conformal 
interface but rather exhibits oscillatory non-convergent behaviour.

\end{abstract}

\end{titlepage}

\renewcommand{\thepage}{\arabic{page}}
\setcounter{page}{1}
\large 

\section{Introduction}
\renewcommand{\theequation}{\arabic{section}.\arabic{equation}}

We are interested in RG flows in two-dimensional Euclidean quantum field theories. We will look at flows that 
originate in a UV fixed point described by a 2D ${\rm CFT}_{\rm UV}$ and arrive at an IR fixed point described by 
a ${\rm CFT}_{\rm IR}$ which may be trivial if  a mass gap develops along the  flow. The flows are triggered by 
perturbations of ${\rm CFT}_{\rm UV}$ by relevant operators $\phi_{i}$. In general it is a hard non-perturbative problem 
 to determine 
the infrared properties of the perturbed theory, in particular to identify  ${\rm CFT}_{\rm IR}$ when it is non-trivial. 
A  technique that allows one to investigate the perturbed theories  numerically  in the  infrared including 
the flows to non-trivial fixed points is the truncated conformal space approach (TCSA) invented in \cite{YZ1}, \cite{YZ2}. 
The basic setup of TCSA is as follows.

In order not to worry about perturbative infrared divergences 
we put the perturbed theory on a cylinder with spacial periodic  coordinate $x\sim x+ R$ and Euclidean time $y$ directed along the axis 
of the cylinder. In the Hamiltonian formalism the perturbed Hamiltonian on a circle at $y=0$ is 
\be\label{Ham_pert}
H=H_{0} + \lambda^{i}\int\!\! dx\, \phi_{i}(x,0) \, .
\ee
Here 
\be
H_{0}=\frac{2\pi}{R}\left( L_{0} + \bar L_{0} - \frac{c_{\rm UV}}{12} \right) 
\ee
is the Hamiltonian of ${\rm CFT}_{\rm UV}$ that is expressed via the Virasoro modes $L_{0}$, $\bar L_{0}$ and the 
central charge $c_{\rm UV}$. The eigenvalues of $H_{0}$ are  scaling dimensions  of ${\rm CFT}_{\rm UV}$ shifted by
$\frac{c_{\rm UV}}{12}$ .
Since the state space ${\cal H}_{0}$ of the unperturbed theory on a cylinder is discrete $H$ can be represented by 
an infinite matrix acting in this space. To regulate the UV divergences we can truncate $H$ to a finite matrix by restricting it 
to a finite dimensional subspace in ${\cal H}_{0}$. This gives rise to a variety of truncated Hamiltonian  techniques in which 
one numerically calculates the eigenvalues and eigenvectors of the truncated Hamiltonian matrix for various values 
of the scale set by the circle length $R$.  In TCSA the truncated subspace is specified by  imposing an upper bound 
on the scaling dimensions of the allowed states.  This upper bound is called the truncation level. The UV divergences show up in the 
dependence of numerics on the truncation level. This dependence and various improvement techniques have been discussed in 
\cite{GW}, \cite{LT_RG}, \cite{HRvR}, \cite{RV}.  

TCSA has been applied to situations in which a perturbed theory arrives to a non-trivial fixed point (see e.g. \cite{Cardy_etal}, \cite{GW}). 
In this case the dimensionless energy eigenvalues $ER/2\pi$ at large $R$ asymptote to  constant values  that give scaling dimensions 
in ${\rm CFT}_{\rm IR}$. Moreover the asymptotic eigenvectors that correspond to scaling states $|v_{i}\rangle$ in ${\rm CFT}_{\rm IR}$ 
are obtained as vectors in the truncated subspace of ${\cal H}_{0}$. The corresponding components  of $|v_{i}\rangle\in {\cal H}_{0}$
 give us  pairings 
 \be \label{TCSA_pairing}
 \langle i | v_{j}\rangle 
 \ee
 between scaling states in both CFT's. 
 
 Pairings of this type are also associated with conformal interfaces.
 A conformal interface ${\cal I}$ between ${\rm CFT}_{\rm UV}$ and ${\rm CFT}_{\rm IR}$ can be described (via the folding trick \cite{BBDO}) 
 as a conformal boundary condition in the tensor product ${\rm CFT}_{\rm UV} \otimes {\rm CFT}_{\rm IR}$. If $\phi_{i}^{\rm UV}$ 
 and $\phi_{j}^{\rm IR}$ are scaling fields in the two theories then we can define a pairing 
 \be \label{conf_pairing}
 {}_{\rm UV}\langle i |j\rangle_{\rm IR} = \langle \phi_{i}^{\rm UV} \phi_{j}^{\rm IR} \rangle_{\cal I}
 \ee
 as a disc one-point function with boundary condition ${\cal I}$. Such pairings are canonically normalised by means of Cardy constraint \cite{Cardy}.

 In TCSA there is no canonical way to fix normalisation of the eigenstates $|v_{j}\rangle$ so the natural observables that are free 
 from this ambiguity are component ratios 
 \be\label{gen_rat}
 \Gamma^{j}_{i,k} = \frac{ \langle i | v_{j}\rangle }{ \langle k | v_{j}\rangle } \, .
 \ee
  In (untruncated) quantum field theory  perturbed states more often than not do not lie in the unperturbed state space. 
  This has many manifestations such as Haag's theorem, orthogonality catastrophe, inequivalent representations of canonical 
  commutation relations, etc (see \cite{H_theorem} for a nice exposition and discussion). 
  Typically\footnote{At least for superrenormalizable theories.} however the problem is with the norm of the perturbed states in interaction representation 
  which is formally 
  infinite, while the component ratios such as (\ref{gen_rat}) are well defined. 
  
To show that the pairing (\ref{TCSA_pairing}) and ratios (\ref{gen_rat}) associated with an RG flow arise from a particular local conformal 
interface we discuss 
states in Lagrangian formalism using wave functionals. Suppose ${\rm CFT}_{\rm UV}$ is described via a fundamental field 
$\varphi(x)$ and an action functional $S_{0}[\varphi]$. States in this theory can be represented by a  wave functional 
$\Psi[\varphi_{0}]$ depending on functions $\varphi_{0}(x)$ defined on the circle  $y=0$. The vacuum state is then given by a renormalised 
functional integral over the left half cylinder $y\le 0$
\be
\Psi_{\rm vac}[\varphi_{0}] = \int\limits_{\varphi(x,0)=\varphi_{0}(x)}\!\!\!\!\! {\cal D}  \varphi \, e^{-S_{0}[\varphi]} \, .
\ee
The functionals describing excited states can be obtained by inserting local operators at positions with $y<0$ inside the above functional integral. 

Suppose now that the perturbed theory is described by an action functional 
\be
S[\varphi] = S_{0}[\varphi] + \iint\! \! dxdy\,  V(\varphi(x,y)) \, 
\ee
where $V(\varphi)$ is some potential. 
The  vacuum 
of the perturbed theory can be represented by the functional integration over the fields defined  
\be\label{perturbed_psi}
\Psi_{\rm vac}^{\rm pert}[\varphi_{0}] = \int\limits_{\varphi(x,0)=\varphi_{0}(x)}\!\!\!\!\! {\cal D}  \varphi \, e^{-S[\varphi]} \, .
\ee 
Expanding inside the functional integral the exponent 
$$\exp\Bigl[{\iint\limits_{y\le 0} \!  dxdy\,  V(\varphi) }\Bigr]$$
in power series we obtain a formal expansion of the perturbed vacuum functional $\Psi_{\rm vac}^{\rm pert}[\varphi_{0}] $ in terms of 
unperturbed states (interaction representation). We can extend this pairing to excited states of the deformed theory by inserting into the functional integral 
(\ref{perturbed_psi}) additional  local operators. This gives a pairing between the states in the two theories. In particular if we follow the perturbed theory to 
the IR fixed point we obtain a pairing of the type (\ref{TCSA_pairing}). We assume in this discussion that all divergences are renormalised including the ones
 that need additional boundary counterterms at $y=0$. 
Renormalisation 
of wave functionals has been discussed in \cite{Symanzik}, \cite{Luscher}, \cite{Minic_Nair} and more recently in \cite{FLP}. (More on renormalisation 
shortly.)

What becomes clear in this picture is that the pairing arises 
by perturbing the UV fixed point on a half space (a half cylinder) and letting the perturbed theory flow to the IR fixed point. 
This gives us a local conformal interface between  ${\rm CFT}_{\rm UV}$ and  ${\rm CFT}_{\rm IR}$ which we call an RG interface. 
The idea to associate such interfaces with RG flows was put forward in \cite{BR} (see also \cite{FQ}). A concrete proposal 
for such an interface for $\psi_{1,3}$-flows between neighbouring minimal models was put forward in \cite{Gaiotto}  and  pairing 
(\ref{conf_pairing}) associated with it was considered\footnote{Unlike \cite{Gaiotto} we restrict our pairing to states only and do not consider issues of 
transport of local operators from UV into IR that we believe to be rather subtle.}.

Note that we do not need to have  fundamental fields and functional integral representation to define the pairing 
in terms of the RG interface. 
To elucidate renormalisation and to relate the above functional integral picture to conformal perturbation and TCSA we 
first  recast (\ref{perturbed_psi}) in the language of  conformal perturbation. Wave functionals correspond to a particular basis 
in state space in which the field operator $\varphi$ is diagonal. We can choose instead the conformal basis of scaling states $|i\rangle$
in ${\rm CFT}_{\rm UV}$. 
Consider the following amplitude on a cylinder of length $L$
\be\label{Z0i}
Z_{0,i}(L) = \langle 0| \exp\Bigl[ \int\limits_{-L\le y\le 0} \lambda^{k}\phi_{k}(x,y) \Bigr]  | i\rangle  
\ee
illustrated on the picture below.
\begin{center}
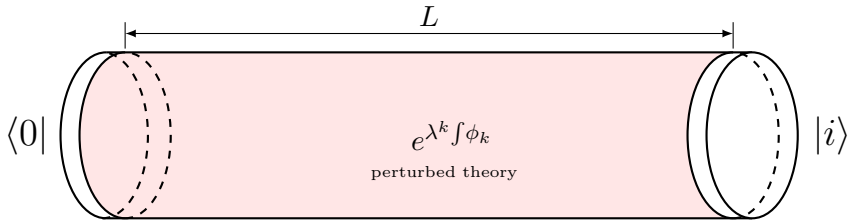
\begin{figure}[H]
\centering

\begin{tikzpicture}[>=latex]

\fill[red!10!white] (-2,0) -- (6,0) arc     ( 270:90: 0.6 and 1.1)-- (-2,2.2) arc (90:270: 0.6 and 1.1)--cycle; 
\draw[thick] (-2.25,0)--(6.25,0);
\draw[thick] (-2.25,2.2)--(6.25,2.2);
\draw[thick] (-2.25,2.2) arc (90: 270: 0.6 and 1.1);
\draw[thick, dashed] (-2.3,0) arc (-90:90:0.6 and 1.1 );

\draw[thick] (-2,2.2) arc (90: 270: 0.6 and 1.1);
\draw[thick, dashed] (-2,0) arc (-90:90:0.6 and 1.1 );

\draw[ thick] (6,0) arc (270:90:0.6 and 1.1);
\filldraw[white,draw=black, thick] (6.25,1.1) circle [x radius=0.6, y radius = 1.1];
\draw[ thick,dashed] (6.0,0) arc (-90:90:0.6 and 1.1);

\draw (7.3,1.1) node {\Large $|i \rangle$};
\draw (-3.3,1.1) node {\Large $\langle 0|$};
\draw (2.3,1.1  ) node {$e^{\lambda^{k}\! \int\! \phi_{k}}$};
\draw (2.2, 0.6) node {\tiny perturbed theory};
\draw (-2,2.25) -- (-2, 2.6);
\draw (6,2.25)--(6,2.6);
\draw[<->] (-2, 2.45) -- (6,2.45);
\draw (2,2.68) node {$L$};

\end{tikzpicture}
\caption{The amplitude $Z_{0,i}(L)$.}
\end{figure}

\end{center}
We have added little collars (depicted white) of unperturbed theory at both ends where the states $\langle 0|$, $|i\rangle$ are attached. The actual width of the collars is 
inessential. The perturbative expansion of (\ref{Z0i}) is 
\be
Z_{0,i}(L) = \sum\limits_{n} \sum_{i_1,\dots, i_{n}}\frac{\lambda^{i_{1}}\dots \lambda^{i_{n}}}{n !} \!\!\!
\int\limits_{-L\le y_1\le 0}\!\! d^2 z_{1} \dots \int\limits_{-L\le y_n\le 0}\!\! d^2 z_{n}\langle 0|  \phi_{i_1}(z_1) \dots   \phi_{i_n}(z_n) |i\rangle 
\ee
The correlators here are  correlators in ${\rm CFT}_{\rm UV}$ and can be calculated by mapping the cylinder onto an annulus on the plane 
 and  inserting  $\phi_{i}$ at infinity. Divergences arise when operators $\phi_{i_{k}}$ collide. Collisions can happen at a point in the bulk of the cylinder 
(annulus) or at a point on the boundary: at $y=0$ or $y=-L$. The former are handled by the usual renormalisation of perturbed theory while 
the latter  may give rise to additional boundary counter terms of the type discussed in 
\cite{Symanzik}\footnote{They result in additional functions present in the Callan-Symanzik equation for $Z_{0,i}$ (and $\Psi_{\rm vac}^{\rm pert}[\varphi_{0}]$)}. 
It is clear from the collisions picture that all divergences are local. Assuming they can be renormalised we obtain a cylinder with two local 
interfaces between ${\rm CFT}_{\rm UV}$ and the perturbed theory with two external states of ${\rm CFT}_{\rm UV}$ attached.
 The perturbed theory  can  be driven towards ${\rm CFT}_{\rm IR}$ that results 
in having a system  with two  conformal interfaces. Taking the limit $L\to \infty$ results in the following asymptotic 
\be
Z_{0,i}(L) \sim e^{-LE_{0}^{\rm IR}} \langle 0|v_{0}\rangle \langle v_{0}|i\rangle 
\ee
where $E_{0}^{\rm IR}=-\pi c_{\rm IR}/6R$ is the vacuum energy of ${\rm CFT}_{\rm IR}$.
Removing the divergent   exponential factor  we obtain up to the overall factor $ \langle 0|v_{0}\rangle$ the overlaps
  $\langle v_{0}|i\rangle $ of the perturbed vacuum $|v_{0}\rangle$ with  the unperturbed scaling states. 
Overlaps with excited states $|v_{i}\rangle$ can be obtained from the subleading terms in the $L\to \infty$ asymptotic.

To summarise the above discussion,  in the Lagrangian formalism the TCSA overlaps (\ref{TCSA_pairing}) are described in terms of a local interface between the perturbed
and unperturbed theories. This has implications that are not evident  in the Hamiltonian picture (and thus in TCSA). 
Thus for flows to a trivial fixed point the RG interface is a conformal boundary condition (the RG boundary). 
In the far infrared the perturbed theory vacuum state is given by the conformal 
boundary state describing this boundary condition (this was previously observed in \cite{TWunpub} and \cite{CEF} for 
particular models). The boundary state at hand
must satisfy  Cardy constraint which   arises from  locality.  There are similar constraints for flows to 
a non-trivial fixed point with local conformal interfaces described by  conformal boundary conditions in the tensor product theories as was already 
mentioned above. 

All massive flows from a given ${\rm CFT}_{\rm UV}$ are then labeled by conformal boundary conditions. They appear to be natural labels of infrared phases of massive theories. The unstable manifold is then broken into regions of possibly different dimensions labeled by conformal boundary conditions with particular conditions on boundaries separating these regions. It seems to be an interesting enterprise  to investigate this structure for particular two-dimensional 
CFTs. For each Virasoro minimal model there is a finite number of fundamental conformal boundary conditions, so that we expect their unstable manifolds to 
be broken into finitely many regions (massive phases). As we will show below superpositions of fundamental boundary conditions may also arise 
for RG flows in the presence of  spontaneous symmetry breaking. 


In this paper we investigate the RG boundaries and interfaces for the simplest  minimal model - the critical Ising model. 
It has three primary fields: ${\bf 1}$, $\epsilon$, 
$\sigma$ and three fundamental conformal boundary conditions with boundary states \cite{Cardy}
\be\label{conf_bcs}
|\pm\rangle\!\rangle = \frac{1}{\sqrt{2}} \Bigl[ |0\rangle\!\rangle + |\epsilon\rangle\!\rangle  \pm 2^{1/4}|\sigma\rangle\!\rangle \Bigr]\, , 
\qquad |F\rangle\!\rangle = |0\rangle\!\rangle - |\epsilon\rangle\!\rangle \, .
\ee
Here $|0\rangle\!\rangle $, $|\epsilon \rangle\!\rangle $, $|\sigma\rangle\!\rangle $ stand for the Ishibashi states built on the  respective primaries: 
${\bf 1}$, $\epsilon$, and 
$\sigma$.
The boundary states $|\pm\rangle\!\rangle$ describe the fixed spin while $ |F\rangle\!\rangle $ describes the free  spin boundary condition. 
The theory has two relevant perturbations - temperature and magnetic field that are described by mass $m$ and magnetic field  coupling $h$ in the free 
fermion theory: 
\be\label{IFT}
S_{\rm IFT}=\frac{1}{2\pi}\int (\psi\bar \partial \psi + \bar \psi \partial \bar \psi + im\bar \psi \psi )\, d^2 x + h\!\int\!\! \sigma\, d^2 x \, .
\ee
For real values of the couplings $m$, $h$ all flows are massive and our results regarding the corresponding RG boundaries are 
 summarised on the diagram presented below. We  parameterise an 
outgoing RG trajectory for Ising field theory by points on a circle on the $m,h$-plane. For all points on the 
upper semicircle except for the points on the $h=0$ axis the RG boundary condition is $|+\rangle\!\rangle$ 
and similarly for all points on the lower semicircle we have $|-\rangle\!\rangle$. The exceptional points are 
those on the $h=0$ axis. For $h=0, m>0$ we obtain a superposition $|+\rangle\!\rangle \oplus |-\rangle\!\rangle$
while for $h=0,m<0$ we obtain $|F\rangle\!\rangle$.

\begin{center}
\begin{figure}[H]  
\centering
\begin{tikzpicture}[>=latex]
\draw[->] (-3,0) --(3.3,0);
\draw[->] (0,-3)--(0,3);
\draw[thick] (0,0) circle [radius=2];
\draw (3.6,0  ) node {$m$};
\draw (0,3.3) node {$h$};
\draw (1.8,1.8) node {$|+\rangle\!\rangle $};
\draw (-1.7,1.8) node {$|+\rangle\!\rangle $};
\draw (-1.7,-1.8) node {$|-\rangle\!\rangle$};
\draw (1.8,-1.8) node {$|-\rangle\!\rangle$};
\draw[->] (3.8,1) -- (2.1,0.1);
\draw (-2,0) node {$\bullet$} ;
\draw (2,0) node {$\bullet$} ;
\draw (3.8,1.2) node {$|+\rangle\!\rangle \oplus |-\rangle\!\rangle$};
\draw (-2.4,0.3) node {$|F\rangle\!\rangle$};
\end{tikzpicture}
\caption{The unstable manifold of IFT and associated RG boundaries.}\label{diagram_real}
\end{figure}
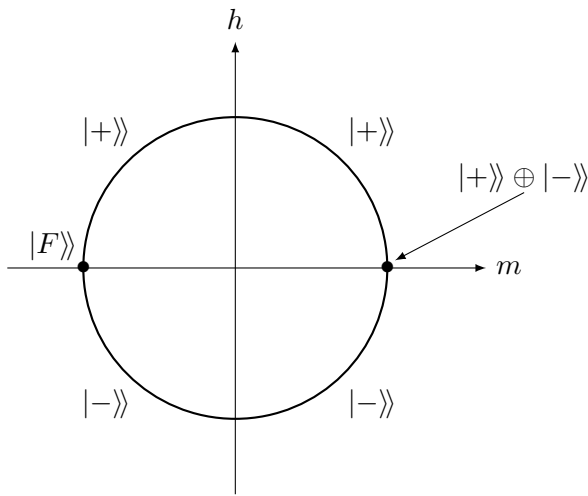
\end{center}
The fact that we obtain a superposition of boundary states for $h=0, m>0$ reflects the spontaneous symmetry 
breaking of the low temperature phase.

We have also investigated the imaginary magnetic field.  
It is convenient to label the RG trajectories by the dimensionless ratio 
\be\label{y_def}
y=\frac{m}{|h|^{8/15}} \, .
\ee
As shown in \cite{FZ} for imaginary magnetic field when $y>y_{\rm cr}=-2.42929$ (the value is taken from \cite{Zam2})
the vacuum energy is complex with  a two-dimensional vacuum subspace spanned by the eigenstates with complex 
conjugated energy eigenvalues. We find that in all such cases the RG boundary is $|+\rangle\!\rangle \oplus |-\rangle\!\rangle$.
For $y<y_{\rm cr}$ the vacuum energy is real and the flows are massive. We argue that in this case the RG boundary does 
not approach a single conformal boundary condition as we go to the far infrared. For large negative values of $y$ the RG flow 
of the vacuum vector can be well approximated by the boundary magnetic field model that is exactly solvable and provides 
further insight into the non-convergent behaviour. 
At $y=y_{\rm cr}$ the RG trajectory 
approaches the Yang-Lee edge singularity \cite{YL1}, \cite{YL2} that is described by a non-unitary conformal minimal 
model  ${\cal M}(2,5)$ \cite{Cardy2}. In this case our findings point as well to the picture in which the RG interface does not approach a single 
conformal interface but demonstrates an oscillatory non-convergent behaviour. The case $y=-\infty$ is the exceptional case of 
no magnetic field and the RG boundary is given by the free boundary condition as before. We summarise these answers on the diagram below.

\begin{center}
\begin{figure}[H]
\centering
\begin{tikzpicture}[>=latex]
\draw[->] (-3,0) --(3.3,0);
\draw[->] (0,-3)--(0,3);
\draw[thick,dashed, red] (-1.41,-1.41) arc (225:135:2);
\draw[thick] (-1.41,-1.41) arc (-135:135:2);
\draw (3.6,0  ) node {$m$};
\draw (0,3.3) node {$ih$};
\draw (1.8,2.1) node {$|+\rangle\!\rangle \oplus |-\rangle\!\rangle$};
\draw (1.8,-2.1) node {$|+\rangle\!\rangle\oplus |-\rangle\!\rangle$};
\draw[->] (-3.6,-1) -- (-2.1,-0.1);
\draw (-2,0) node {$\bullet$} ;
\draw (-1.3,2.2) node {\mbox{\small $y=y_{\rm cr}$}};
\draw (-3.95,-1) node {$|F\rangle\!\rangle$};
\draw[thick,red] (0,0) parabola (-1.7,2.04);
\draw[thick,red] (0,0) parabola (-1.7,-2.04);

\end{tikzpicture}
\caption{RG boundaries for  IFT with imaginary magnetic field. Everywhere on the dashed red line there is no limit 
except for the $h=0$ point. This includes the $y=y_{\rm cr}$ trajectory flowing to Yang-Lee theory. Everywhere on the solid 
black line the RG boundary is  $|+\rangle\!\rangle \oplus |-\rangle\!\rangle$.}
\end{figure}
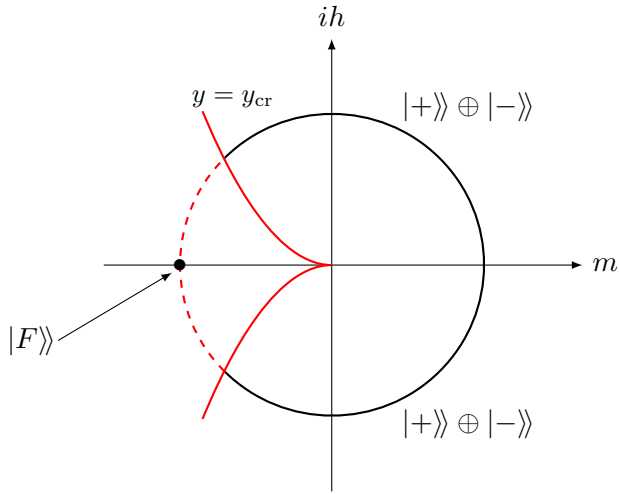
\end{center}

The main body of the paper is organised as follows. In section 2 we work out in detail the analytically solvable case 
of vanishing magnetic field. In section 3 we explain the basic TCSA setup for the perturbed Ising model using free fermions.
In section 4 we present numerical results obtained using TCSA for real valued couplings $m$ and $h$. In particular we give
plots for some component ratios of the type given by (\ref{gen_rat}). The numerical results together with the analytic ones from section 2 lead to 
the diagram on Fig.   \ref{diagram_real}  which is further discussed in section 4.2.
While TCSA works really well for the real couplings, for imaginary magnetic field when the vacuum energy is real another numerical method - 
truncated free fermion space approach (TFFSA) of \cite{FZ} works much better. This method which we explain in section 5 allows one 
to treat the mass coupling non-perturbatively. In section 6 we present our numerical results obtained using TFFSA for the case of 
imaginary magnetic field and the region $y>y_{\rm cr}$ where the vacuum energy is complex. In section 7 we discuss the region $y\le y_{\rm cr}$.
We  first consider  the massive flows with $y<y_{\rm cr}$ in section 7.1. Before discussing in section 7.3 the numerics for component ratios related to the 
flow to the Yang-Lee fixed point at $y=y_{\rm cr}$ we describe all conformal interfaces between the Ising and Yang-Lee models in section 7.2. 
We finish off with a brief  discussion of some open problems in section 8.


\section{RG flows at zero magnetic field}
\setcounter{equation}{0}
\subsection{Preliminaries. Massive and massless free fermions.}\label{preliminaries}
The Ising model at zero magnetic field is a theory of free fermions with Euclidean action 
\be
\frac{1}{2\pi}\int (\psi\bar \partial \psi + \bar \psi \partial \bar \psi + im\bar \psi \psi )\, d^2 x
\ee
Here the coupling $m$ measures the deviation from the critical temperature of the classical model: 
 $m=(T_{c}-T)/T_{c}$. 
 
 We are going to put this theory on an infinite cylinder of radius $R$. Let $x\sim x + R$ be the periodic 
 coordinate and $y$ - the coordinate along the cylinder. Two sectors arise - with periodic 
 (Neveu-Schwarz or NS-sector for brevity) and anti periodic (Ramond or R-sector). 
 Quantizing on the spacial $x$-circle we get  the single-particle  frequencies  
\be
\omega_n = \sqrt{m^2 + \left( \frac{2\pi n}{R}\right)^2 }
\ee
where $n\in {\mathbb Z}$ in the R-sector and $n\in 1/2 + {\mathbb Z}$ in the NS-sector. Below we will use the index $n$ (or $n_1, n_2, \dots$) 
for integers and $k$ (or $k_1, k_2, \dots$) for half-integers. 

Using the complex coordinate $z=x+iy$ the  mode expansion in the NS sector is introduced as 
\be
\psi(x,y)=\sum_{k\in 1/2 + {\mathbb Z}}  \left(-\frac{\pi i}{R}\right)^{1/2} \sqrt{1- \frac{2\pi k}{R\omega_{k}}} \Bigl[  \epsilon(k)b_{k}e^{-y\omega_{k}-ixk/2\pi} 
+  \epsilon^{*}(k)b_{k}^{\dagger}e^{y\omega_{k}+ixk/2\pi} 
  \Bigr]
\ee
\be
\bar \psi(x,y)=\sum_{k\in 1/2 + {\mathbb Z}}  \left(\frac{\pi i}{R}\right)^{1/2} \sqrt{1+ \frac{2\pi k}{R\omega_{k}}} \Bigl[  \epsilon^{*}(-k)b_{k}e^{-y\omega_{k}-ixk/2\pi} 
+  \epsilon(-k)b_{k}^{\dagger}e^{y\omega_{k}+ixk/2\pi} 
  \Bigr]
\ee
where 
\be
\epsilon(k) =  \left \{
\begin{array}{r@{\quad}l}
1\, ,  &k<0
\\[1ex]
-i{\rm sign}(m)\, , &k>0
\end{array}
\right . 
\ee 
We also have similar expressions in the R-sector in terms of creation and annihilation operators $b^{\dagger}_{n}$, $b_{n}$  which we omit for brevity. 
The physical state space on a circle ${\cal H}_{m}$ is spanned by the states 
\be \label{mbasis1}
|k_1, \dots, k_{N}\rangle_{\rm NS} = b^{\dagger}_{k_1}\dots  b^{\dagger}_{k_N}|0\rangle_{\rm NS} \, , \quad k_{i}\in 1/2 + {\mathbb Z}
\ee
with $N$ even  and by the states 
\be \label{mbasis2}
|n_1, \dots, n_{M}\rangle_{\rm R} = b^{\dagger}_{n_1}\dots  b^{\dagger}_{n_M}|0\rangle_{\rm R}\, , \quad n_{i}\in  {\mathbb Z}
\ee
where when $m>0$ the number of oscillators $M$ is even and when $m<0$ $M$  is odd.

The Hamiltonian blocks are 
\be\label{HFF1}
H^{\rm NS} =  \sum_{n=-\infty}^{\infty} \omega_{n+1/2}b^{\dagger}_{n+1/2}b_{n+1/2} + E_{0}^{({\rm NS})}\, , 
\ee
\be\label{HFF2}
H^{\rm R} = \sum_{n=-\infty}^{\infty} \omega_{n}b^{\dagger}_{n}b_{n} + E_{0}^{({\rm R})}\,
\ee
where the vacuum energies are 
\be
E_{0}^{({\rm NS})}=\frac{m^2 R}{4\pi}\ln\left( \frac{|m|}{\mu}\right) -|m|\int\limits_{-\infty}^{\infty} \frac{d\theta}{2\pi} 
\cosh(\theta) \ln\left(  1 + e^{-|m|R\cosh(\theta)} \right)\, ,
\ee
\be
E_{0}^{({\rm R})}=\frac{m^2 R}{4\pi}\ln\left( \frac{|m|}{\mu}\right) -|m|\int\limits_{-\infty}^{\infty} \frac{d\theta}{2\pi} 
\cosh(\theta) \ln\left(  1 - e^{-|m|R\cosh(\theta)} \right)\, .
\ee
Here $\mu$ is a subtraction scale for the logarithmic divergence that arises from the OPE of two energy operators 
and is responsible for the Onsager singularity of free energy.

The theory with an arbitrary value of $m$ can be described in the $m=0$ state space using a Bogolyubov
transformation. The massless fermionic fields which we denote $\chi(z)$, $\bar \chi(\bar z)$ have mode expansions 
\bea
&& \chi(z) = \left( -i\frac{2\pi}{R} \right)^{1/2} \sum_{k\in 1/2+{\mathbb Z}} a_{k}e^{i(x+iy)\frac{2\pi}{R}k} \, , \\
&& \bar \chi(\bar z) = \left( i\frac{2\pi}{R} \right)^{1/2} \sum_{k\in 1/2+{\mathbb Z}} \bar a_{k}e^{-i(x-iy)\frac{2\pi}{R}k} \, .
\eea
The operators $a_{k}, \bar a_{k}, k>0$ are the annihilation operators and $a^{\dagger}_{k}=a_{-k}$, 
$\bar a^{\dagger}_{k}=\bar a_{-k}$,  $k>0$ are 
the corresponding creation operators. Similarly in the Ramond sector the expansions are 
\bea
&& \chi(z) = \left( -i\frac{2\pi}{R} \right)^{1/2} \sum_{n\in {\mathbb Z}} a_{n}e^{i(x+iy)\frac{2\pi}{R}n} \, , \\
&& \bar \chi(z) = \left( i\frac{2\pi}{R} \right)^{1/2} \sum_{n\in {\mathbb Z}} \bar a_{n}e^{-i(x-iy)\frac{2\pi}{R}n}
\eea
where the zero modes satisfy
\be
a_{0}^2 = \frac{1}{2}\, , \qquad \bar a_{0}^2 = \frac{1}{2}\, .
\ee
We choose conventions in which the zero modes act on the spin and disorder primary states according to
\be
a_{0}|\sigma\rangle = \frac{1}{\sqrt{2}}|\mu\rangle \, , \qquad \bar a_{0}|\sigma\rangle = -\frac{i}{\sqrt{2}}|\mu\rangle \, .
\ee
The physical state space ${\cal H}_{0}$ is spanned by the states 
\be
\bar a_{k_1}^{\dagger}\dots \bar a_{k_p}^{\dagger}a_{k_{p+1}}^{\dagger}\dots a_{k_N}^{\dagger}|0\rangle \, , \qquad k_{i}\in 1/2 + {\mathbb Z}\, , \qquad N - \mbox{ even}
\ee
in the NS sector and by the states 
\be
\bar a_{n_1}^{\dagger}\dots \bar a^{\dagger}_{n_q}a^{\dagger}_{n_{q+1}}\dots a_{n_M}^{\dagger}|\sigma \rangle \, , \qquad n_{i}\in {\mathbb Z}, , \qquad M - \mbox{ even}
\ee
 in the R sector. Above $|0\rangle$ stands for the conformal vacuum state.

The Bogolyubov transformation relating the massive and massless theories is 
\be \label{BogNS1}
b_{k}^{\dagger} = f(k)\cdot   \left \{
\begin{array}{l@{\qquad}l}
(\bar a^{\dagger}_{k} - i\displaystyle{\frac{\Delta \omega_k}{m}} a_{k}) \, , \enspace k>0
\\[3ex]
(a_{-k}^{\dagger} + i\displaystyle{\frac{\Delta \omega_k}{m}} \bar a_{-k} )\, , \enspace k<0
\end{array}
\right .
\ee
in the NS sector and 
\be \label{BogR1}
b_{n}^{\dagger} = f(n)\cdot   \left \{
\begin{array}{l@{\qquad}l}
(\bar a^{\dagger}_{n} - i\displaystyle{\frac{\Delta \omega_n}{m}} a_{n}) \, , \enspace n>0
\\[3ex]
(a_{-n}^{\dagger} + i\displaystyle{\frac{\Delta \omega_n}{m}} \bar a_{-n} )\, , \enspace n<0
\end{array}
\right .
\ee
\be \label{BogR2}
b_{0}^{\dagger} = \frac{1}{\sqrt{2}}\cdot   \left \{
\begin{array}{l@{\qquad}l}
( a_{0} + i\bar a_{0}) \, , \enspace m>0
\\[3ex]
(a_{0} - i\bar a_{0} )\, , \enspace m<0
\end{array}
\right . \, , \quad  
b_{0} = \frac{1}{\sqrt{2}}\cdot   \left \{
\begin{array}{l@{\qquad}l}
( a_{0} - i\bar a_{0}) \, , \enspace m>0
\\[3ex]
(a_{0} + i\bar a_{0} )\, , \enspace m<0
\end{array}
\right .
\ee
in the R-sector. In the  above 
\be
\Delta\omega_k = \omega_{k} -  \frac{2\pi |k|}{R} \, , \qquad f(k) = \frac{1}{\sqrt{1 + (\Delta\omega_k/m)^2}}
\ee
where $k$ is a half-integer and similar formulas hold with $k$ replaced by an integer $n$ for the R-sector.

\subsection{The interface between the massless and massive theories}
\label{mass_interface}

 Bogolyubov transformation (\ref{BogR1}), (\ref{BogR2}) is a proper Bogolyubov transformation in the terminology of \cite{Berezin} 
 which means that it is implemented by a unitary operator $U: {\cal H}_{0}\to {\cal H}_{0}$ so that 
 \bea
 U \bar a^{\dagger}_{k} U^{*}  & = & f(k)(\bar a^{\dagger}_{k} - i\displaystyle{\frac{\Delta \omega_k}{m}} a_{k})\, , \\
 U a^{\dagger}_{k}U^{*}&=&f(k)(a_{k}^{\dagger} + i\displaystyle{\frac{\Delta \omega_k}{m}} \bar a_{k} )\, 
 \eea
 and similarly in the R-sector with the zero modes transforming as 
 \be
 U a_{0} U^{*} = a_{0} \, , \qquad U \bar a_{0} U^{*} = {\rm sign}(m)\bar a_{0} \, .
 \ee
 
 The Bogolyubov transformation at hand allows one to embed the states from ${\cal H}_{m}$ into ${\cal H}_{0}$ and vice versa. 
 More precisely the first embedding is realised by means of
  an interface operator 
 \be
 \hat D_{m} : {\cal H}_{0} \to {\cal H}_{m} \, , \qquad \hat D_{m} = {\cal O} U
 \ee
where 
 \be
 {\cal O}\, \bar a^{\dagger}_{k_1} \dots  a^{\dagger}_{k_n}\dots |0\rangle = 
  b^{\dagger}_{k_1} \dots  b^{\dagger}_{-k_n}\dots |0\rangle_{\rm NS}
 \ee
 and similar relations are satisfied in the R-sector with the zero modes transforming as 
  \be
 {\cal O}\, \bar a^{\dagger}_{0} \dots  a^{\dagger}_{0}\dots |\sigma \rangle = 
  \frac{i(b_{0}-b_{0}^{\dagger})}{\sqrt{2}}  \dots  \frac{(b_{0}+b_{0}^{\dagger})}{\sqrt{2}} \dots |0\rangle_{\rm R}\, .
 \ee

  The inverse operator 
 \be
 \hat D_{m}^{-1} : {\cal H}_{m} \to {\cal H}_{0} \, , \qquad \hat D_{m}^{-1} = {\cal O}^{-1} V
 \ee
where $V:{\cal H}_{m} \to {\cal H}_{m}$ is a unitary operator that realises the inverse Bogolyubov transformation. 

A general formula for operators $U$, $V$ for a given Bogolyubov transformation is known (see e.g. formula (5.15) from Berezin's book \cite{Berezin}). 
Using it we find the following 
explicit formulas for the two blocks of  $\hat D_{m}^{-1}$:
\be
\hat D_{m}^{-1} = \left( \begin{array}{cc}
\hat D^{\rm NS} & 0\\
0& \hat D^{\rm R} 
\end{array} \right) \, , 
\ee
\be \label{IONS}
\hat D^{\rm NS} = {\cal N}(\nu)|0\rangle\langle 0|_{\rm NS} \prod_{k>0}\exp\left(  i\frac{\Delta\omega_{k}}{m}(a^{\dagger}_{k}\bar a^{\dagger}_{k}   - b_{-k}b_{k}) 
+ f^{-1}(k)(a^{\dagger}_{k}b_{-k} + \bar a^{\dagger}_{k}b_{k}) \right)\, , 
\ee
\be \label{IOR}
\hat D^{\rm R} = {\tilde {\cal N}}(\nu)\hat \Pi \prod_{n=1}^{\infty}\exp\left(  i\frac{\Delta\omega_{n}}{m}(a^{\dagger}_{n}\bar a^{\dagger}_{n}   - b_{-n}b_{n}) 
+ f^{-1}(n)(a^{\dagger}_{n}b_{-n} + \bar a^{\dagger}_{n}b_{n}) \right)
\ee
where 
\be
\hat \Pi  =   \left \{
\begin{array}{l@{\qquad}l}
|\sigma\rangle \langle 0|_{\rm R}  +  |\mu\rangle \langle 0|_{\rm R}b_{0} \, , \enspace m>0
\\[3ex]
|\sigma\rangle \langle 0|_{\rm R} b_{0} +  |\mu\rangle \langle 0|_{\rm R}\, , \enspace m<0
\end{array}
\right .
\ee
and all creation operators $a_{n}^{\dagger}, \bar a_{m}^{\dagger}$ are understood to act on $|0\rangle$ and $|\sigma\rangle$ from the left. 
The normalisation factors  ${\cal N}(\nu) $, $ {\tilde {\cal N}}(\nu)$ are given in terms of infinite products
\be
{\cal N}(\nu) = \prod_{n=0}^{\infty} f(n+1/2)\, , \qquad {\tilde {\cal N}}(\nu) = \prod_{n=1}^{\infty} f(n) \, .
\ee
where we denoted $\nu = |m|R$ - the dimensionless mass. 

Taking the logarithms of  ${\cal N}(\nu) $, $ {\tilde {\cal N}}(\nu)$ and applying Abel-Plana summation formula we obtain the following 
integral expressions 
\be \label{N1}
{\cal N}(\nu) = \exp\Bigl[\nu\left(\frac{2-\pi}{8\pi}\right) +\frac{\nu}{2\pi}\int\limits_{0}^{1}\frac{\arcsin(x)}{e^{\nu x} +1}dx   \Bigr]\, ,
\ee
\be\label{N2}
\tilde {\cal N}(\nu) = 2^{1/4}\exp\Bigl[\nu\left(\frac{2-\pi}{8\pi}\right) - \frac{\nu}{2\pi}\int\limits_{0}^{1}\frac{\arcsin(x)}{e^{\nu x} -1}dx   \Bigr]\, ,
\ee

The ratio of the normalisation factors $\tilde {\cal N}/{\cal N}$ interpolates between $1$ at $\nu=0$ and $2^{1/4}$ at $\nu=\infty$. 
Below is a plot of 
\be\label{Fnu}
f(\nu) = \frac{\tilde {\cal N}(\nu)}{2^{1/4}{\cal N}(\nu)}
\ee

\begin{center}
\begin{figure}[h]
\centering
\includegraphics[scale=0.55]{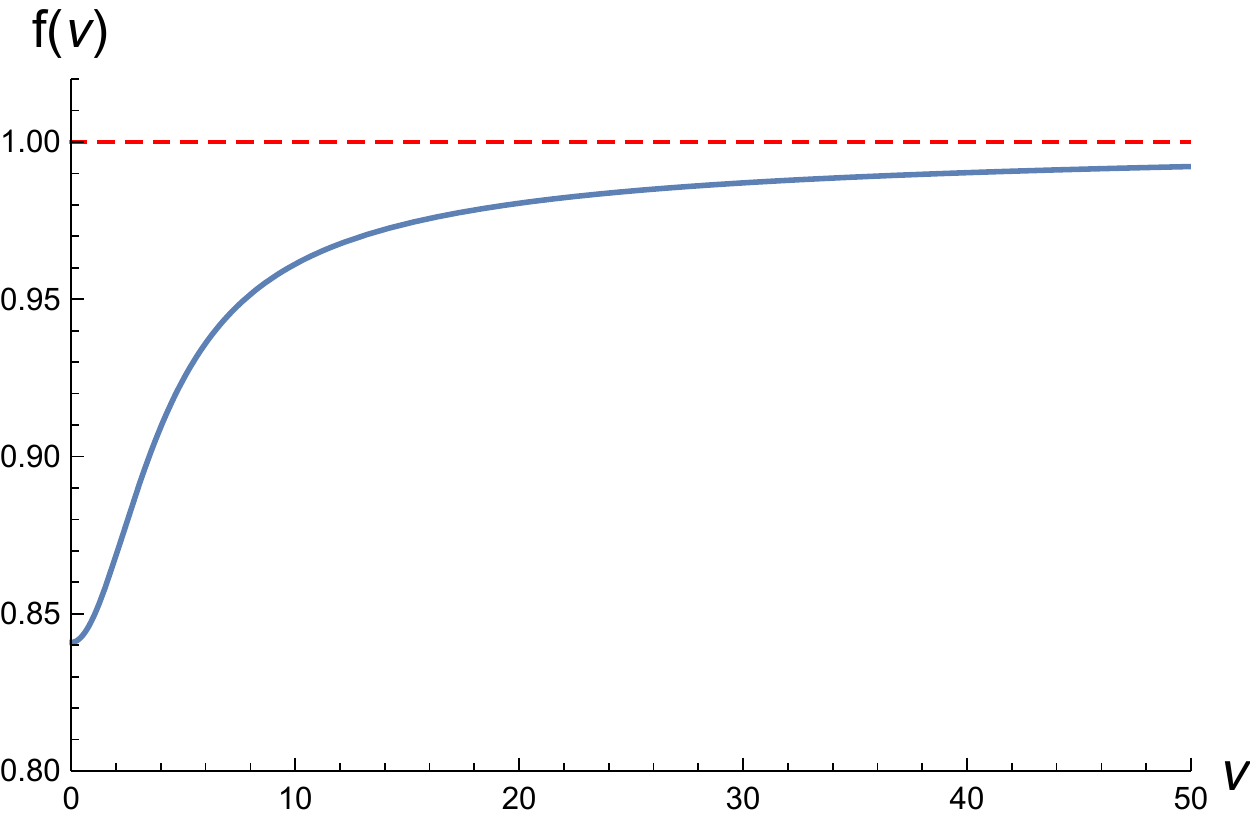}
\caption{The normalised  ratio of NS- and R-sector normalisation factors}
\end{figure}

\end{center}

This function monotonically increases from the value $2^{-1/4}\approx 0.84$ to 1. It passes the value $0.9$  around $\nu=3.5$ and the value 
$0.99$ around $\nu = 40$.

The interface operator can be alternatively described in terms of a local gluing condition at $y=0$:
\be
\chi(x,0)=\psi(x,0)\, , \qquad \bar \chi(x,0)=\bar \psi(x,0) \, . 
\ee
This description is manifestly  local in the $x$-direction.  

\subsection{RG flows and the corresponding conformal boundary conditions}

In the decompactification limit  $\nu \to \infty $ all excited states in the spectrum acquire infinite energy above the vacuum. 
As conformal symmetry is restored we expect the vacuum states of the infinitely massive theory to be described as  conformal 
boundary states in the massless theory ${\cal H}_{0}$.

For $m>0$ the system is in the low temperature ordered phase. As $m$ goes to plus infinity the vacuum becomes doubly degenerate.
This happens exponentially fast in $\nu$. More precisely 
 the difference of vacuum energies in the NS- and R- sectors is 
\be
\Delta e \equiv (E_{0({\rm R})} - E_{0({\rm NS})} )\frac{R}{2\pi} = \nu\!\! \int\limits_{-\infty}^{\infty}\frac{d\theta}{(2\pi)^2} \cosh(\theta) 
\ln\left[ {\rm cotanh}\left( \frac{\nu \cosh(\theta)}{2}\right) \right]
\ee
that for  $\nu\to \infty$ behaves as 
\be
\Delta e \sim \pi^{-3/2}\sqrt{\frac{\nu}{2}}e^{-\nu} \, . 
\ee
 At $m = +\infty$ only a two-dimensional vacuum space is left in the spectrum. 
Using (\ref{IONS}), (\ref{IOR}) we can find the expressions for the asymptotic vacuum states. As they have infinite norm we 
should think of them, as we usually do for boundary states,
as vectors in the dual space ${\cal H}_{0}^{*}$.

The details are as follows. 
The normalisation factors ${\cal N}(\nu)$, $\tilde {\cal N}(\nu)$ each goes to zero due to the Casimir energy 
factors 
\be\label{Casimir}
\exp\Bigl[\nu\left(\frac{2-\pi}{8\pi}\right)\Bigr]
\ee
 in (\ref{N1}), (\ref{N2}). This reflects the fact that the conformal boundary states 
to which the massive vacuum  approaches have infinite norm. Stripping off this vanishing factor 
we obtain the following asymptotic representations of the vacuum sectors of ${\cal H}_{\infty}$ in ${\cal H}_{0}$
\be
|0\rangle\!\rangle_{\rm NS}^{+}\equiv \lim_{m\to +\infty} \exp\Bigl[\nu\left(\frac{\pi-2}{8\pi}\right)\Bigr]
\hat D_{\rm NS}|0\rangle_{\rm NS} = \prod_{n=0}^{\infty}\exp(ia^{\dagger}_{n+1/2}\bar a^{\dagger}_{n+1/2})|0\rangle \, , 
\ee
\be
|0\rangle\!\rangle_{\rm R}^{+}\equiv \lim_{m\to +\infty} \exp\Bigl[\nu\left(\frac{\pi-2}{8\pi}\right)\Bigr]
\hat D_{\rm R}|0\rangle_{\rm R} = 2^{1/4} \prod_{n=1}^{\infty}\exp(ia^{\dagger}_{n}\bar a^{\dagger}_{n})|\sigma \rangle \, .
\ee
These states can be decomposed into conformal Ishibashi states\footnote{The states with an even number of chiral oscillators correspond 
to the Virasoro tower of the identity while the ones with an odd number to that of the energy primary $\epsilon$.}  \cite{Cardy} using  the identification of 
the energy primary state 
\be\label{eps_state}
 |\epsilon\rangle = ia^{\dagger}_{1/2}\bar a_{1/2}^{\dagger}|0\rangle \, . 
\ee
We obtain 
\be
|0\rangle\!\rangle_{\rm NS}^{+} = |0\rangle\!\rangle  + |\epsilon\rangle\!\rangle \, , 
\ee
\be
|0\rangle\!\rangle_{\rm R}^{+} = 2^{1/4} |\sigma \rangle\!\rangle \, .
\ee
We observe that these states already have correct relative normalisations to be combined into  conformal boundary states. By multiplying them by $1/\sqrt{2}$ and 
choosing the appropriate relative phases between the NS- and R- sector components 
we combine them into two Cardy states which represent local boundary conditions corresponding to fixed boundary spin: 
$|+\rangle\!\rangle$, $|-\rangle\!\rangle$ given in (\ref{conf_bcs}).
Hence the RG boundary state that corresponds to the $m\to +\infty$ flow is a superposition 
$$
|{\rm RG}\rangle\!\rangle = |+\rangle\!\rangle \oplus |-\rangle\!\rangle \, .
$$

   For $m<0$ the system is in the high temperature disordered phase. In the R-sector the lowest energy state is the Fock space one-particle 
   excitation  $b_{0}^{\dagger}|0\rangle_{\rm R}$ so that all R-sector states disappear from the spectrum as $m\to -\infty$. The remaining 
   vacuum state is 
\be
|0\rangle\!\rangle_{\rm NS}^{-}\equiv \lim_{m\to -\infty} \exp\Bigl[\nu\left(\frac{\pi-2}{8\pi}\right)\Bigr]
\hat D_{\rm NS}|0\rangle_{\rm NS} = \prod_{n=0}^{\infty}\exp(-ia^{\dagger}_{n+1/2}\bar a^{\dagger}_{n+1/2})|0\rangle \, , 
\ee
which as above can be decomposed into the Ishibashi states as 
\be
|0\rangle\!\rangle_{\rm NS}^{-} = |0\rangle\!\rangle - |\epsilon \rangle\!\rangle 
\ee
that already corresponds to a normalised Cardy boundary state $|F\rangle\!\rangle = |0\rangle\!\rangle - |\epsilon\rangle\!\rangle $
 giving the free boundary condition.
 Thus in this case 
\be
|{\rm RG}\rangle\!\rangle = |F\rangle\!\rangle \, .
\ee  

In the presence of  magnetic field  (the case we take up in the next section) the main quantities we will calculate numerically 
are component ratios of the type (\ref{gen_rat}) discussed in the introduction. 
Specifically for the massive flows of the Ising field theory it is convenient to introduce two ratios: 
\be\label{Ising_ratios}
\Gamma_{\epsilon}^{0} =  \frac{\langle \epsilon|v_{0}\rangle}{\langle 0|v_{0}\rangle} \, , \qquad 
\Gamma_{\sigma}^{0} =  \frac{\langle \sigma |v_{0}\rangle}{\langle 0|v_{0}\rangle}\, .
\ee
Here $|v_{0}\rangle$ stands for the vacuum of perturbed theory as it appears inside the critical Ising state space. 

It is instructive to look at these component  ratios  for the $h=0$ flows where we can get the exact formulas for them. 
The (unnormalised) vacuum  of the perturbed theory (at finite value of $m$) is 
$$|v_{0}\rangle=\hat D^{\rm NS}|0\rangle_{\rm NS} $$
that  always lies in the NS-sector. From (\ref{IONS}) we find  
\be
\Gamma_{\sigma}^{0}=0\, , \qquad \Gamma_{\epsilon}^{0} =  \frac{\langle \epsilon|v_{0}\rangle}{\langle 0|v_{0}\rangle} = \frac{\Delta\omega_{1/2}}{m}= {\rm sign}(m) g(\nu)
\ee
where 
\be\label{gnu}
g(\nu)= 
\sqrt{1+ \left(\frac{\pi}{\nu}\right)^2} - \frac{\pi}{\nu}\, .
\ee
The function $g(\nu)$  asymptoticaly goes to 1 very slowly, as a power function. Here is a plot of $g(\nu)$ 
\begin{center}
\begin{figure}[h]\label{g_fig}
\centering
\includegraphics[scale=0.55]{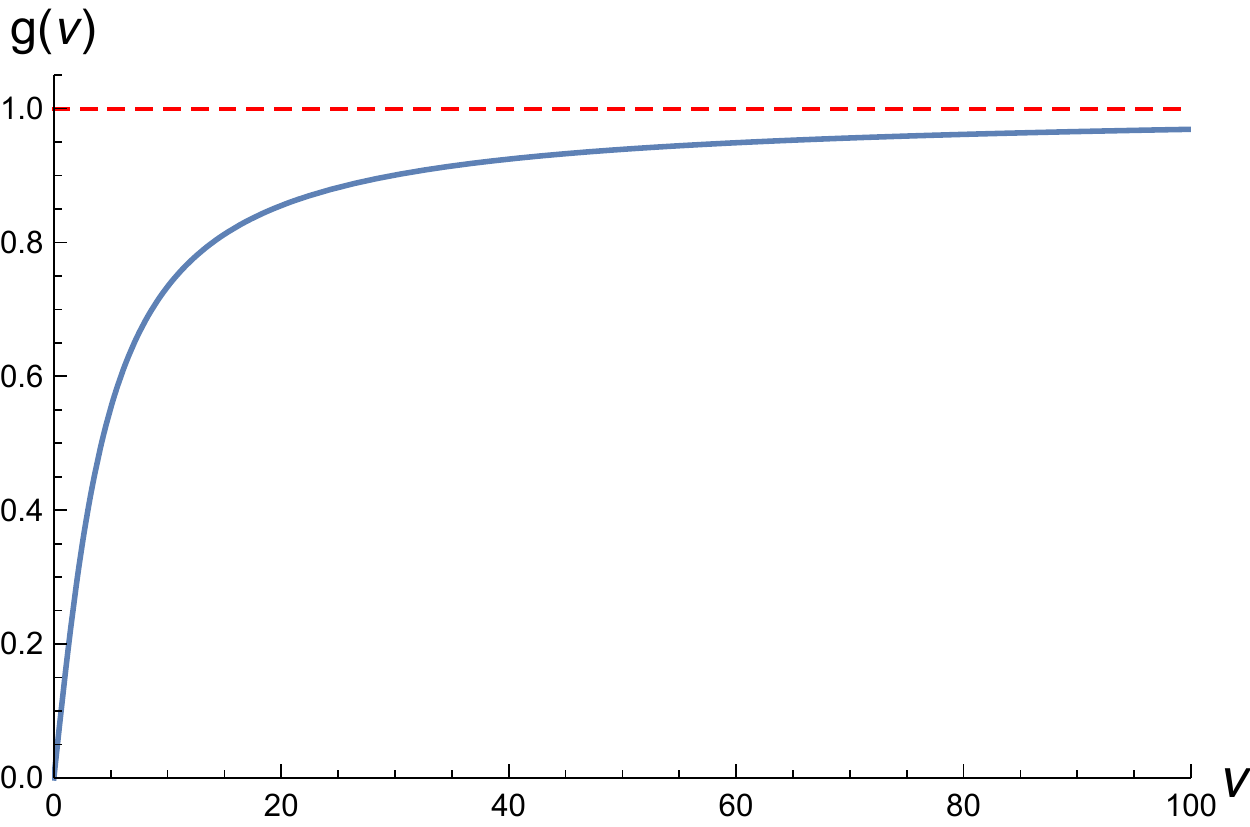}
\caption{The component ratio $\Gamma_{\epsilon}^{0}$ for zero magnetic field, $m>0$}
\end{figure}
\end{center}
This function monotonically increases and becomes larger than $0.9$  past $\nu=30$ and larger than $0.99$ past $\nu=313$.
These values of the dimensionless mass are very large. This  shows that it may be much harder to read 
off the component ratios from TCSA numerics than say the energy spectrum for which the convergence is typically exponential. 
The further we have to go into the infrared 
the greater the TCSA errors can be for a fixed truncation level.  We will see that while the perturbative corrections grow the numerics  
demonstrates that the total truncation error remains bounded.

For $m\to +\infty$ the first excited vector remains in the spectrum. It is $|v_{1}\rangle=\hat D^{\rm R}|0\rangle_{\rm R} $ 
for which we can define a ratio of components   
\be
\left| \frac{\langle \sigma |L_{1}\bar L_1| v_{1}\rangle}{\langle \sigma |v_{0}\rangle}\right| =\frac{1}{8} g(\nu) \, 
\ee
that again converges as slowly as $\Gamma_{\epsilon}^{0}$.

\section{TCSA for Ising field theory}
\setcounter{equation}{0}

We would like now to consider a generic perturbation of the critical Ising model with $m\ne 0 $ and $h\ne 0$. 
For real-valued $h\ne 0$ the system flows to a trivial theory with a single vacuum. We can study numerically the 
ratios $\Gamma_{\epsilon,\sigma}^{0}$ using  TCSA. 

We work in the massless fermion physical space ${\cal H}_{0}$. 
The perturbed Euclidean action of the Ising field theory is given in (\ref{IFT}).
Let us choose an orthonormal basis  in ${\cal H}_{0}$ described in section \ref{preliminaries}. In the NS sector we choose 
\be\label{basis1}
|\bar k_1, \dots \bar k_{p}; k_{1}, \dots k_{q}\rangle_{\rm NS} = 
\bar a_{\bar k_1}^{\dagger}\dots \bar a_{\bar k_p}^{\dagger}a_{k_{1}}^{\dagger}\dots a_{k_q}^{\dagger}|0\rangle 
\ee
where 
\be
k_{i}, \bar k_{j} \in 1/2 + {\mathbb Z}\, , \quad    k_i>0\, , \bar k_j > 0 \, , \quad p+q - \mbox{ even}
\ee
and 
\be
\bar k_1 > \bar k_2 >\dots > \bar k_p \, , \quad 
k_{1}> k_{2} >\dots > k_{q} \, .
\ee
In the R-sector we choose  
\be\label{basis2}
|\bar n_1, \dots \bar n_{p}; n_{1}, \dots n_{q}\rangle_{\rm R} =
\bar a_{\bar n_1}^{\dagger}\dots \bar a^{\dagger}_{\bar n_p}a^{\dagger}_{n_{1}}\dots a_{n_q}^{\dagger}|\sigma \rangle \,  , \qquad p + q - \mbox{ even}
\ee
and 
\be\label{basis3}
|\bar n_1, \dots \bar n_{p}; n_{1}, \dots n_{q}\rangle_{\rm R} =
\bar a_{\bar n_1}^{\dagger}\dots \bar a^{\dagger}_{\bar n_p}a^{\dagger}_{n_{1}}\dots a_{n_q}^{\dagger}|\mu \rangle \,  , \qquad p + q - \mbox{ odd}\, .
\ee
Here we assume 
\be
\qquad n_{i}, \bar n_j \in {\mathbb Z}\, , \quad n_{i}>0, \bar n_j>0
\ee
and 
\be
\bar n_1 > \bar n_2 >\dots > \bar n_p \, , \quad 
n_{1}> n_{2}> \dots > n_{q} \, .
\ee
We are going to work in a finite-dimensional truncated space ${\cal H}_{0}^{\rm tr}$  is spanned by  
(\ref{basis1}), (\ref{basis2}), (\ref{basis3})  satisfying the 
following additional constraints  
\be
\sum_{i=1}^{p}\bar k_i = \sum_{i=1}^{q}k_i \le n_c \, , \qquad \sum_{i=1}^{p}\bar n_i = \sum_{i=1}^{q}n_i \le n_c
\ee
where $n_c$ is the integer that sets the  level truncation\footnote{This means that $n_c$ truncates the total Virasoro weight. 
Note that in some 
TCSA schemes the truncation is done in the level of descendants only rather than in the total weight.}. 
We choose the space ${\cal H}_{0}^{\rm tr}$ to contain only zero spin vectors that is relevant for identifying RG boundary states.

In the table below we give dimensions of truncated spaces for a range of values of $n_{c}$. In the last row 
we give dimension of the truncated subspace of diagonal states - this one is spanned by states of the form 
(\ref{basis1}) with $p=q$ and $k_i=\bar k_i$, $i=1, \dots p$ in the NS sector and similarly 
by  states (\ref{basis2}), (\ref{basis3}) with $n_{i}=\bar n_i$ in the R-sector. These states are of particular interest because 
the conformal Ishibashi states are linear combinations of  such states. 

\begin{center}
\begin{figure}[h!] 
\centering
\begin{tabular}{|p{4cm}|c|c|c|c|c|c|c|c|c|}
\hline
$n_c$ & 8&10&11&12&13&14&16&17&18\\
\hline
 dimension of truncated NS space& 91&226&354&556&844&1296&2838&4139&6069\\
 \hline
 dimension of truncated R space &97&261&405&630&954&1438&3191&4635&6751\\
 \hline 
 Total dimension of  truncated space&188&487&759&1186&1798&2734&6029&8774&12820\\
 \hline
 Dimension of diagonal subspace&58&99&127&162&204&256&393&482&590\\
 \hline
 \end{tabular}
\caption{Dimensions of truncated spaces  } \label{table}
\end{figure}
\end{center}

In operator quantisation the matrix elements of $\sigma$ factorise into holomorphic and antiholomorphic factors
\bea\label{sigma_matrix}
&& {}_{\rm NS}\langle \bar k_1, \dots \bar k_p,  k_1 \dots  , k_{q}|\sigma(0,0)|\bar n_1,\dots \bar n_r, n_1,\dots , n_{s} \rangle_{\rm R} \nonumber \\ 
&&  = 
\left(\frac{2\pi}{R}\right)^{1/8}(-1)^{q(q-1)/2 + s(s-1)/2 + r(s+1)} G(\bar k_1, \dots \bar n_1, \dots  ,\bar n_{r}  )  \nonumber \\
 && \times G(k_1, \dots , k_q, n_1, \dots , n_s ) 
\eea
where 
\bea \label{G}
&& G( k_1, \dots  k_p,n_1, \dots  ,n_{r}  ) = \prod_{j=1}^{p}  g_{\rm NS}( k_j) \prod_{i=1}^{r}g_{\rm R}( n_i) \left(\prod_{1\le i<j\le p} \frac{ k_i -  k_j}{ k_i +  k_j} \right)
\nonumber \\
&& \times  \left(\prod_{1\le i<j \le q} \frac{ n_i -  n_j}{ n_i +  n_j} \right)\left(\prod_{1\le i \le p; 1\le j\le r} 
\frac{k_i + n_j}{k_i - n_j}\right)
\eea
where 
\be
 g_{\rm NS}(k) = \frac{\Gamma(k)}{\sqrt{2\pi}\Gamma(k+1/2)}\, , \qquad 
g_{\rm R}(n) =  \frac{\Gamma(n+1/2)}{\sqrt{2\pi}\Gamma(n)}
\ee
are the massless leg factors. The antiholomorphic factor $G(\bar k_1, \dots \bar n_1, \dots  ,\bar n_{r}  )$ is given by the same formula as (\ref{G}) 
with $n_i, k_j$ mode numbers replaced by $\bar n_i, \bar k_j$.

The RG trajectories of the Ising field theory (\ref{IFT}) are labeled by a dimensionless parameter 
\be
y=\frac{m}{|h|^{8/15}}
\ee
for which we use the same notation as \cite{FZ}. We also use a dimensionless distance scale 
\be
r=R|h|^{8/15} \, .
\ee
In terms of these two scaling variables the dimensionless mass can be expressed as 
\be
\nu = R|m| = |y| r \, .
\ee
The  TCSA Hamiltonian  is 
\be\label{H_TCSA}
H_{\rm TCSA} = \frac{2\pi}{R}\Bigl[  H_{0} +  r^{15/8} (2\pi)^{-7/8} B   +  \frac{y r}{2\pi} M \Bigr] \, .
\ee
Here $H_{0}$, $B$ and $M$ are matrices not containing any dimensional parameters or scaling variables $y$, $r$. They are defined as 
follows. We have 
\be
H_{0} = \sum_{n=0}^{\infty} (n + 1/2)[a^{\dagger}_{n+1/2}a_{n+1/2} + \bar a^{\dagger}_{n+1/2}\bar a_{n+1/2}] - \frac{1}{24} 
\ee
in the NS-sector and 
\be
H_{0} = \sum_{n=0}^{\infty} n[ a^{\dagger}_{n}a_{n} + \bar a^{\dagger}_{n}\bar a_{n}] + \frac{1}{12}
\ee
in the R-sector.
The matrix $B$ has the same matrix elements as (\ref{sigma_matrix}) with the factor $\left(\frac{2\pi}{R}\right)^{1/8}$ removed. 
The matrix $M$ is the matrix for the operator $i\psi\bar \psi$ in the basis given by (\ref{basis1}), (\ref{basis2}), (\ref{basis3}).

If $E(r)$ is an eigenvalue of $H_{\rm TCSA}$ we also define  dimensionless eigenvalues 
\be
e(r) = \frac{R}{2\pi} E(r)
\ee
and 
\be
\tilde E(r) \equiv \frac{E(r)}{|h|^{8/15}} = \frac{2\pi e(r)}{r} \, . 
\ee
For a flow to Yang-Lee fixed point the differences of the eigenvalues $e(r)$ interpolate between the differences of scaling dimensions at the fixed points (see a detailed discussion 
in section \ref{ycr_sec}).

To identify RG boundaries for the massive flows we are going to calculate numerically the component ratios $\Gamma^{0}_{\epsilon}$, 
$\Gamma^{0}_{\sigma}$ defined in (\ref{Ising_ratios}). Also later we will consider the RG flow to Yang-Lee fixed point which occurs at imaginary 
magnetic field. In that case we will also look at the ratios 
\be \label{Ising_ratios2}
\Gamma_{\epsilon}^{i} =  \frac{\langle \epsilon|v_{i}\rangle}{\langle 0|v_{i}\rangle} \, , \qquad 
\Gamma_{\sigma}^{i} = \frac{\langle \sigma |v_{i}\rangle}{\langle 0|v_{i}\rangle} \, 
\ee
corresponding to the $i$-th excited energy eigenstates $|v_{i}\rangle$, $i=1,2,\dots$. 

TCSA results at finite $n_{c}$ of course differ from the exact renormalised QFT results  at $n_{c}=\infty$. 
Some of the corrections correspond to redefining the coupling constants and can be taken into account using a certain  RG scheme \cite{GW}, 
while there are also non-local corrections \cite{HRvR}, \cite{RV}.  The redefined coupling constants corresponding to the RG scheme 
of \cite{GW} can be approximated by  running couplings that depend on continuous $t=1/n_{c}$. 
Let $\tau=yr={\rm sign}(y)\nu$.
The RG equations for the running couplings $\tau(t)$, $r(t)$ then read 
\be
\frac{dr}{dt}=-Ar\tau \, , \qquad \frac{d\tau}{dt}=-sBt^{7/4}r^{15/4} 
\ee
where 
\be 
A = \frac{2}{15\pi^2}\approx 0.01 \, , \qquad B=\frac{1}{4(2\pi)^{3/4}\Gamma^{2}(-3/8)}\approx 0.004 \, 
\ee
and $s=1$ for real magnetic field and $s=-1$ for the imaginary one.
These equations are to be solved with the bare couplings from (\ref{H_TCSA}) taken as initial condition at $t=0$. 
Both couplings change quite slowly. For example for $y=y(0)=-3$ and $s=-1$ (close to the Yang-Lee trajectory) we find numerically 
for $n_c =12$ that for $r=r(0)=12$ the effective values are $r(1/12)\approx 12.49$,  $\tau(1/12)\approx -35.98$.
The mass coupling $\tau$ 
changes particularly slowly.

The method for dealing with non-local perturbative corrections to energy eigenvalues has been worked out  in \cite{HRvR}, \cite{RV}.
One strategy to implement such corrections in practice is by varying $n_{c}$ and by fitting the variations in numerical results to 
combinations of negative powers of    $n_{c}$. For the Ising field theory this has been done in \cite{LT_Potts}, \cite{Takacs_quenches}
(in particular see Appendix B in \cite{Takacs_quenches} where RG corrections and fittings in both TCSA and TFFSA are discussed 
in detail). Perturbative truncation corrections to eigenvectors (which would be of interest for calculating the component 
ratios $\Gamma^{i}_{\epsilon, \sigma}$) have not been so far systematically investigated. Moreover as  
Figure \ref{g_fig} shows one can expect a very slow approach of $\Gamma^{i}_{\epsilon, \sigma}$ to their IR asymptotics that does not 
make perturbative corrections useful for finding their IR values.  Nevertheless our TCSA  numerical results, that we present in the next section, 
show that the component ratios for real magnetic field asymptotically  approach constant values with (non-perturbative) truncation errors 
staying bounded.
Since the running couplings corrections are really tiny and do not change any qualitative conclusions and also to make direct comparison 
with the results of \cite{FZ}, \cite{Zam1}, \cite{Zam2} possible, everywhere in the paper we present uncorrected (raw) TCSA and TFFSA data.


\section{Massive flows. Real magnetic field.}
\setcounter{equation}{0}
\subsection{TCSA data }\label{TCSA_section}

The component ratios we are going to find numerically are $\Gamma^{0}_{\epsilon}$ and $\Gamma^{0}_{\sigma}$ defined 
in (\ref{Ising_ratios}). The sign of $\Gamma^{0}_{\sigma}$  is correlated with the sign of $h$. For simplicity we will always 
choose $h$ so that $\Gamma^{0}_{\sigma}$ is positive. 

When a magnetic field perturbation is present it is natural to expect the $|\sigma\rangle\!\rangle$ component to be present in 
the RG boundary state. As we don't expect any vacuum degeneration we expect the RG boundary state to be $|+\rangle\!\rangle$ or 
$|-\rangle\!\rangle$ depending on the sign of the coupling $h$. This means that $\Gamma_{\epsilon}^{0}(r)$ should approach 1 and 
$\Gamma_{\sigma}^{0}(r)$ should approach $2^{1/4}\approx 1.18921$. Indeed this is what we observe in TCSA numerics presented below 
for a range of values of $y$. The approach to asymptotic values is always quite slow - of the type the exact solution for $\Gamma_{\epsilon}^{0}$ 
demonstrates in the $h=0$ case. On the graphs for $\Gamma_{\sigma}^{0}$ below the red dashed line corresponds to the  value $2^{1/4}$.
The results presented below are not very sensitive to the the cutoff level $n_c$ (see a more detailed discussion below). For illustration we chose $n_c=12$. 
\begin{center}
\begin{figure}[h!]

\begin{minipage}[b]{0.5\linewidth}
\centering
\includegraphics[scale=0.5]{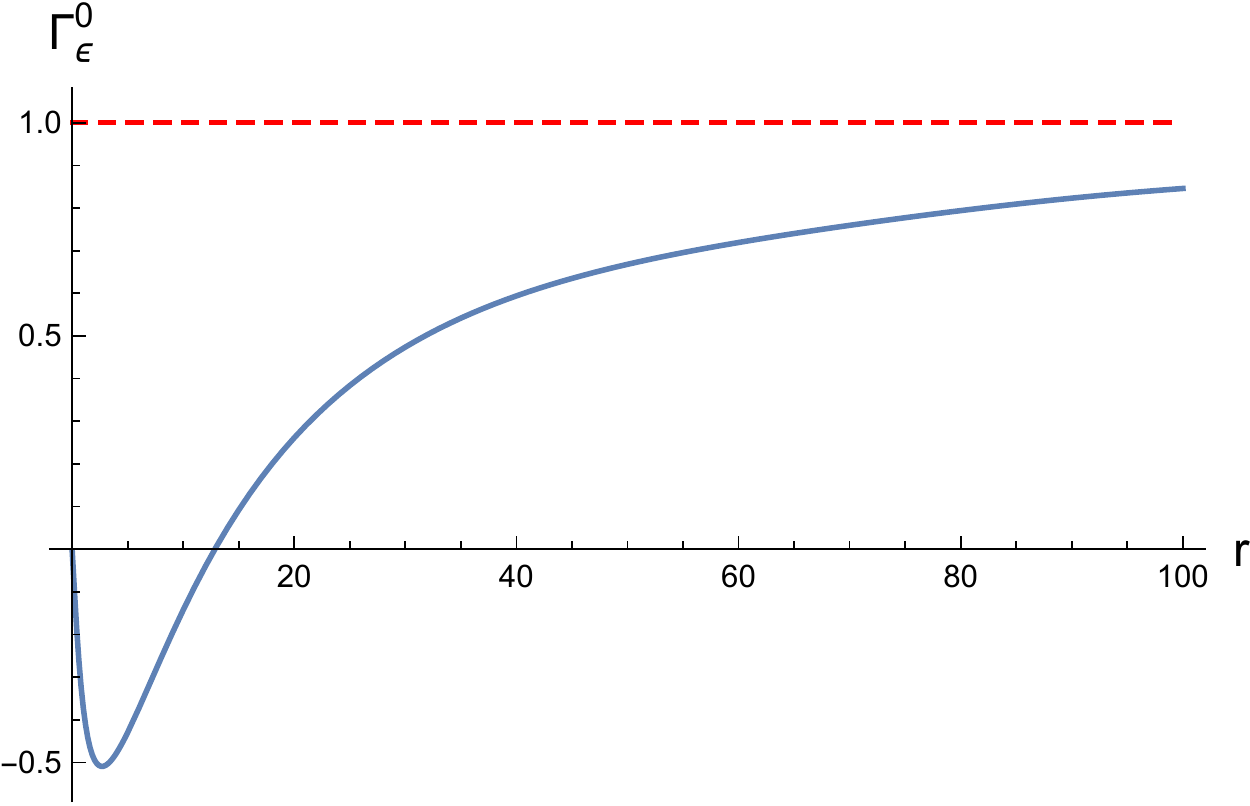}
\end{minipage}%
\begin{minipage}[b]{0.5\linewidth}
\centering
\includegraphics[scale=0.5]{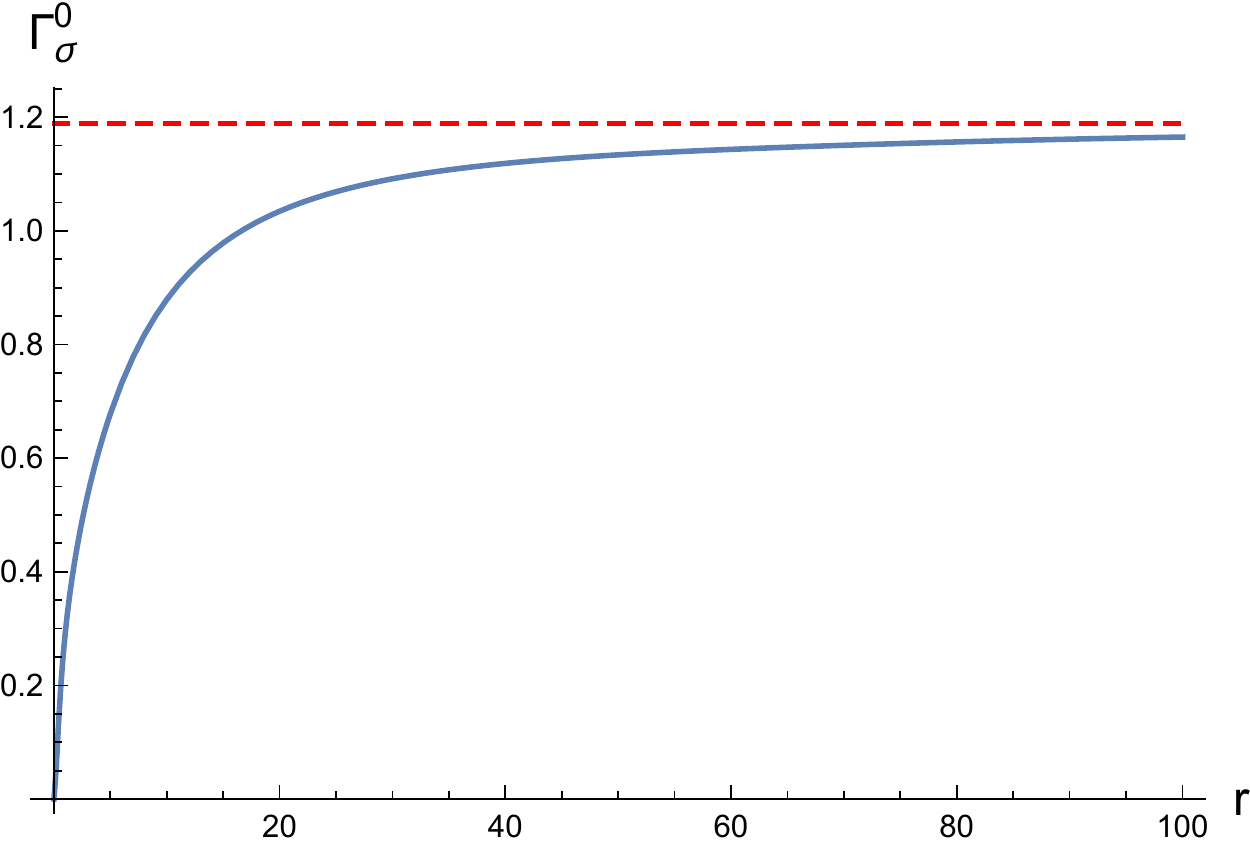}
\end{minipage}
\caption{$y=-3$}
\end{figure}
\end{center}

\begin{center}
\begin{figure}[h!]

\begin{minipage}[b]{0.5\linewidth}
\centering
\includegraphics[scale=0.5]{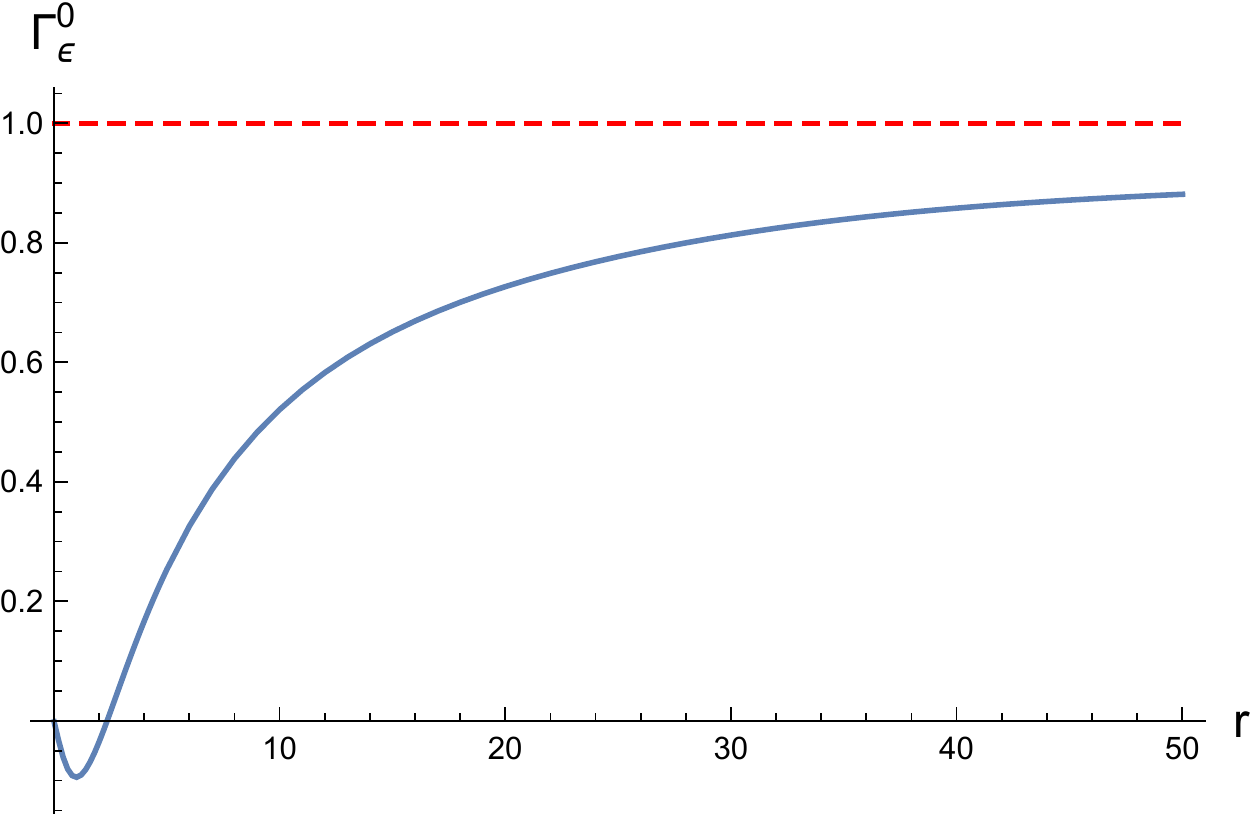}
\end{minipage}%
\begin{minipage}[b]{0.5\linewidth}
\centering
\includegraphics[scale=0.5]{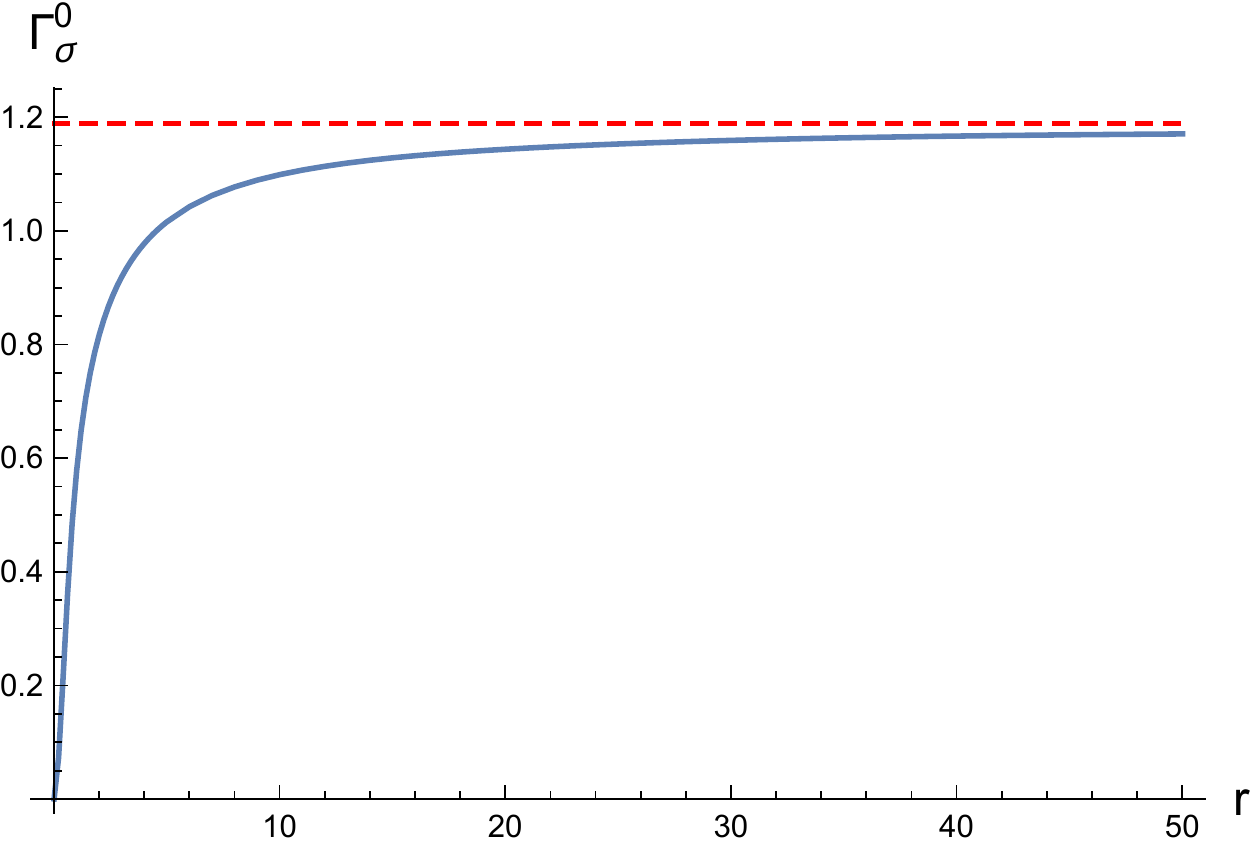}
\end{minipage}
\caption{$y=-1$}
\end{figure}
\end{center}

\begin{center}
\begin{figure}[h!]

\begin{minipage}[b]{0.5\linewidth}
\centering
\includegraphics[scale=0.5]{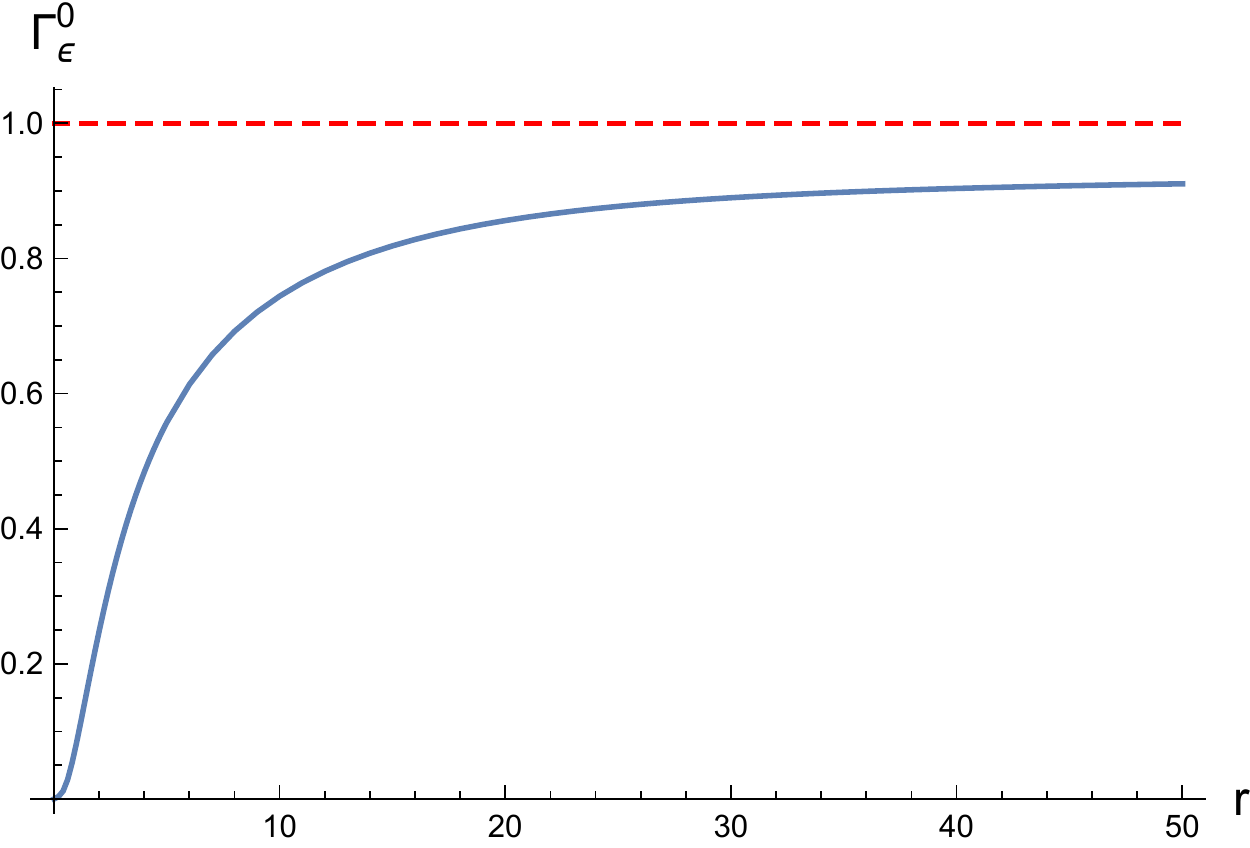}
\end{minipage}%
\begin{minipage}[b]{0.5\linewidth}
\centering
\includegraphics[scale=0.5]{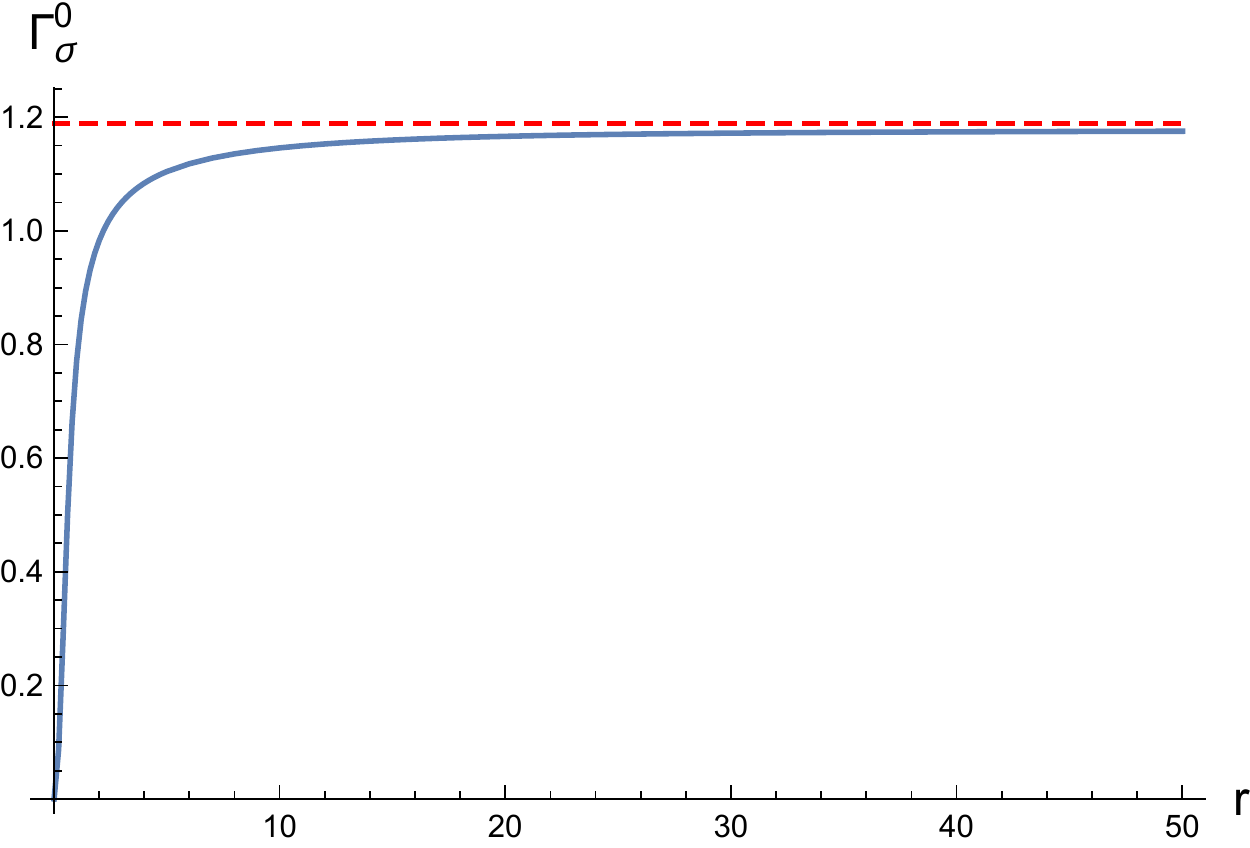}
\end{minipage}
\caption{$y=0.1$}
\end{figure}
\end{center}

\begin{center}
\begin{figure}[H] 

\begin{minipage}[b]{0.5\linewidth}
\centering
\includegraphics[scale=0.5]{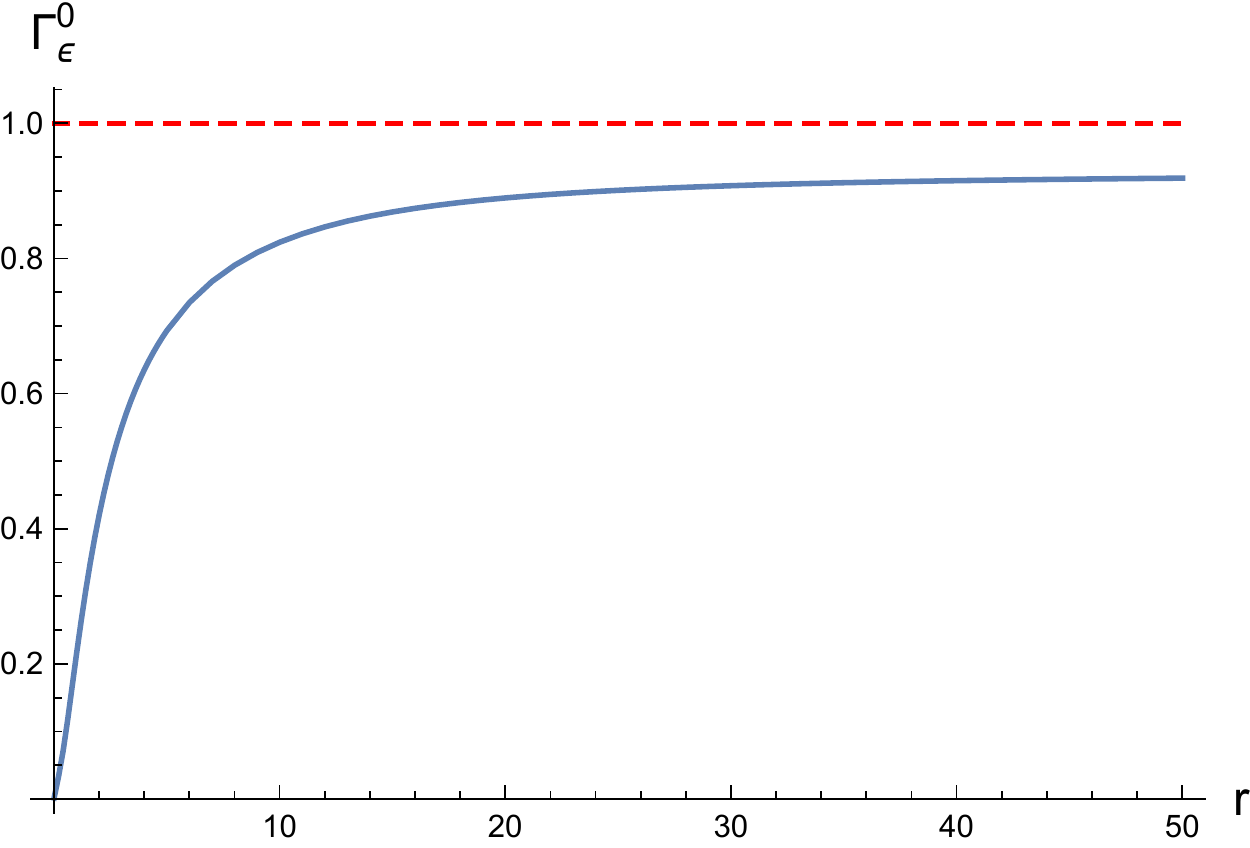}
\end{minipage}%
\begin{minipage}[b]{0.5\linewidth}
\centering
\includegraphics[scale=0.5]{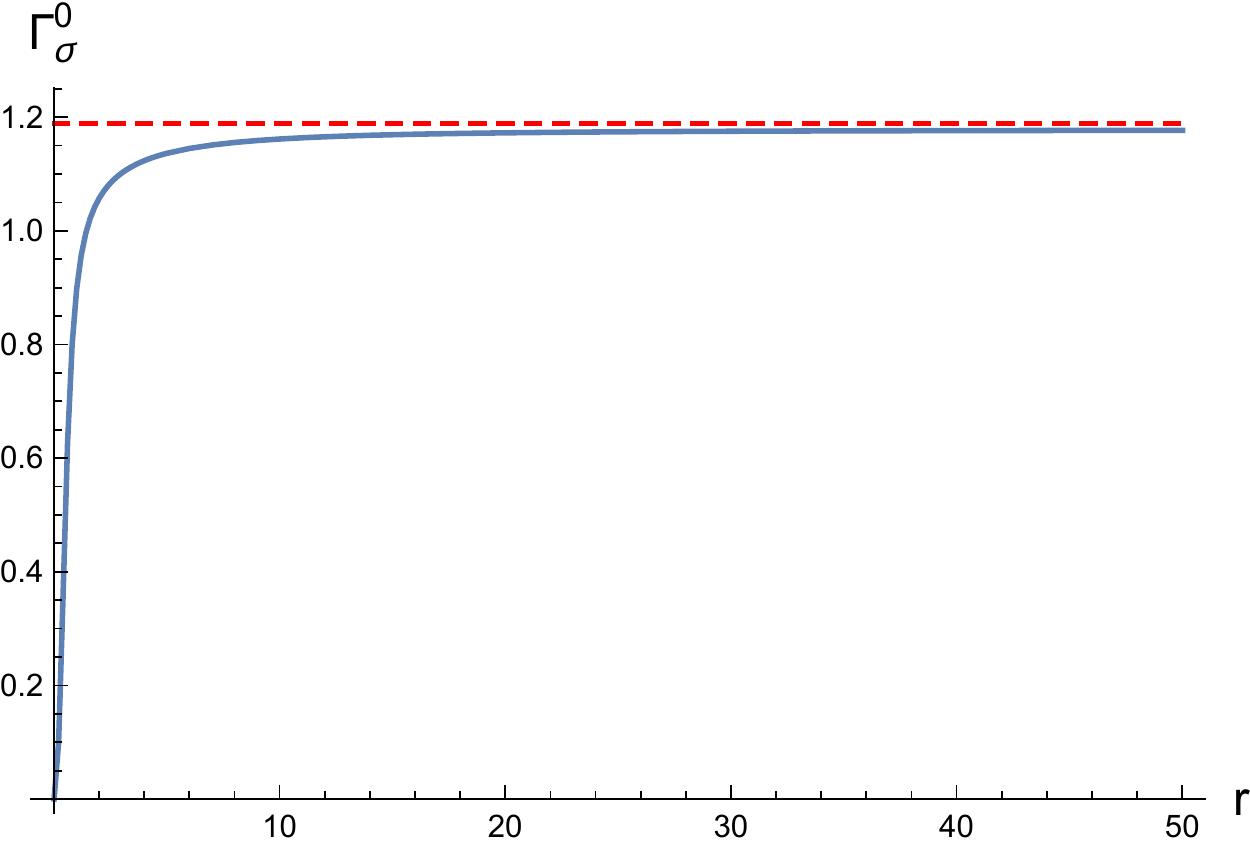}
\end{minipage}
\caption{ $y=1$}
\end{figure}

\end{center}

\begin{center}
\begin{figure}[h!]

\begin{minipage}[b]{0.5\linewidth}
\centering
\includegraphics[scale=0.5]{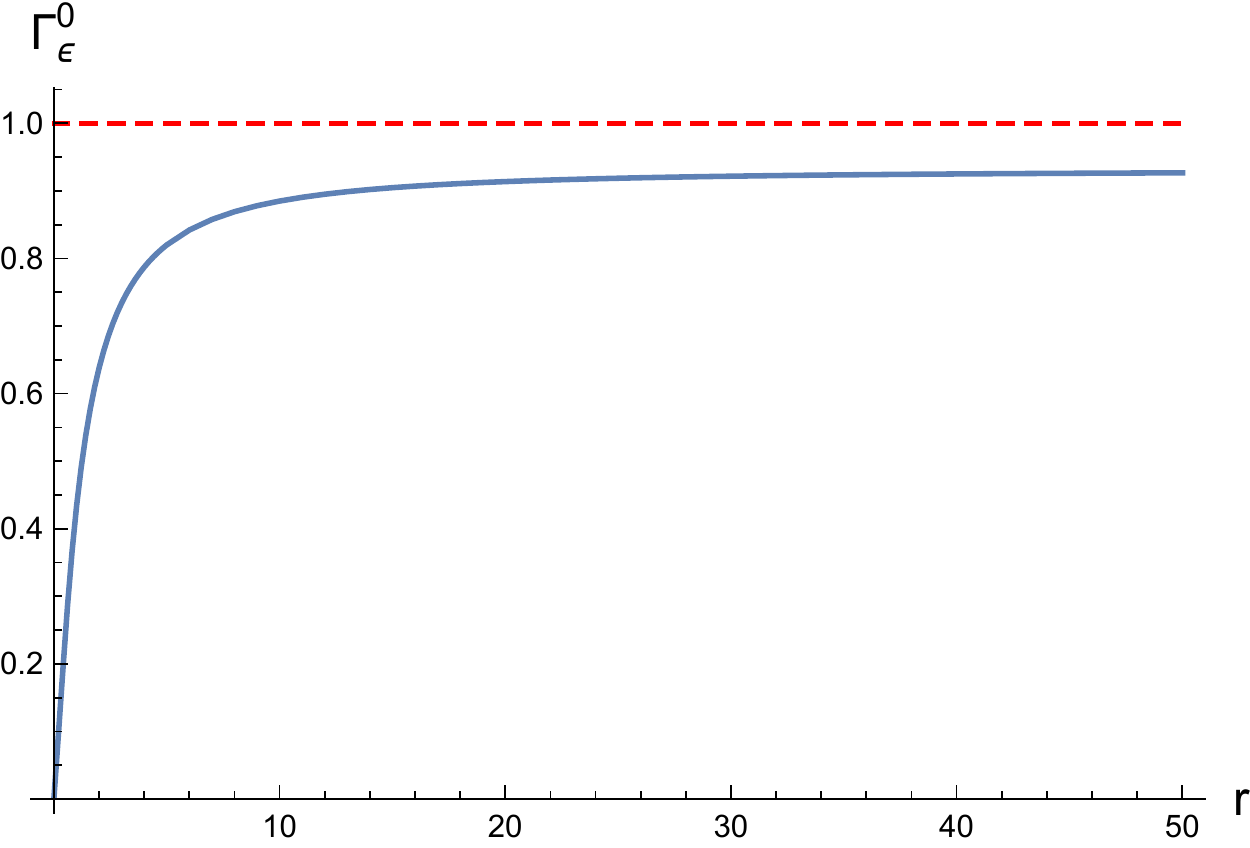}
\end{minipage}%
\begin{minipage}[b]{0.5\linewidth}
\centering
\includegraphics[scale=0.5]{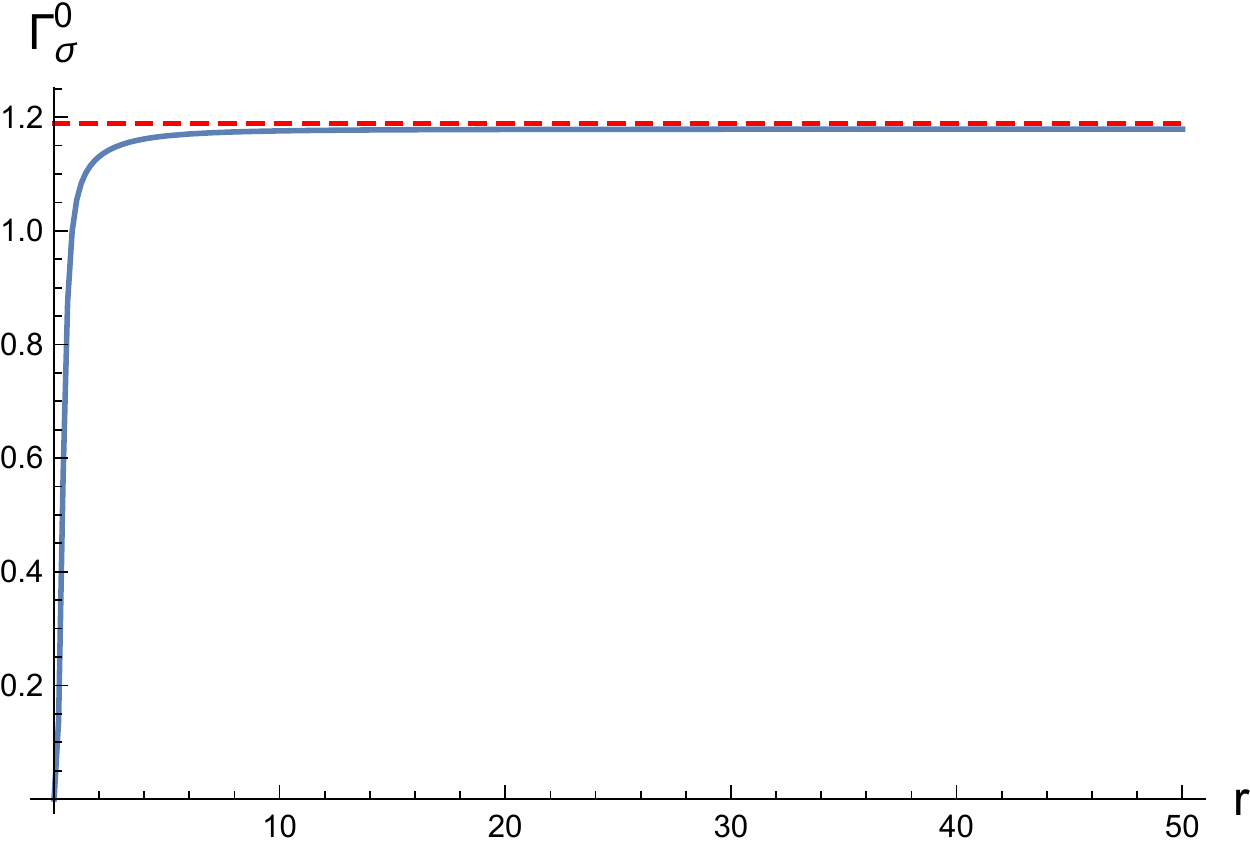}
\end{minipage}
\caption{\normalsize { $y=3$}}
\end{figure}
\end{center}


We observe that the ratio $\Gamma_{\sigma}^{0}$ converges to its expected limiting value with much better accuracy 
than $\Gamma_{\epsilon}^{0}$. While for the values of $r\sim 50$ we have  $\Gamma_{\sigma}^{0}\approx 1.17$ for $\Gamma_{\epsilon}^{0}$ 
 typical asymptotic values are near $0.9$.
In the disordered phase, where $y<0$, the mass term dominates at small distances so 
that $\Gamma_{\epsilon}^{0}$ starts out with negative values. The magnetic field takes over at sufficiently large distances with
 $\Gamma_{\epsilon}^{0}$ passing through zero and monotonically increasing to a positive asymptotic value that is less than 1. 
 Thus for $y<0$ the convergence of $\Gamma_{\epsilon}^{0}$ to its asymptotic value is slower than in the ordered phase. 
 
 It should be noted that at the large values of $r$ we considered: $30<r<100$,   the values of $\nu/n_{c}$ and 
$hr^{15/8}/n_{c}$ are large. This means that any perturbative corrections are large in that region. However the 
numerics shows that the perturbative corrections must sum up to a form $f(r)/n_{c}$ where $f(r)$ is a bounded function of $r$. 
The following plot shows how the numerical values of $\Gamma_{\epsilon}^{0}$ taken for $y=1$ change when we change the 
truncation level $n_{c}$.

\begin{center}
\begin{figure}[h!]
\centering
\includegraphics[scale=0.8]{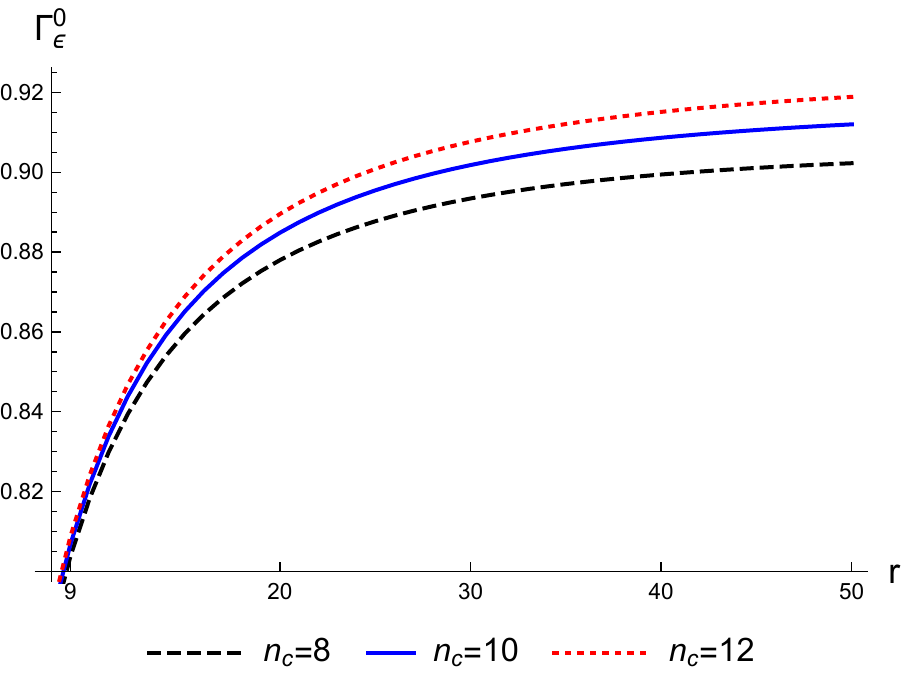}
\caption{The  $\Gamma_{\epsilon}^{0}$ numerics at $y=1$ for different  values of $n_c$}
\end{figure}
\end{center}

It is not clear to us why the total truncation correction remains bounded for such large values of the scale but this is certainly 
crucial for getting reasonable numerical results for the ratios $\Gamma_{\epsilon, \sigma}^{0}$ due to their extremely slow 
asymptotics.

We focused on two particular ratios $\Gamma^{0}_{\epsilon, \sigma}$  for convenience and for practical reasons (these ratios are related to the 
lowest weight components and we expect them to be best approximated by TCSA). But in general we expect all components of the asymptotic 
vacuum vector to arrange themselves according to the components of  Cardy boundary states (or their superposition). 
To get a better measure of that we can plot all vacuum components. 
More precisely let $|i\rangle_{\rm NS}$, $|j\rangle_{\rm R}$ stand for the basis elements in the critical Ising model that correspond to scaling 
fields of dimension $\Delta_{i}^{\rm NS}$, $\Delta_{j}^{\rm R}$ in 
the NS and R-sectors respectively. Up to phases the basis elements are given by formulae   (\ref{basis1}), (\ref{basis2}), (\ref{basis3}). 
We fix the phases by state-operator correspondence (see (\ref{eps_state}) and label these basis vectors in such a way that 
the conformal weights are monotonically non-decreasing: $\Delta_{i}^{\rm NS}\le \Delta_{j}^{\rm NS}$ if $i\le j$ and 
similarly in the R-sector. Let $C^{\rm NS, R}(i)$ be the vacuum vector components relative to this basis:
\be\label{conf_basis}
|0\rangle = \sum_{i} C^{\rm NS}(i)|i\rangle_{\rm NS} + \sum_{j} C^{\rm R}(j)|j\rangle_{\rm R} \, .
\ee
For $n_{c}=12$ we have 556 basis vectors in the NS sector out of which 92 are diagonal. On the graph below all components 
are plotted, out of which only the diagonal ones are clearly visible while the rest of the components are much smaller and on the 
graph they are concentrated on the $i$-axis. We  see that the diagonal components have the same phase and decrease in 
amplitude with increasing conformal weight practically forming a continuous curve\footnote{The basis elements come in batches. 
Vectors in each batch have the same conformal weight. 
On the graph the red numbers mark the first component with the indicated conformal 
weight.}.
  The decrease is a truncation effect. In the untruncated theory we would expect all components to be the same according to 
the composition of the Cardy states (\ref{conf_bcs}).

\begin{center}
\begin{figure}[h!]
\centering
\includegraphics[scale=0.65]{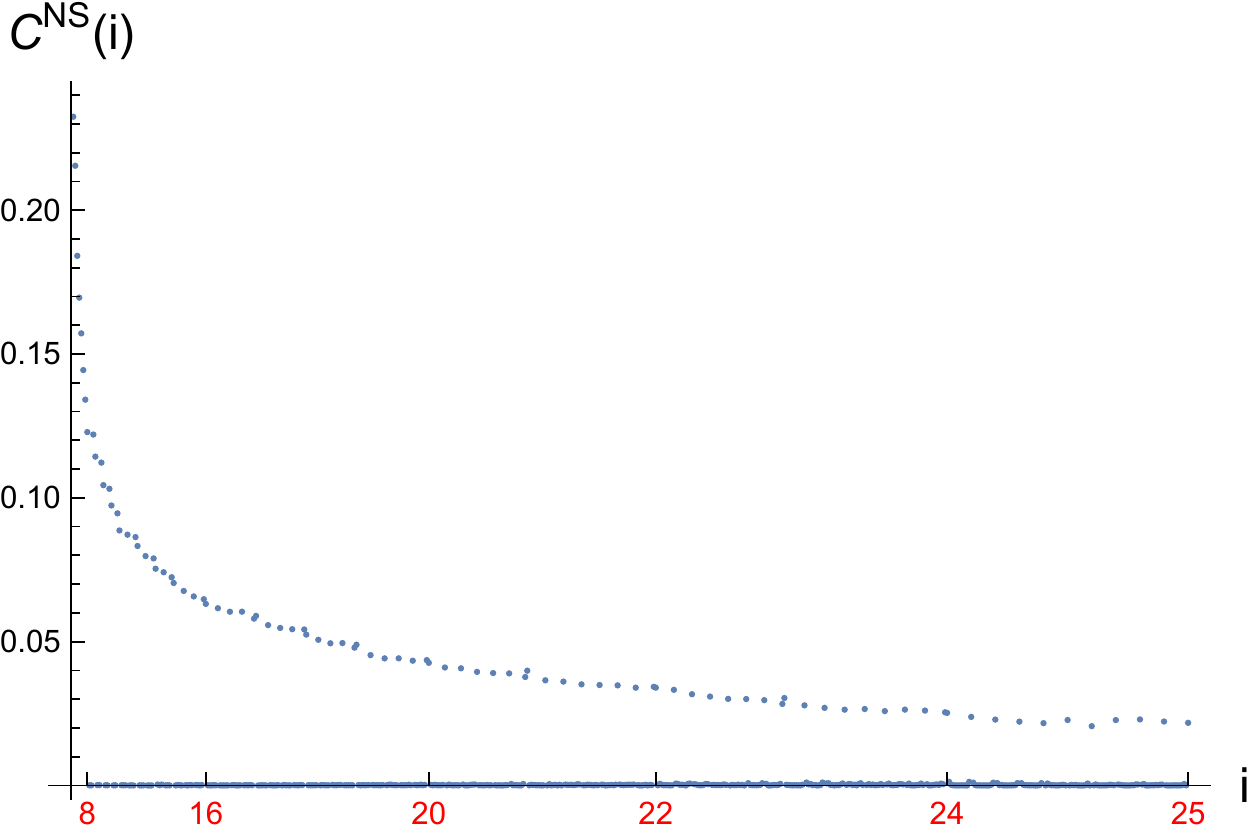}
\caption{Components of the vacuum state in the NS sector for $y=3$, $r=50$, $n_c=12$. The red numbers mark the conformal weights of the components.
Only diagonal components are distinguishable above the $i$-axis.}
\end{figure}
\end{center}

The Ramond components form a similar pattern with the diagonal components decoupled from the much smaller non-diagonal ones. 

\begin{center}
\begin{figure}[h!]
\centering
\includegraphics[scale=0.65]{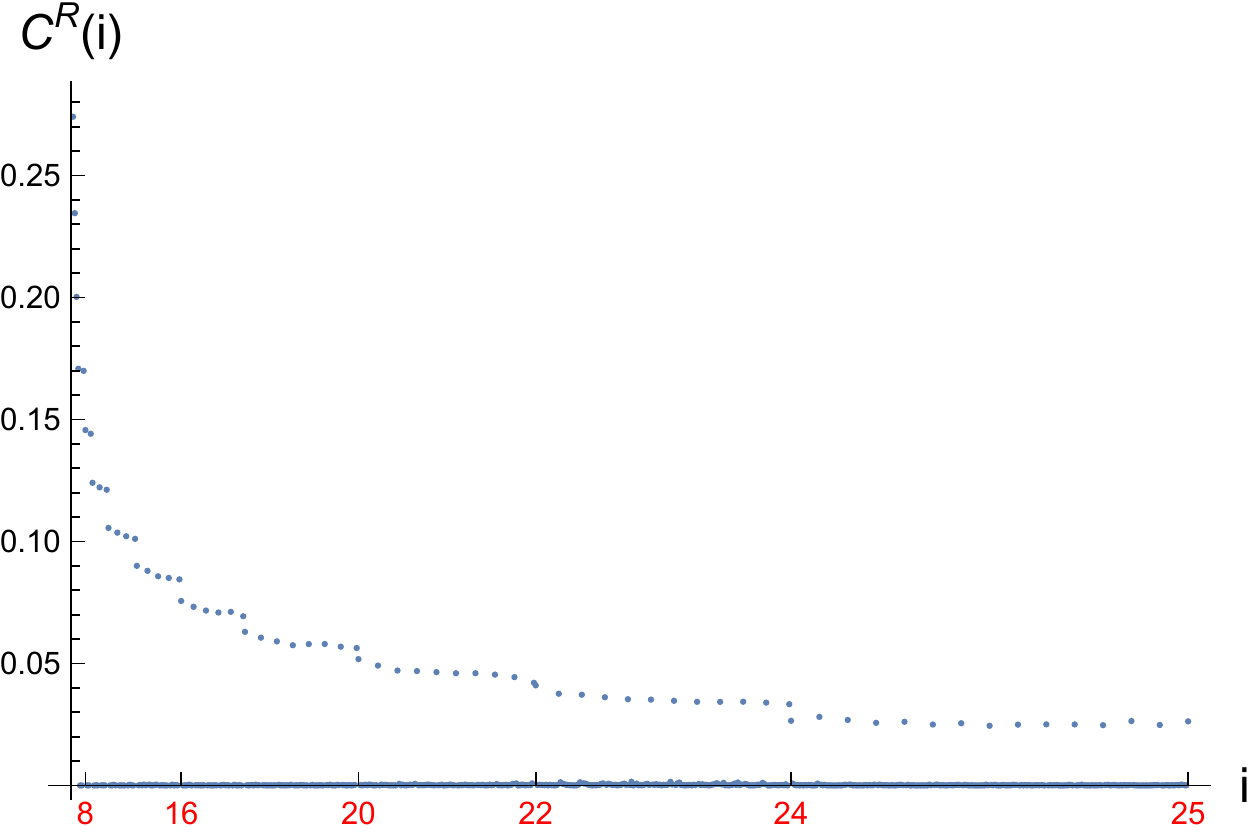}
\caption{Components of the vacuum state in the R sector for $y=3$, $r=50$, $n_c=12$. The red numbers mark the conformal weights of the components.}
\end{figure}
\end{center}

Another way to see the domination of the diagonal states in the vacuum is by calculating the share of the diagonal 
states in the square of the norm:
\be
S=\frac{\sum\limits_{i\in {\rm diagonal}} |C(i)|^2 }{\sum\limits_{i} |C(i)|^2}\, .
\ee
We found that numerically $S$ is above $0.9999$ for a range of $y$, $n_{c}$ and $r$ between $0$ and $50$.
Below is a sample graph
\begin{center}
\begin{figure}[h!]
\centering
\includegraphics[scale=0.65]{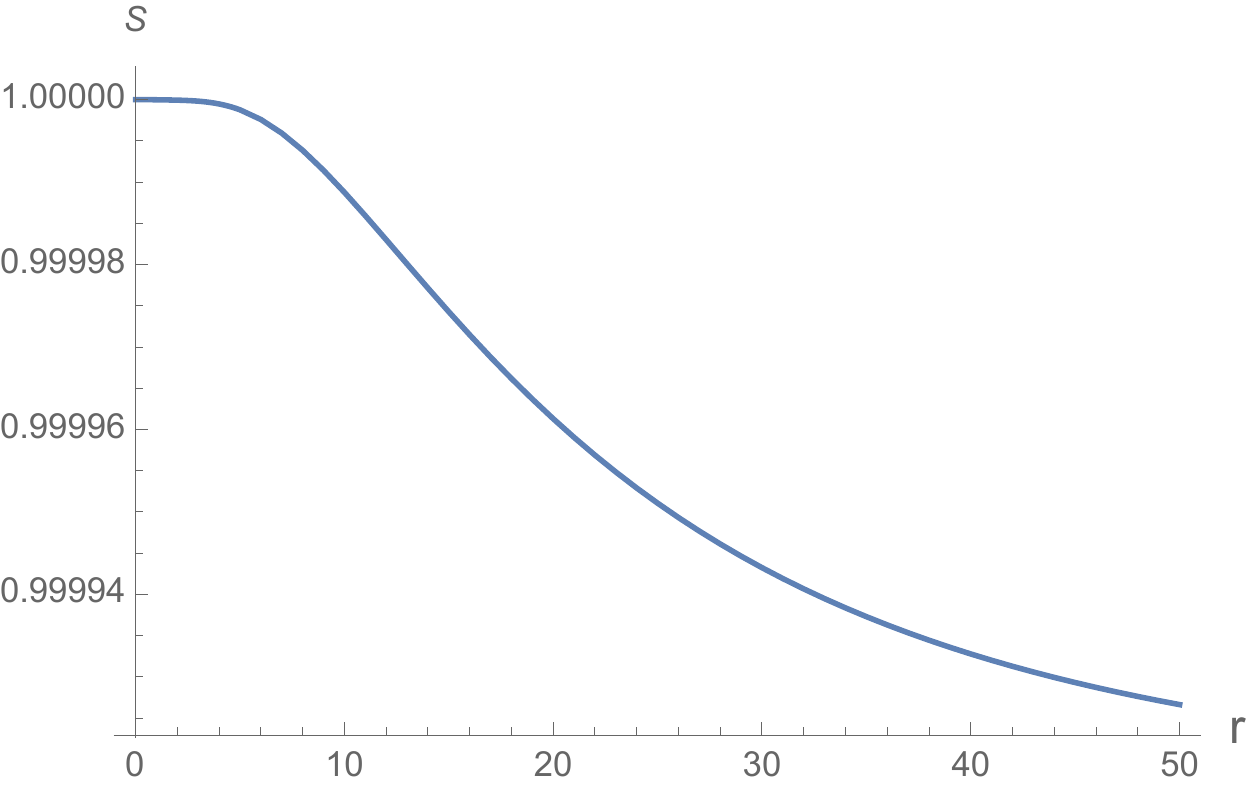}
\caption{Share of the diagonal states in the norm of the vacuum, $y=3$, $n_{c}=12$.}
\end{figure}
\end{center}
As can be seen from the table given on Fig. \ref{table} the number of diagonal states grows significantly slower with $n_c$ than 
the total number of states. It is remarkable that these states dominate the vacuum with such a high proportion. 
Even more unexpectedly we found that the first and second excited states have about the same high proportion of the diagonal states 
in their norm. It remains to be seen whether this observation can be put to use to improve the numerics.

\subsection{Summary}\label{massive_summary}
Our analytical results for the vanishing magnetic field together with the numerical results for measuring
 $\Gamma_{\epsilon}^{0}, \Gamma_{\sigma}^{0}$ are  summarised
 on the  diagram depicted on Fig. \ref{diagram_real}. 
We observe that the two exceptional RG boundaries: $|+\rangle\!\rangle \oplus |-\rangle\!\rangle$, $|F\rangle\!\rangle$, are 
unstable. There are boundary RG flows from each of them that end up either with $|+\rangle\!\rangle$ or with $|-\rangle\!\rangle$. 
These end points are precisely the RG boundary conditions corresponding to the one-dimensional regions joined at the exceptional 
points. This fact has the following physical explanation. Suppose we are at an exceptional point 
that belongs to a submanifold separating higher-dimensional regions labeled by different RG boundaries. Far in the infrared the RG 
interface is almost non-transparent and can be well approximated  by a conformal boundary condition.
 If we add now a small bulk perturbation on the massive side of the interface that moves us away from the separating submanifold 
 this will result in an effective perturbation of the RG  boundary and a subsequent effective RG flow to a new RG boundary condition. 
 (More precisely this is still a bulk plus boundary flow, however in this representation low energy degrees of freedom survive only on the boundary 
 so it is effectively described by a pure boundary flow.)

 In the Ising field theory case we can be more quantitative in describing these effective boundary RG flows that start from the 
 exceptional RG boundaries. These flows are triggered by the $\sigma$-perturbation taken at a large dimensionless 
 mass $\nu$. The matrix elements of $\sigma$ in a massive Fock space are given explicitly in formula (\ref{matrix_mass}).
 In the limit $\nu \to \infty$ all of these matrix elements go to zero or to a finite constant except for the vacuum-vacuum and the vacuum-one-particle 
 ones that diverge.  This means that for $m>0$ effectively the $\sigma$ perturbation at large mass will act as a boundary identity 
 field  that mixes the two vacua, while for $m<0$ the one-particle matrix elements  give rise to a boundary magnetic field perturbation.

\section{Truncated Free Fermion Space Approach and component ratios}
\setcounter{equation}{0}
In this section we outline how the truncated free fermion space approach (TFFSA) invented in \cite{FZ} can be 
used to find numeric approximations to the ratios $\Gamma_{i}$. This method has the advantage over the TCSA 
in treating the mass coupling non-perturbatively and also in having greater  control over the large $r$ asymptotics. 
However it has its own subtleties related  to UV  divergences which we are going to discuss as well.

In TFFSA one uses the massive fermion physical space ${\cal H}_{m}$ described in section \ref{preliminaries}.
We write the Ising field theory Hamiltonian as 
\be\label{Ham2}
H = H_{\rm FF} + h\!\int\limits_{0}^{R}\! \sigma(x,0)\, dx 
\ee
where $H_{\rm FF}$ is the free massive fermion Hamiltonian given in (\ref{HFF1}), (\ref{HFF2}) and the matrix elements 
of the magnetic field perturbation are given by \cite{FZ}
\bea\label{matrix_mass}
&& {}_{\rm NS}\langle k_1, k_2,\dots, k_{N}|\sigma(0,0)|n_1,n_2, \dots , n_{M}\rangle_{\rm R}= i^{\left[\frac{N+M}{2}\right]}\, \bar \sigma S(R)
\prod_{i=1}^{N}\tilde g(\theta_{k_i}) \prod_{j=1}^{M} g(\theta_{n_j})\nonumber \\
&& \times \prod_{1\le i<j<N}\tanh\left(\frac{\theta_{k_i}-\theta_{k_j}}{2}\right)  \prod_{1\le p<q<M}\tanh\left(\frac{\theta_{n_p}-\theta_{n_q}}{2}\right) 
\nonumber \\
&& \times \prod_{1\le r\le N; 1\le s\le M}\coth\left(\frac{\theta_{k_r}-\theta_{n_s}}{2}\right) \, . 
\eea
Here $\theta_{n}, \theta_{k}$ are finite size rapidities in the R-  and NS-sectors:
\be
\sinh (\theta_{n})  = \frac{2\pi n}{\nu} \, , \enspace n\in {\mathbb Z} \qquad  \sinh (\theta_{k}) = \frac{2\pi k}{\nu} \, , \enspace k\in \frac{1}{2} + {\mathbb Z} \, , 
\ee 
the functions $g(\theta), \tilde g(\theta)$ are the leg factors defined as 
\be
g(\theta) = \frac{e^{\kappa(\theta)}}{\sqrt{\nu\cosh (\theta)}}\, , \qquad \tilde g(\theta) = \frac{e^{-\kappa(\theta)}}{\sqrt{\nu\cosh (\theta)}}
\ee
where 
\be
\kappa(\theta) = \frac{1}{2\pi}\int\limits_{-\infty}^{+\infty}\frac{d\theta'}{\cosh(\theta-\theta')}\ln[ \tanh \left(\frac{\nu\cosh(\theta')}{2}\right)]\, .
\ee
The overall factor $ \bar \sigma S(R)$ in (\ref{matrix_mass}) is the vacuum-vacuum matrix element 
\be
\bar \sigma S(R)=    \left \{
\begin{array}{l@{\qquad}l}
{}_{\rm NS}\langle 0|\sigma(0,0)|0\rangle_{\rm R}\, ,  &m>0 \, , \\[1ex]
{}_{\rm NS}\langle 0|\mu(0,0)|0\rangle_{\rm R}\, ,  &m<0   \, .
\end{array}
\right .
\ee 
We have the following explicit expressions
\be
\bar \sigma = |m|^{1/8}2^{1/12}e^{-1/8}A^{3/2}
\ee
where $A$ is Glaisher-Kinkelin constant, and 
\be
\ln S(R) = \frac{1}{2}\left(\frac{\nu}{2\pi}\right)^2 \iint\!\! d\theta_1d\theta_2 
\frac{\sinh(\theta_1)\sinh(\theta_2)  \ln\!|\!\coth\left(\frac{\theta_1-\theta_2}{2}\right)\!|}{\sinh[\nu\cosh(\theta_1)]\sinh[\nu\cosh(\theta_2)]}\, .
\ee

The Hamiltonian (\ref{Ham2}) is restricted to a truncated space ${\cal H}_{m}^{\rm tr}\subset {\cal H}_{m}$ that is spanned 
by vectors (\ref{mbasis1}), (\ref{mbasis2}) satisfying 
\be
\sum_{i=1}^{N} k_i = \sum_{j=1}^{M}n_j = 0 \, ,  
\ee
and 
\be
\sum_{i=1}^{N}|k_i| \le 2n_c \, , \quad \sum_{j=1}^{M} |n_j| \le 2n_c \, .
\ee
Here $n_c$ is an integer that controls the truncation.  In contrast with the TCSA, it no longer is related to the energy of the 
unperturbed Hamiltonian $H_{\rm FF}$. The total dimensions of truncated spaces are the same as the ones given in Figure \ref{table}.

While the TFFSA eigenvectors lie in ${\cal H}_{m}$ we can use the interface operator $\hat D_{m}^{-1}: {\cal H}_{m}\to {\cal H}_{0}$ 
to obtain their image in ${\cal H}_{0}$. This can be formally thought of as combining (or fusing) two perturbation interfaces:  
the mass interface $\hat D^{-1}_{m}$ and the magnetic field interface $\hat D^{-1}_{m,h}$. The last one corresponds to perturbing the free massive theory 
by the magnetic field.  

\begin{center}
\begin{figure}[H]
\centering
\begin{tikzpicture}[>=latex]
\fill[orange!10!white] (3.4,0) -- (5.8,0) arc   (270:90:0.6 and 1.1)  -- (3.4, 2.2) arc (90:270: 0.6 and 1.1)--cycle;
\fill[red!10!white] (5.8,0) -- (9,0) arc   (-90:90:0.6 and 1.1 )  -- (5.8, 2.2) arc (90:270: 0.6 and 1.1)--cycle;
\draw[thick, dashed] (0,0) arc (270:90:0.6 and 1.1);
\draw[thick, dashed] (0,0) arc (-90:90:0.6 and 1.1 );
\draw[ thick] (0,0)--(9,0);
\draw[thick] (0,2.2 cm )--(9, 2.2 cm);
\draw[thick, dashed] (9,0) arc (270:90:0.6 and 1.1);
\draw[thick, dashed] (9,0) arc (-90:90:0.6 and 1.1);
\draw[very thick] (3.4,0) arc (270:90:0.6 and 1.1); 
\draw[very thick, dashed] (3.4,0) arc (-90:90:0.6 and 1.1);
\draw[very thick] (5.8,0) arc (270:90:0.6 and 1.1);
\draw[very thick, dashed] (5.8,0) arc (-90:90:0.6 and 1.1 );

\draw (3.6,-0.35) node {${ D}_{m}^{-1}$};
\draw (6.0,-0.4) node {${ D}_{m,h}^{-1}$};
\draw (1.5,1.2  ) node {$m=0 $};
\draw (1.5, 0.85) node {$h=0$};
\draw (4.6,1.2 ) node {$m\ne 0 $};
\draw (4.65,0.85 ) node {$ h=0$};
\draw (7.5,1.2 cm) node {$m\ne 0$};
\draw (7.5,0.85 cm) node {$ h\ne 0$};
\draw (3.4, 2.25) -- (3.4,2.6);
\draw (5.8, 2.25) -- (5.8,2.6);
\draw[<->] (3.4, 2.45) -- (5.8,2.45);
\draw (4.55,2.64) node {$\epsilon$};
\end{tikzpicture}
\caption{Truncated free fermion method as a fusion of defects: the mass defect and the magnetic 
field perturbation defect. Fusion corresponds to the $\epsilon \to 0$ limit. }
\end{figure}
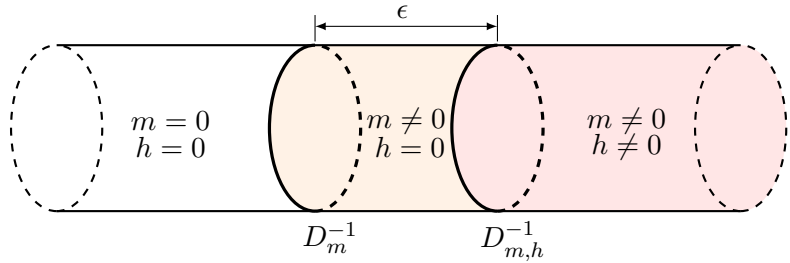
\end{center}

The mass interface operator was constructed analytically  in section \ref{mass_interface} while we can use the TFFSA numerics to 
obtain information on the second interface. Fusion of conformal  interfaces usually contains  multiplicative divergences 
(see e.g. the discussion in \cite{BBDO}, \cite{BB}). This divergence is regulated by the truncation of ${\cal H}_{m}$ present in  TFFSA. 
However one may still worry whether the fusion procedure gives the same interface as the TCSA one in the $n_{c}\to \infty$ limit. 
While we cannot exclude this situation with definiteness we have not observed anything in the numerical results to be presented below 
that would suggest this scenario\footnote{For example the fusion could trigger an RG flow on the interface. If that was the case we 
would expect some additional sensitivity to truncation level in the TFFSA numerics which we have not observed.}.

Let $P_{n_c}: {\cal H}_{m} \to {\cal H}_{m}^{\rm tr}$ be the projector implementing TFFSA truncation at level $n_c$. 
Then we can write for the component ratios (\ref{Ising_ratios}), (\ref{Ising_ratios2}) 
\be
\Gamma_{\epsilon}^{i} =  \frac{\langle \epsilon| \hat D_{\rm NS}P_{n_c} |v_{i}\rangle}{\langle 0| \hat D_{\rm NS}P_{n_c} |v_{i}\rangle } \, ,
\ee
\be
\Gamma_{\sigma}^{i} = \frac{\langle \sigma| \hat D_{\rm R}P_{n_c} |v_{i}\rangle}{\langle 0| \hat D_{\rm NS}P_{n_c} |v_{i}\rangle } \, .
\ee
Here and below $i=0,1,2,\dots$.
Using (\ref{IONS}), (\ref{IOR})  we calculate 
\be
\Gamma_{\epsilon}^{i}=  -\,{\rm sign}(m) \frac{T_1(v_i,\nu) -  g^2(\nu)T_2(v_i,\nu)}{g(\nu)[T_1(v_i,\nu) + T_2(v_i,\nu)]}  \, , 
\ee
\be
\Gamma_{\sigma}^{i} = 2^{1/4}f(\nu)  \frac{U(v_i,\nu)}{T_1(v_i,\nu) + T_2(v_i,\nu)}
\ee
where  $f(\nu)$ and $g(\nu)$ are given by  (\ref{Fnu}), (\ref{gnu}), and
\be 
T_1(v,\nu)= {}_{\rm NS}\langle 0|\left( -i g(\nu) b_{-1/2}b_{1/2} \right) \prod_{n=1}^{n_c-1} \left( 1 - i\frac{\Delta \omega_{n+1/2}}{m}
b_{-n-1/2}b_{n+1/2}\right) |v\rangle \, ,
\ee
\be 
T_2(v,\nu)= {}_{\rm NS}\langle 0| \prod_{n=1}^{n_c-1} \left( 1 - i\frac{\Delta \omega_{n+1/2}}{m}
b_{-n-1/2}b_{n+1/2}\right) |v\rangle \, ,
\ee
\be
U(v,\nu) = {}_{\rm R}\langle 0| (b_{0})^{p} \prod_{n=1}^{n_c}\left( 1 - i\frac{\Delta \omega_{n}}{m} b_{-n}b_{n}\right) |v\rangle
\ee
where $p=1$ if $y<0$ and $p=0$ when $y>0$. 

We see from these expressions that part of the scale dependence of $\Gamma_{\epsilon, \sigma}^{i}$ comes from 
the functions $f(\nu)$ and $g(\nu)$ and from $\Delta \omega_{k}$ which we know analytically. 
In practice we observed a  faster rate of convergence of $\Gamma_{\sigma}^{i}$ to its asymptotic value than that of  $\Gamma_{\epsilon}^{i}$.
This can be at least partially attributed to the faster convergence of $f(\nu)$ than that of  $g(\nu)$ to its asymptotic value 1.
Having noted this we can  formally define quantities in which $\hat D_{\rm NS}$, $\hat D_{\rm R}$  are taken at $\nu = \infty$:
\be
\tilde \Gamma_{\epsilon}^{i}=-{\rm sign}(m)  \frac{T_1(v_i,\infty) -  T_2(v_i,\infty)}{T_1(v_i,\infty) + T_2(v_i,\infty)}  \, , 
\ee
\be
\tilde \Gamma_{\sigma}^{i} = 2^{1/4} \frac{U(v_i,\infty)}{T_1(v_i,\infty) + T_2(v_i,\infty)}\, .
\ee
In practice we found that using $\tilde \Gamma_{\epsilon, \sigma}^{i}$ gives a  small improvement in the 
 convergence rate towards the asymptotic values.  As these quantities do not have a clear physical meaning 
 (the eigenvector $|v_{i}\rangle$  is still taken at finite $\nu$) we are not going to present the numerical results for them   in the paper.

Applied to the flows triggered by real $h$ the TFFSA method here described gives data very similar to the TCSA one presented 
in section \ref{TCSA_section}. 
The method however has a significant advantage when applied to the imaginary magnetic field flows discussed in the forthcoming sections.


\section{Imaginary magnetic field. Complex vacuum energy.} \label{Complex_vac_sec}
\setcounter{equation}{0}
The Ising field theory (\ref{IFT}) taken at imaginary values of $h$ is not unitary. 
However the corresponding Hamiltonian $H$ enjoys the following symmetry 
\be\label{Ssym}
SHS=H^{\dagger}
\ee
where $S$ is the operator that multiplies any Ramond sector vector by $-1$ and leaves any NS sector vector intact. To see the implications 
of this symmetry consider $H$ as an operator in ${\cal H}_{0}$ and choose 
a basis   in which $H$ is  (complex) symmetric. 


For example we can take the conformal basis  described before equation (\ref{conf_basis}).  
In this basis the matrices $M$ and $B$ in (\ref{H_TCSA}) are symmetric and hence  $H$ is symmetric as well. 
As before denote these basis vectors as $|i\rangle_{\rm R}$, $|j\rangle_{\rm NS}$. If 
\be
|v_{\lambda}\rangle = \sum_{i} C^{\rm NS}(i)|i\rangle_{\rm NS} + \sum_{j} C^{\rm R}(j)|j\rangle_{\rm R}
\ee
is an eigenvector of $H$ with eigenvalue $\lambda$ then 
\be
|\bar v_{\lambda}\rangle \equiv \sum_{i} \bar C^{\rm NS}(i)|i\rangle_{\rm NS} + \sum_{j} \bar C^{\rm R}(j)|j\rangle_{\rm R}
\ee
is an eigenvector of $\bar H$ with eigenvalue $\bar \lambda$. (The bar stands for complex conjugation everywhere.) 
Then (\ref{Ssym}) in this particular basis implies that 
\be 
HS|\bar v_{\lambda}\rangle = \bar \lambda S|\bar v_{\lambda}\rangle
\ee
that is 
\be
|v_{\bar \lambda}\rangle \equiv  \sum_{i} \bar C^{\rm NS}(i)|i\rangle_{\rm NS} - \sum_{j} \bar C^{\rm R}(j)|j\rangle_{\rm R}
\ee
is an eigenvector of $H$ with eigenvalue $\bar \lambda$. Thus the energy eigenvalues are either real or form a pair of 
complex conjugated values.
Moreover the above implies that if the vacuum of $H$ has real energy and is non-degenerate then the vacuum vector must be of the form 
\be
|v_{0}\rangle = |0\rangle + C^{\rm NS}|\epsilon\rangle + iC^{\rm R}|\sigma\rangle + \dots 
\ee
where $C^{\rm NS}$, $C^{\rm R}$ are real and we show only the three lowest components. 
Hence in this case $\Gamma_{\epsilon}^{0}$ is real and  $\Gamma_{\sigma}^{0}=iC^{\rm R}$ is imaginary. 

Alternatively if the vacuum energy is complex and the vacuum space is a two-dimensional subspace  corresponding 
to two conjugate eigenvalues then the corresponding eigenvectors can be written as 
\bea\label{pair_eig}
&& |v_{0}\rangle = |0\rangle + C^{\rm NS}e^{i\chi}|\epsilon\rangle + C^{\rm R} e^{i\phi} |\sigma\rangle + \dots \nonumber \\
&& |\bar v_{0}\rangle = |0\rangle + C^{\rm NS}e^{-i\chi}|\epsilon\rangle - C^{\rm R} e^{-i\phi} |\sigma\rangle + \dots
\eea
where $C^{\rm NS}$, $C^{\rm R}$ are real and hence $\Gamma_{\epsilon}^{0}=C^{\rm NS}e^{\pm i\chi}$ $\Gamma_{\sigma}^{0}=\pm C^{\rm R}e^{\pm i\phi}$ 
are both complex in general. 
Since none of the conformal boundary states (\ref{conf_bcs}) 
has a complex value of $\Gamma_{\sigma}$ 
we see\footnote{The reality of $\Gamma_{\sigma}$ for conformal boundary states can be seen as a consequence of 
locality of the boundary condition.} that in the first case {\it as long as the vacuum eigenvalue remains real the vacuum vector has no chance of  approaching 
a conformal boundary state as we move along an RG trajectory\footnote{Strictly speaking this leaves out the possibility of approaching 
the free  boundary condition. This would be however highly unlikely in the view of the perturbation explicitly breaking the spin reversal symmetry.}.} In the second case  the only way  $\Gamma_{\epsilon}$ and 
$\Gamma_{\sigma}$ can become real at the 
end of the RG flow is if the phases $e^{i\chi}$, $e^{i\phi}$ tend to 1 or -1 as $r\to \infty$. If this is the case the eigenvectors  given 
in (\ref{pair_eig}) should either tend to the pair of fixed boundary states $|\pm\rangle\!\rangle$ or to $|F\rangle\!\rangle\oplus |F\rangle\!\rangle$.
We will present evidence that supports the first possibility.  

We now focus on the case of complex vacuum energy. The spectrum of the Ising field theory for imaginary magnetic field and arbitrary mass 
was investigated in \cite{FZ},  \cite{Zam1}, \cite{Zam2} using  TFFSA as well as analytic results. It was shown in those papers that
for $y>y_{\rm cr}$ where $y_{\rm cr}\approx -2.429$ the vacuum energy becomes complex for sufficiently large values of $r$. 
The vacuum vectors form a pair with conjugated energy values. As discussed in \cite{FZ} for positive $y$ and large enough $r$ 
the vacuum and excited  states form complex pairs with ${\rm Re} \tilde E(r)$ asymptotically approaching   constant values and ${\rm Im} \tilde E(r)$ 
asymptoting to straight lines with equal slopes (free energy density) thus furnishing  particle-like excitations with complex masses.

Here we present  plots of the numerical results (using TFFSA) for the ratios $\Gamma_{\epsilon}$, $\Gamma_{\sigma}$ calculated for the
 two vacuum vectors 
at $y=2$. 
\begin{center}
\begin{figure}[H]

\begin{minipage}[b]{0.5\linewidth}
\centering
\includegraphics[scale=0.5]{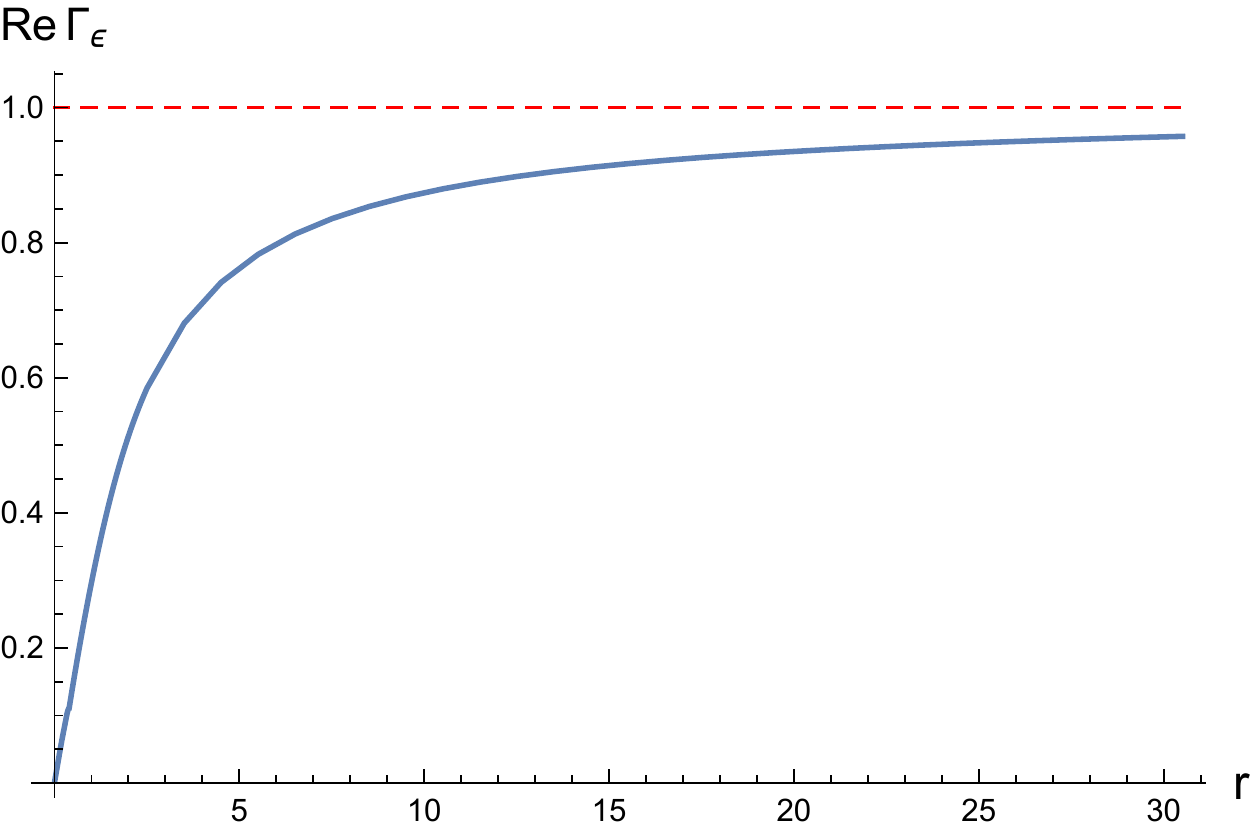}
\end{minipage}%
\begin{minipage}[b]{0.5\linewidth}
\centering
\includegraphics[scale=0.5]{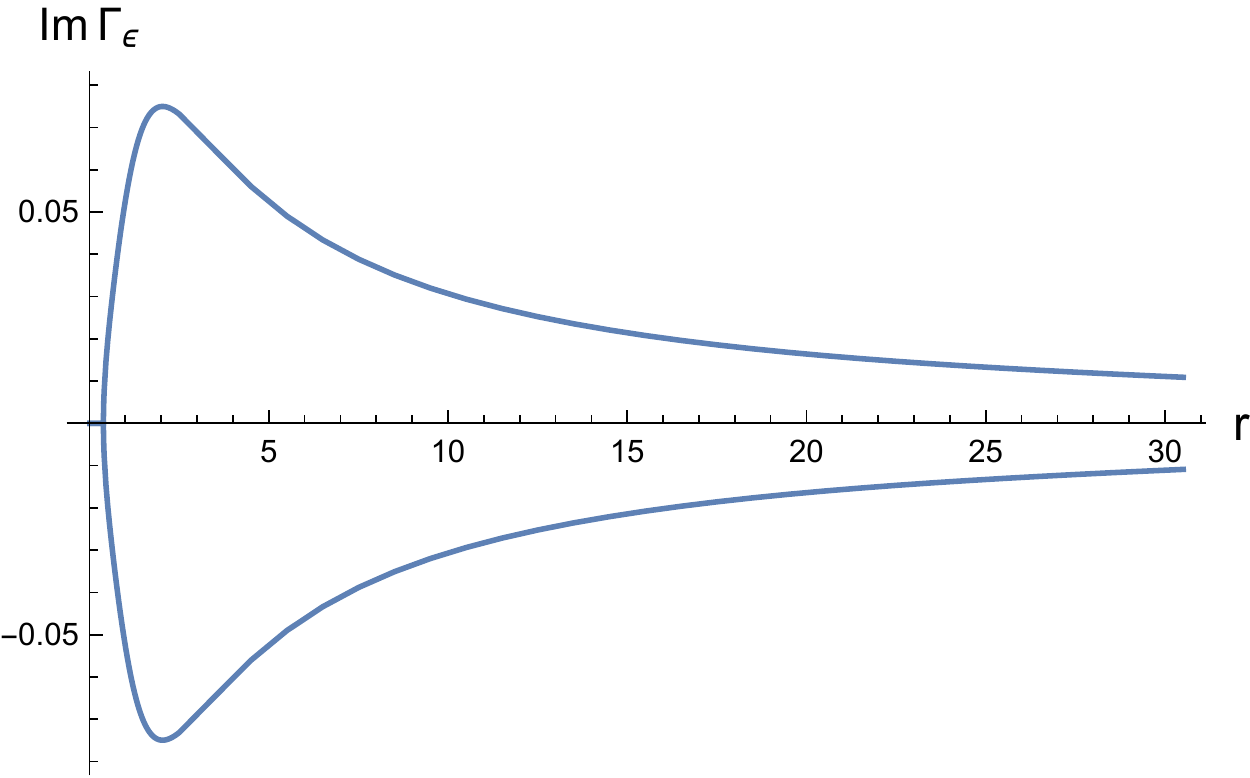}
\end{minipage}
\caption{Imaginary magnetic field. Real and Imaginary parts of $\Gamma_{\epsilon}^{0}$ for the two vacuum vectors at $y=2$, $n_{c}=11$.}
\end{figure}

\end{center}
\begin{center}
\begin{figure}[H]

\begin{minipage}[b]{0.5\linewidth}
\centering
\includegraphics[scale=0.55]{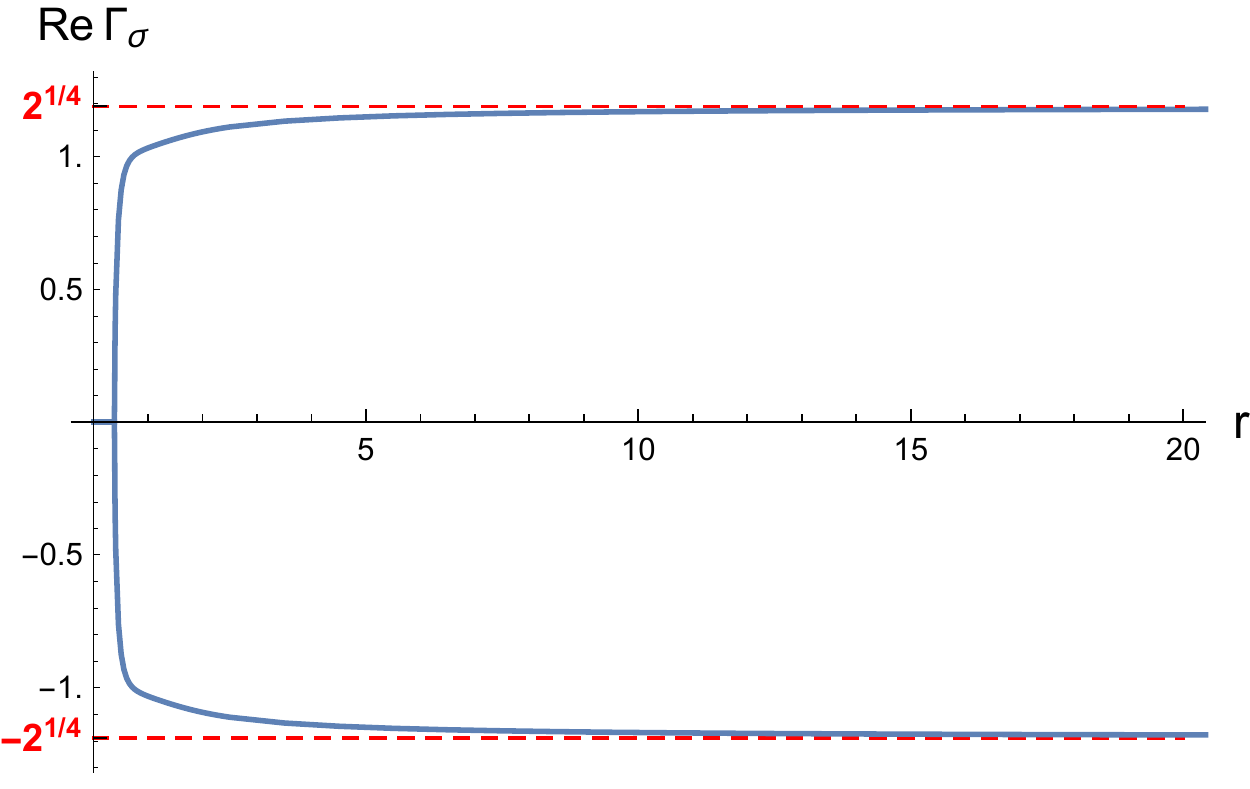}
\end{minipage}%
\begin{minipage}[b]{0.5\linewidth}
\centering
\includegraphics[scale=0.55]{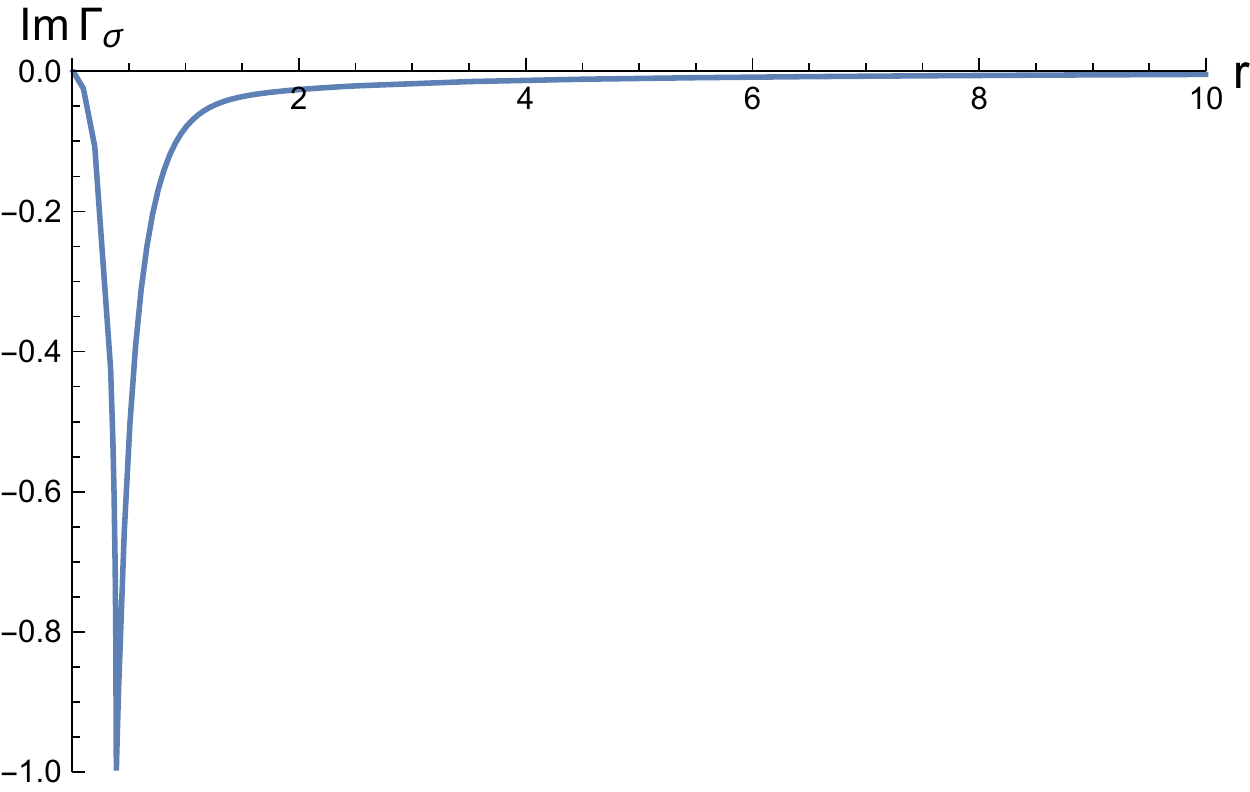}
\end{minipage}
\caption{Imaginary magnetic field. Real and Imaginary parts of $\Gamma_{\sigma}^{0}$ for the two vacuum vectors at $y=2$, $n_{c}=11$.}
\end{figure}
\end{center}
Comparing these plots to the ratios in conformal boundary conditions (\ref{conf_bcs}) we find that the RG boundary is a superposition 
$|+\rangle\!\rangle \oplus |-\rangle\!\rangle$.

The case of large positive $y$ can be understood in terms of boundary RG flows (see the discussion in section 
\ref{massive_summary}). This regime corresponds to starting with a large $m>0$ and adding to it a small 
imaginary magnetic field perturbation. Thus with a good approximation we start with a vacuum described by 
  the conformal boundary condition $|+\rangle\!\rangle \oplus |-\rangle\!\rangle$ and the perturbation 
  is just the identity field between the two boundary conditions multiplied by an imaginary coupling. 
  The only effect of this is a relative phase factor which keeps rotating as we change the scale but does not change any 
  physical quantities.  

We have also checked numerically the region  $y_{\rm cr}<y<0$. As long as the vacuum complex pair does not experience 
collisions with upper level eigenvalues the picture is qualitatively the same as presented on the above plots. Such collisions 
are sensitive to the truncation level so strictly speaking we do not have a proof that the above picture remains 
 in the extreme $r\to \infty$ asymptotics at infinite truncation level. But it looks highly plausible to us. 

To summarise the numerical results support the RG boundary being 
$|+\rangle\!\rangle \oplus |-\rangle\!\rangle$ for all cases when the vacuum energy is complex and we have a pair of vacuum vectors. 

\section{Imaginary magnetic field. Real vacuum energy.} 
\setcounter{equation}{0}

\subsection{Massive flows}\label{Im_massive}
For $y\le y_{\rm cr}$ numerical studies \cite{FZ} show that the vacuum energy remains  real until a large positive value if $r$. The latter value at which the 
vacuum eigenvalue collides with the first excited eigenvalue and forms a conjugate pair is numerically the larger the larger $n_c$ is and 
presumably such collision is a numerical artefact. We found that TFFSA method works much better than TCSA in this regime. In particular the 
lowest eigenvalue stays real for much larger value of $r$ when using the TFFSA method.

We start discussing the numerical results for this region by taking a large negative $y$ where 
the vacuum energy stays real for large $r$'s and where we can use the boundary RG picture discussed in section \ref{massive_summary}.
Below we present  numerical data for a sample point $y=-3.5$. One finds that the behaviour of the component ratios is very different here from the situations 
discussed before. Namely the overlap of the perturbed vacuum with the unperturbed one goes 
to zero and both $\Gamma^{0}_{\epsilon}$ and $\Gamma^{0}_{\sigma}$ go to infinity at a finite value of RG scale $r$. 
The ratio $\Gamma^{0}_{\epsilon}/\Gamma^{0}_{\sigma}$ remains finite.
Define 
\be 
T=\frac{\left|\langle 0|v_{0}\rangle\right|}{\| v_{0} \|} \, , \qquad \Gamma=\left|\frac{\Gamma_{\epsilon}^{0}}{\Gamma_{\sigma}^{0}}\right| \, .
\ee
Here is a plot of these ratios for $y=-3.5$

\begin{center}
\begin{figure}[H]

\begin{minipage}[b]{0.5\linewidth}
\centering
\includegraphics[scale=0.75]{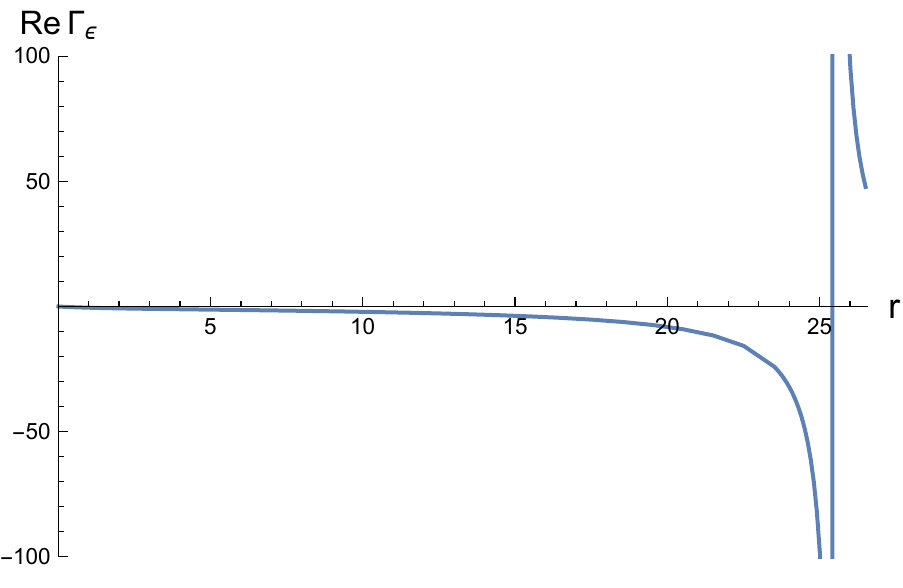}
\end{minipage}%
\begin{minipage}[b]{0.5\linewidth}
\centering
\includegraphics[scale=0.75]{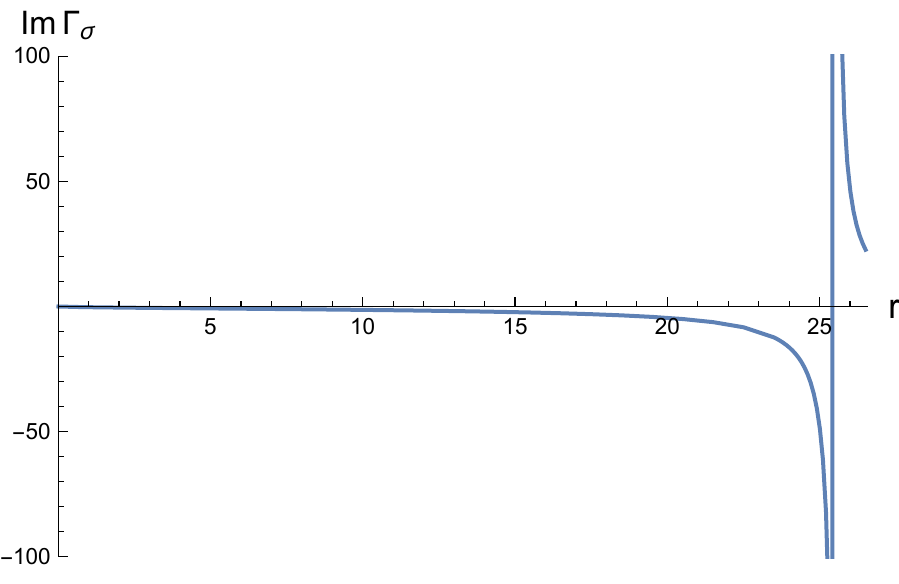}
\end{minipage}
\caption{Imaginary magnetic field. The component ratios $\Gamma_{\epsilon}^{0}$ and $\Gamma_{\sigma}^{0}$ at $y=-3.5$, $n_{c}=12$.}
\label{data1}
\end{figure}
\end{center}
\begin{center}
\begin{figure}[H]

\begin{minipage}[b]{0.5\linewidth}
\centering
\includegraphics[scale=0.55]{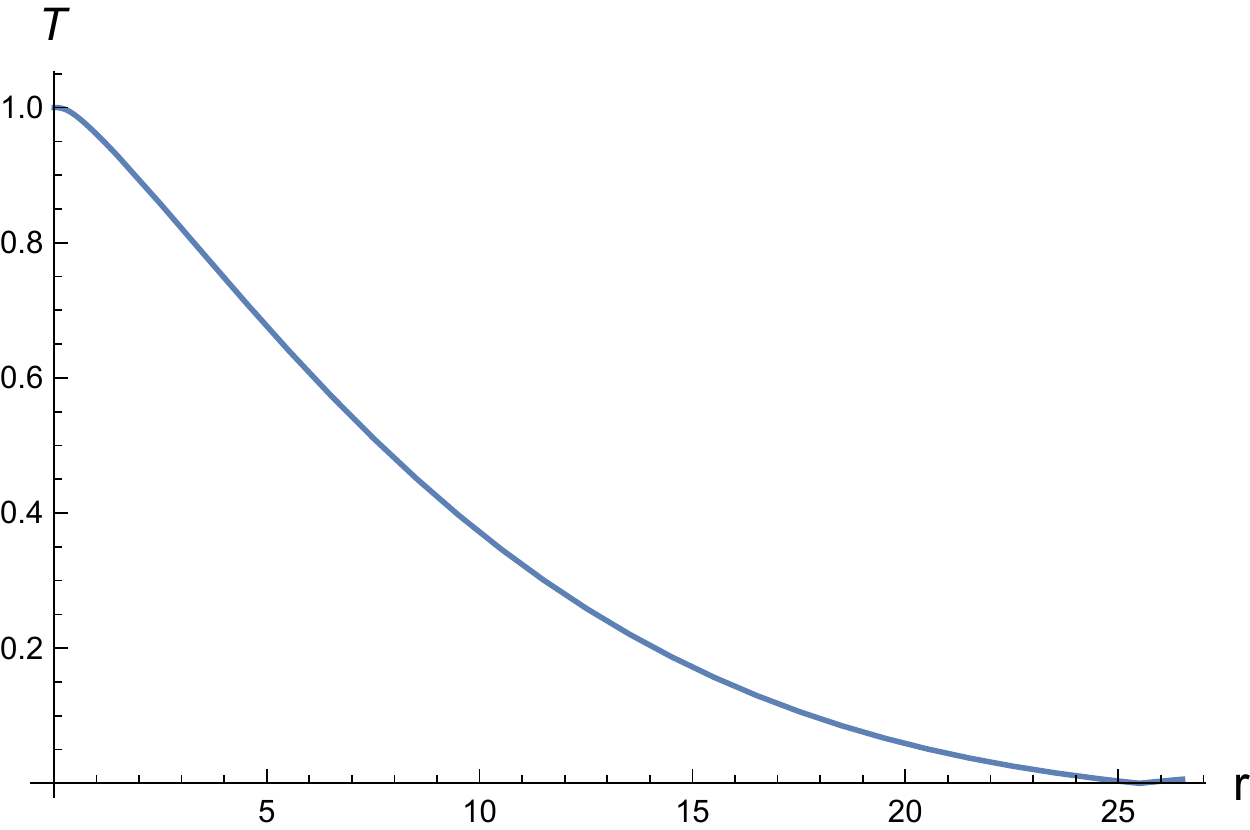}
\end{minipage}%
\begin{minipage}[b]{0.5\linewidth}
\centering
\includegraphics[scale=0.55]{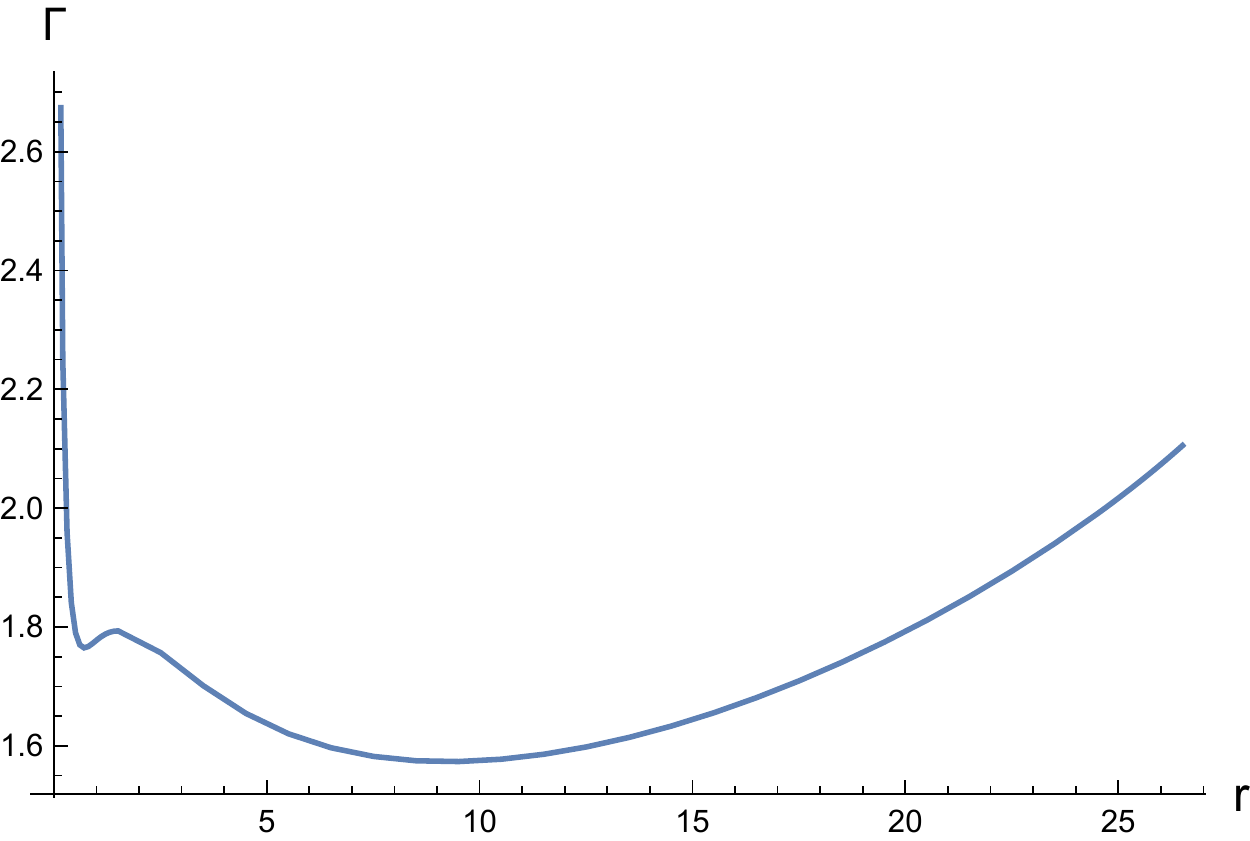}
\end{minipage}
\caption{Imaginary magnetic field. The component ratios $T$ and $\Gamma$ at $y=-3.5$, $n_{c}=12$.}
\label{data2}
\end{figure}
\end{center}

We find that $T$ vanishes and $\Gamma^{0}_{\epsilon,\sigma}$ blow up at $r=r_{1}\approx 25.5$. The ratio $\Gamma$ remains finite 
at this point with the value $\Gamma(25.5)=\Gamma^{*}_{\rm num} \approx 2.0445$. 

For large negative values of $y$ we can approximate the flow of the vacuum vector  by an RG boundary flow.
For large negative values of $m$  the vacuum is well approximated by the boundary state $|F\rangle\!\rangle$ corresponding to free boundary spin.
Switching on a small imaginary magnetic field in the bulk perturbed this boundary condition by an imaginary boundary magnetic field. 
The critical Ising model with a boundary magnetic field is a Gaussian theory and can be solved exactly \cite{CZ}, \cite{Chat}.   
It is not hard to extend this solution 
to imaginary magnetic field. 

The action functional on a cylinder for this model reads 
\be \frac{1}{2\pi}\iint (\psi\bar \partial \psi + \bar \psi \partial \bar \psi  )\, d^2 x + \int\! \left(\frac{i}{4\pi}\psi\bar\psi + \frac{1}{2}a\dot a
+h_{\rm b}(\omega \psi + \bar \omega \bar \psi ) a \right)dy
\ee
where the boundary is located at $x=0$, $\omega = e^{i\pi/4}$ and the boundary magnetic field coupling is 
taken here to be  $ih_{\rm b}$ with $h_{\rm b}$ - real. 
The boundary fermion $a=a(y)$ accounts for 
the double degeneracy of the vacuum.

Adopting the  boundary state found in \cite{Chat} for the case of imaginary magnetic field we obtain  
\bea\label{b_state_Chat}
&& |h_{\rm b}\rangle\!\rangle = \sqrt{\pi}e^{-\alpha\ln(R\mu )}\Bigl[ \frac{1}{\Gamma(1/2-\alpha)}\exp\left(- i\sum_{n=0}^{\infty} 
\frac{n+1/2 + \alpha}{n+1/2-\alpha} a^{\dagger}_{n+1/2}\bar a^{\dagger}_{n+1/2}\right) |0\rangle  \nonumber \\
&& \pm i \frac{2^{1/4}\sqrt{\alpha}}{\Gamma(1-\alpha)}\exp\left( -i\sum_{n=1}^{\infty} 
\frac{n + \alpha}{n-\alpha} a^{\dagger}_{n}\bar a^{\dagger}_{n}\right) |\sigma \rangle \Bigr]
\eea
where $\alpha = 2h^2_{\rm b}R$ and the sign in front of the Ramond component is fixed by the sign of $h_{\rm b}$. 
This exact solution gives us the following component ratios
\be\label{GexactC}
\Gamma_{\epsilon}^{0} = -\frac{1/2+\alpha}{1/2-\alpha}\, , \qquad \Gamma_{\sigma}^{0}= \pm i 2^{1/4}\frac{\sqrt{\alpha}\Gamma(1/2-\alpha)}{\Gamma(1-\alpha)}\, , 
\ee
\be\label{Gh}
\Gamma = \frac{\Gamma(1-\alpha)(1/2+\alpha)}{2^{1/4}\sqrt{\alpha}\Gamma(3/2-\alpha)} \, .
\ee
The vacuum component of $|h_{\rm b}\rangle\!\rangle$ vanishes linearly at $\alpha = 1/2$ that is $T\sim 1/2-\alpha$. 
This results in $\Gamma^{0}_{\epsilon}$ 
and $\Gamma^{0}_{\sigma}$ having a simple pole at this point while their ratio is finite and given (in absolute value) by 
 function (\ref{Gh}) a portion of which which we depict below 

\begin{center}
\begin{figure}[H]
\centering
\includegraphics[scale=0.7]{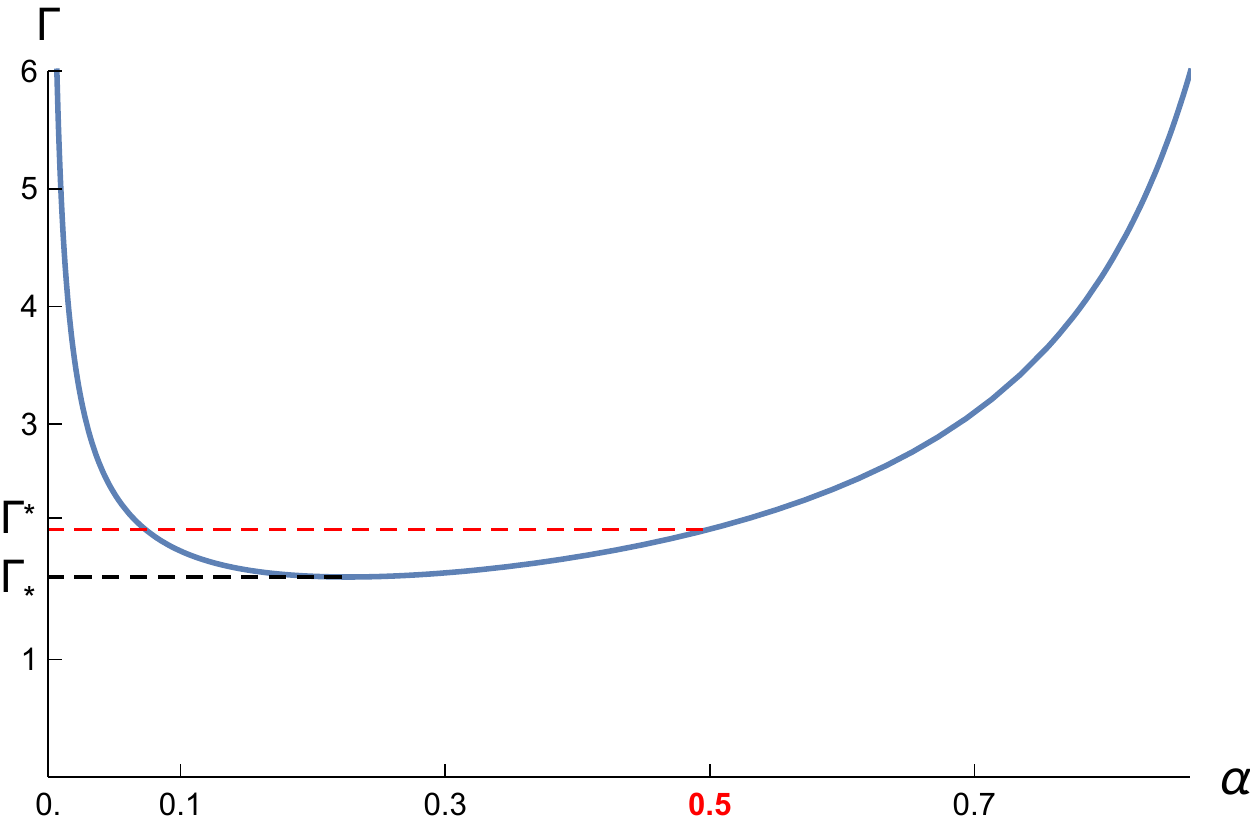}
\caption{The ratio $\Gamma(\alpha)$ for imaginary boundary magnetic field model.}
\end{figure}
\end{center}

Comparing (\ref{GexactC}), (\ref{Gh})  with Fig. \ref{data1} and Fig. \ref{data2} we find qualitatively exactly the same picture. 
We find that the phases of all ratios are the same as in (\ref{b_state_Chat}) and that at   $r=r_{1}$ in the NS 
sector all components are small except for the components including the $a_{1/2}^{\dagger}\bar a_{1/2}^{\dagger}$ oscillators as 
in (\ref{b_state_Chat}).
  Since we do not know how the effective boundary $h_{\rm b}$ depends on $y$ and $r$ we cannot match the two curves for 
  $\Gamma$ 
 (they are parameterised differently). However
we can still match quantitatively the two special values: $\Gamma_{*}\approx  1.70$ - the minimal value and 
$\Gamma^{*}=2^{1/4}\sqrt{\pi}\approx 2.107$ - the value at the 
point at which $\Gamma^{0}_{\epsilon,\sigma}$ blow up. The numerical TFFSA values: 
  $\Gamma_{*}^{\rm num}$, $\Gamma^{*}_{\rm num}$  are approximately   1.573 and  2.044 respectively. These values are within few percent from those 
 predicted by the boundary magnetic field model. The match gets better for larger values of $|y|$ and larger $n_c$.  

In the numerics for $y=-3.5$ the data continues past the blow up point $r=r_1$ with $\Gamma$ continuing to grow. 
In the boundary magnetic field model going past the first special point $\alpha=1/2$ the ratio $\Gamma$ blows up at $\alpha=1$.
In the numerics we do not quite get to that second point as the vacuum and the first excited energy levels collide and 
form a complex pair. The collision point is sensitive to the truncation level  $n_c$ and is moved towards larger values of $r$ 
as $n_c$ is increased. So presumably  in the untruncated theory 
the vacuum remains real at all scales. As explained at the beginning of section \ref{Complex_vac_sec} 
as long as the vacuum energy remains real and $\Gamma_{\sigma}^{0}$ does not vanish (and thus remains imaginary) 
we cannot approach  a local conformal boundary condition. So, what happens? Given a good match at the onset of the flow  
with the imaginary boundary magnetic field model  we are going to rely on it in describing what happens 
as we continue increasing the scale. 
As we can clearly see from (\ref{b_state_Chat}) the model keeps going through a sequence of special points 
at which $\alpha_{n}$ is a (positive)  integer or  half integer. At half integer points $\alpha_{n}^{*}=n+1/2$
all boundary state components with level less than $2n+1$  in the NS sector vanish and in the higher weight components 
only those including $a_{n+1/2}^{\dagger}\bar a_{n+1/2}^{\dagger}$ oscillators survive. For $n$ large the low energy 
components in the R-sector approach those of the $|\sigma\rangle\!\rangle$ Ishibashi state. 
At integer points $ \alpha_{*}^{n} = n$  similarly the low energy components in the R-sector are wiped out while those 
in the NS-sector approach those of the $|0\rangle\!\rangle + |\epsilon\rangle\!\rangle$ Ishibashi states combination. 
If we focus on the low lying components with level smaller than  $\alpha$ then asymptoticaly  (up to an overall factor) we have
\be
|h_{\rm b}\rangle\!\rangle \sim \cos(\pi \alpha) [|0\rangle\!\rangle + |\epsilon\rangle\!\rangle ] \pm i\sin(\pi \alpha) 2^{1/4}|\sigma\rangle\!\rangle 
+ \dots 
\ee 
where the ellipsis stands for terms that contain components of level larger than $\alpha$.
Thus as $\alpha$ goes to infinity we will see a never ending rotation of
the two  combinations of Ishibashi states. This cyclic behaviour is 
of course  in violation of $g$-theorem which however is possible here because we are in a non-unitary situation.


\subsection{Conformal interfaces between Ising and Yang-Lee models}

At any real value of $h$ the Ising field theory flows to a trivial fixed point. For pure imaginary $h$ with 
a certain value of the scaling parameter $y=y_{\rm cr}$ the theory flows to the Yang-Lee  fixed point which is 
a non-unitary minimal model ${\cal M}(2,5)$ with central charge  $c=-22/5$. It has two primary fields: the identity and 
a field $\phi$ with scaling dimension $\Delta_{\phi}=-2/5$ and correspondingly two conformal boundary conditions 
with Cardy states $|1\rangle\!\rangle_{\rm YL}\, , |\phi\rangle\!\rangle_{\rm YL} $. 
The value of $y_{cr}$ was most recently estimated  numerically  \cite{Zam2}
to be  $y_{cr}=-2.42929(2) $.

All conformal interfaces between the Ising and Yang-Lee models are known \cite{QRW} due to a remarkable fact that 
the tensor product $({\rm Ising})\otimes (\mbox{Yang-Lee})$ is itself a minimal model  ${\cal M}(5,12)$ with $E_{6}$ modular invariant.  
The interfaces are described as boundary conditions in the tensor product theory ${\cal M}(5,12)$ which were found in \cite{BPPZ}. 
The  ${\cal M}(5,12)$ theory has 12 primary states 
and thus 12 elementary conformal boundary conditions. We will use the same conventions as in \cite{QRW}, \cite{BPPZ}
 in which  chiral primaries $\phi_{r,s}$
are labelled by  $r\in \{1,3\}$ and $s\in \{1,4,5,7,8,11\}$ - the set of $E_6$ exponents. 
The tensor products of primaries are identified as 
\be
{\bf 1}\otimes {\bf 1} = \phi_{1,1} \, , \quad {\bf 1}\otimes \epsilon = \phi_{1,5} \, , \quad 
{\bf 1}\otimes \sigma = \phi_{1,4} \, , 
\ee
\be
\phi\otimes {\bf 1} = \phi_{3,7} \, , \quad \phi\otimes \epsilon = \phi_{3, 5} \, , \quad 
\phi\otimes \sigma = \phi_{3,8} \, .
\ee
The conformal boundary states 
$|\widetilde{(r,a)}\rangle\!\rangle$ are labelled by a pair $r \in \{ 1,3\}$, $a\in \{ 1,2,3,4,5,6   \}$ (the labels of nods of the $E_6$ Dynkin 
diagram). The decomposition into the Ishibashi states can be written as\footnote{Formula (\ref{Cardy_E6}) was derived in 
\cite{BPPZ} with the assumption that the matrix $\Psi$ is unitary. A more general derivation in which this assumption was not made was later 
presented in \cite{TFT}. }  \cite{BPPZ}
\be \label{Cardy_E6}
|\widetilde{(r,a)}\rangle\!\rangle = \sum_{r', s'}\frac{\Psi_{r,a}^{(r',s')}}{\sqrt{S_{11, r's'}}} |(r',s')\rangle\!\rangle \, 
\ee
where 
\be \label{Psi}
\Psi_{r,a}^{(r',s')} = \sqrt{2}S_{rr'}\psi^{s'}_{a}  
\ee
and $\psi^{s'}_{a}$ is the matrix made of the eigenvectors of the $E_6$ adjacency matrix. Explicitly we have 
\be\label{matrix}
(\psi^{s}_{i}) = \left( \begin{array}{rrrrrr}
a& \frac{1}{2} & b&b & \frac{1}{2}& a\\
b& \frac{1}{2}& a& -a & -\frac{1}{2}& -b\\
c&0&-d&-d& 0 & c \\
b& -\frac{1}{2}& a&-a& \frac{1}{2}&-b\\
a&-\frac{2}{2}& b & b & -\frac{1}{2} & a \\
d& 0 & -c&c & 0 &-d 
\end{array} \right) 
\ee
where 
\be
a =\frac{\sqrt{3 - \sqrt{3}}}{\sqrt{24}} \, , \quad b=\frac{\sqrt{3 + \sqrt{3}}}{\sqrt{24}}\, , 
\ee
\be
c =\frac{\sqrt{3 + \sqrt{3}}}{\sqrt{12}} \, , \quad d=\frac{\sqrt{3 -\sqrt{3}}}{\sqrt{12}}\, .
\ee
In (\ref{matrix}) the row index $i\in \{1,2,3,4,5,6\}$ labels the boundary states and the column index $s\in \{1,4,5,7,8,11\}$ 
labels the primaries. The $S$-matrices present in (\ref{Cardy_E6}), (\ref{Psi}) are 
\be
S_{rs,r' s'} = \sqrt{\frac{2}{15}} (-1)^{(r+s)(r'+s')} \sin\left( \pi r r' \frac{7}{5}\right) \sin\left( \pi s s' \frac{7}{12}\right) \, ,
\ee 
\be
S_{rr'} =\sqrt{\frac{2}{5}} \sin\left(rr' \frac{\pi}{5}\right) \, .
\ee

Using the above we calculate the component ratios $\Gamma_{\epsilon}^{i}$, $\Gamma_{\sigma}^{i}$ as ratios 
of one-point functions of factored primaries in the $E_6$ theory. 
As they turn out to be independent of the first index - $r$ labelling the boundary states, we present the answers 
in the form of 6-vectors with components labeled by the second index - $a$. 
\be \label{InterfaceG1}
\Gamma_{\epsilon}^{0}  = \frac{\langle \phi \otimes \epsilon \rangle_{(r,a)}}{\langle \phi \otimes {\bf 1}\rangle_{(r,a)}} = 
\left( \begin{array}{r}
1\\
-1 \\
1\\
-1 \\
1\\
-1
\end{array}\right)\, , \quad 
\Gamma_{\sigma}^{0}  = \frac{\langle \phi \otimes \sigma \rangle_{(r,a)}}{\langle \phi \otimes {\bf 1}\rangle_{(r,a)}} = 
\left( \begin{array}{r}
\tilde \beta\\
\beta \\
0\\
-\beta \\
-\tilde \beta\\
0
\end{array}\right) \, , 
\ee

\be \label{InterfaceG2}
\Gamma_{\epsilon}^{1}  = \frac{\langle {\bf 1}\otimes \epsilon \rangle_{(r,a)}}{\langle {\bf 1}\otimes {\bf 1}\rangle_{(r,a)}} = 
\left( \begin{array}{r}
\gamma \\
1\\
-1\\
1 \\
\gamma\\
-\gamma
\end{array}\right) \, , \qquad \Gamma_{\sigma}^{1}  = \frac{\langle {\bf 1}\otimes \sigma \rangle_{(r,a)}}{\langle {\bf 1}\otimes {\bf 1}\rangle_{(r,a)}} = 
\left( \begin{array}{r}
\alpha \\
\beta\\
0\\
-\beta \\
-\alpha\\
0
\end{array}\right) 
\ee
where 
\be
\beta=2^{1/4}\approx 1.189\, , \qquad   \tilde \beta = 2^{1/4}\left(\frac{\sqrt{3} - 1}{\sqrt{3}+1}\right)^{1/2}\approx 0.615 \, , 
\ee
\be
\alpha = 2^{1/4}\left(\frac{\sqrt{3} + 1}{\sqrt{3}-1}\right)^{1/2} \approx 2.297\, , \qquad \gamma = \frac{\sqrt{3}+1}{\sqrt{3}-1} \approx 3.732\, .
\ee
We observe that knowing the above four component ratios determines the index $a$ uniquely. 
Furthermore the factorised interfaces correspond to $a=2,3,4$. Since a magnetic perturbation is present 
on the Ising to Yang-Lee trajectory we expect a non-vanishing $|\sigma\rangle$-component in the vacuum. 
This means that there are essentially two types of candidates  for the Ising - Yang-Lee RG interface -  
the factorised defects $|\pm\rangle\!\rangle\langle\!\langle \phi, 1|_{\rm YL}$ 
and  the non-factorisable defects corresponding to $a=1,5$. Note that the pair $a=1,5$ forms a doublet 
 under $Z_2$ spin reversal symmetry so that there is essentially one non-factorizable candidate interface. 
To determine the index $r$ one would need to know other ratios of one-point functions e.g. the ratio
\be
\frac{\langle \phi \otimes \epsilon \rangle_{(1,a)}}{\langle {\bf 1} \otimes \epsilon \rangle_{(1,a)}}  
= i \left( \frac{5 + \sqrt{5}}{5-\sqrt{5}}\right)^{3/4} \, , \qquad \frac{\langle \phi \otimes \epsilon \rangle_{(3,a)}}{\langle {\bf 1} \otimes \epsilon \rangle_{(3,a)}} = 
 -i \left( \frac{5 - \sqrt{5}}{5+\sqrt{5}}\right)^{1/4}\,  
\ee
which is independent of $a$.
However this requires knowing how the Ising eigenvector $|\epsilon\rangle$  is realised in the Yang-Lee state space 
that is inaccessible by TCSA method which only gives the infrared theory eigenvectors in the UV theory space.
In the next section we will argue that the RG interface for the Ising to Yang-Lee flow does not asymptote to any 
single conformal interface.

As a final remark about the conformal interfaces considered in this section we note that using the Cardy constraint we can also 
find the spectrum of interface fields. We found that all interfaces contain 
relevant fields (and are thus unstable) except for the factorizable  ones:  $|\pm\rangle\!\rangle\langle\!\langle  1|_{\rm YL}$.

\subsection{RG interface for the flow to Yang-Lee fixed point}\label{ycr_sec}
The approach to a non-trivial fixed point at $y=y_{\rm cr}$ is marked  by the appearance of a 
"nose" on the plot of dimensionless energies $\tilde E(r)$ depicted below: the vacuum and the first excited level 
energies get asymptoticaly close without collision. The relative energies $e_{i}-e_{0}$ level to constant values 
equal to the conformal dimension difference $\Delta_{i}-\Delta_{0}$. We find numerically (see the second graph below) 
that at truncation level 
$n_{c}=12$ the difference $e_{1}-e_{0}$ levels at the value $0.391$ while $e_{2}-e_{0}$ approaches $1.90$. 
These values approximate the dimension gaps for the Yang-Lee operators ${\bf 1}$ and $L_{-1}\bar L_{-1}\phi$  
respectively (the lowest dimension state is $\phi$ with dimension -0.4). 
 
\begin{center}
\begin{figure}[H]
\begin{minipage}[b]{0.5\linewidth}
\centering
\includegraphics[scale=0.6]{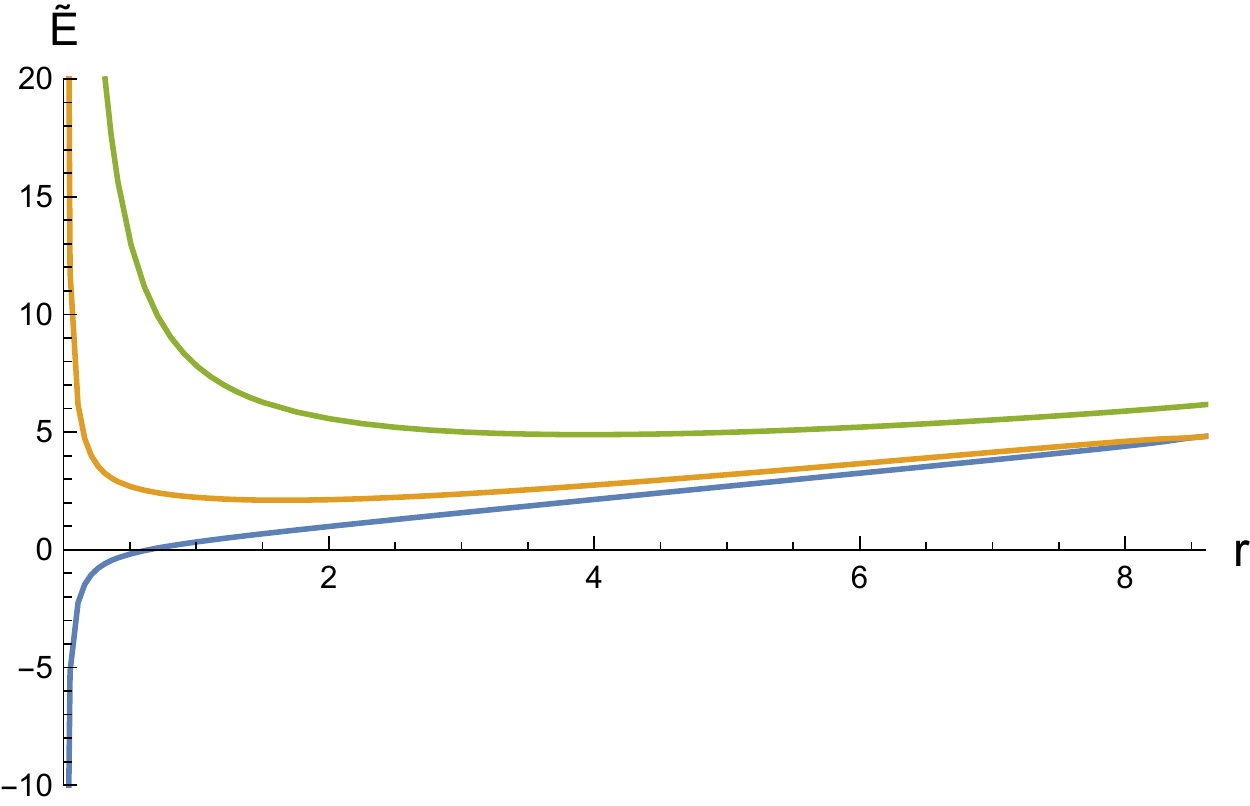}
\end{minipage}%
\begin{minipage}[b]{0.5\linewidth}
\centering
\includegraphics[scale=0.6]{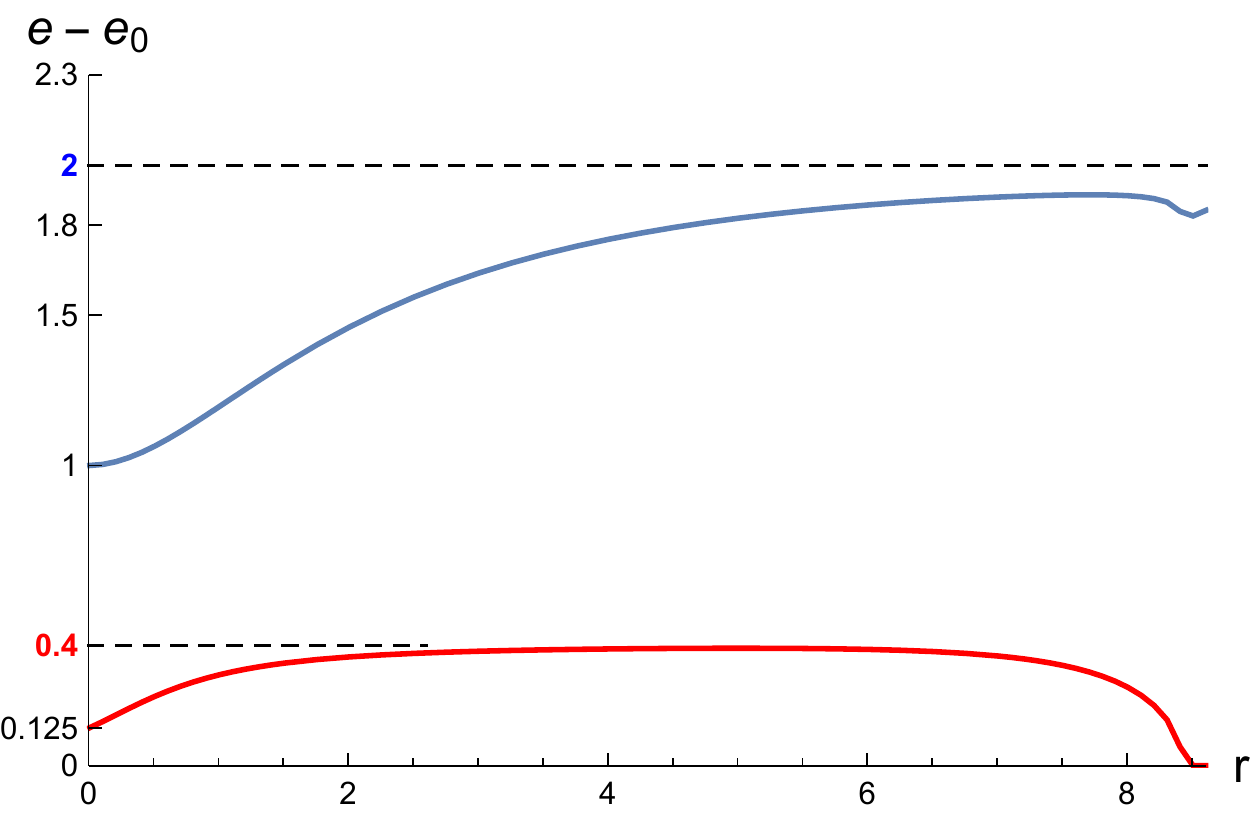}
\end{minipage}
\caption{The lowest 3 energy eigenvalues at $y=y_{\rm cr}$, $n_{c}=12$.}
\label{nose}
\end{figure}
\end{center}

For the truncation level $n_c=12$ we find numerically that the lowest two levels merge into a complex pair 
at $r\approx 8.5$. Looking at the behaviour at different values of $n_{c}$ we find that the merging point moves to higher $r$ as 
we increase $n_c$.  So as in the case $y<y_{\rm cr}$ such a merger seems to be merely a finite precision artefact  
which however limits the domain of $r$ we can investigate numerically. 

In the previous section we found explicitly the component ratios $\Gamma^{0}_{\epsilon,\sigma}$, 
$\Gamma^{1}_{\epsilon,\sigma}$ for all conformal interfaces between the critical Ising and Yang-Lee models 
(see formulai (\ref{InterfaceG1}), (\ref{InterfaceG2})). Crucially all possible values are real while, 
as discussed in section  \ref{Complex_vac_sec},  for as long as an eigenvalue (vacuum or excited) stays real 
the corresponding ratio $\Gamma_{\epsilon}$ is real and $\Gamma_{\sigma}$ is  imaginary. 
This leaves us with two possibilities: {\it either the limiting conformal interface does not exist as 
in the case of massive flows discussed before, or  the RG interface 
approaches a conformal interface with} $\Gamma_{\sigma}^{0}=\Gamma_{\sigma}^{1}=0$ (i.e. an interface symmetric under spin reversal). 

The numerical data for the component ratios of the vacuum and the first two  excited states is shown on 
the graphs below.

\begin{center}
\begin{figure}[H]
\begin{minipage}[b]{0.5\linewidth}
\centering
\includegraphics[scale=0.5]{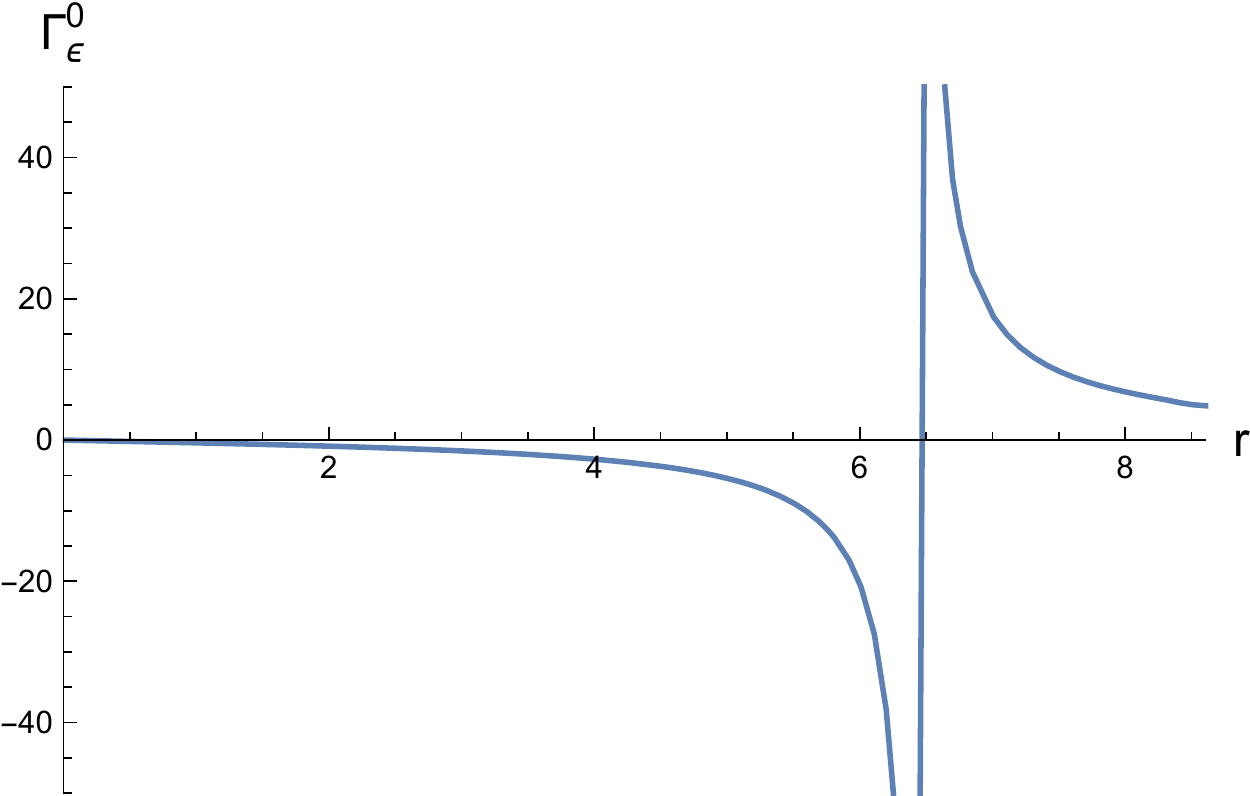}
\end{minipage}%
\begin{minipage}[b]{0.5\linewidth}
\centering
\includegraphics[scale=0.5]{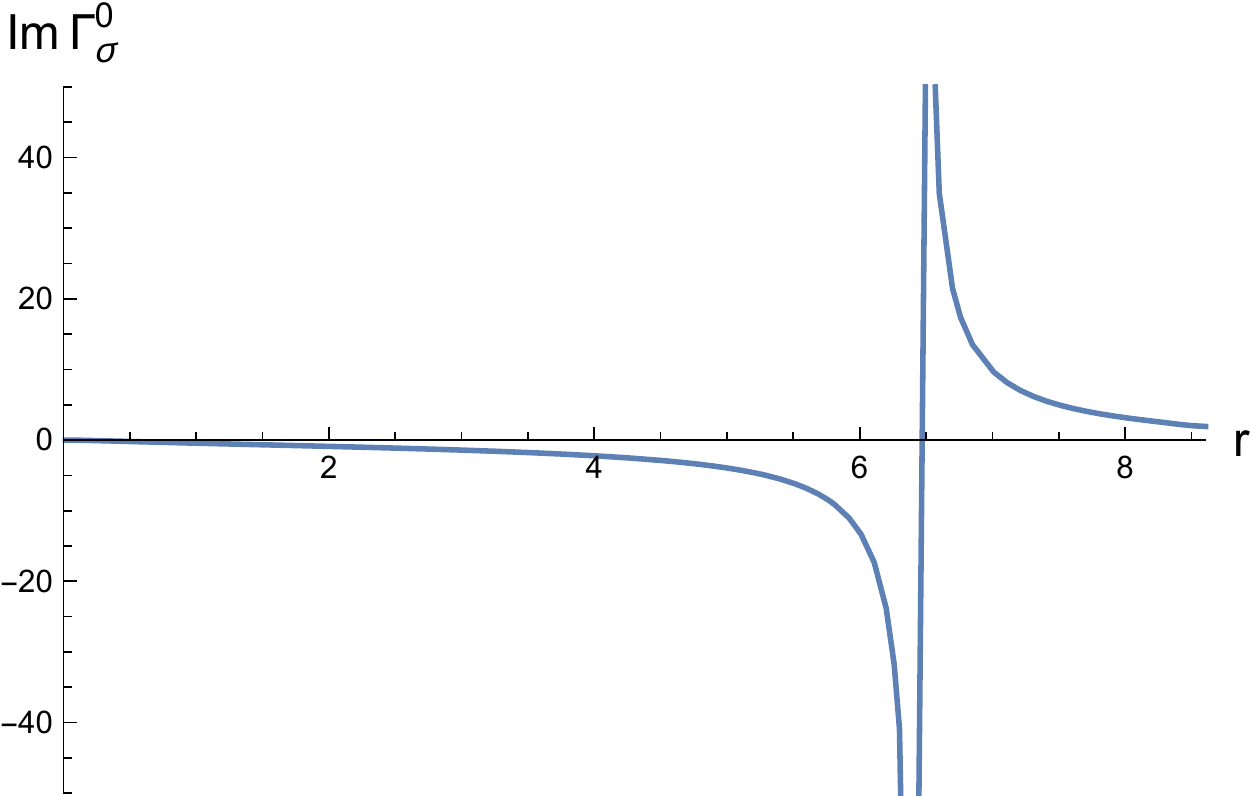}
\end{minipage}
\caption{The component ratios $\Gamma^{0}_{\epsilon,\sigma}$ at $y=y_{\rm cr}$, $n_{c}=12$.}
\end{figure}
\end{center}
\begin{center}
\begin{figure}[H]
\begin{minipage}[b]{0.5\linewidth}
\centering
\includegraphics[scale=0.5]{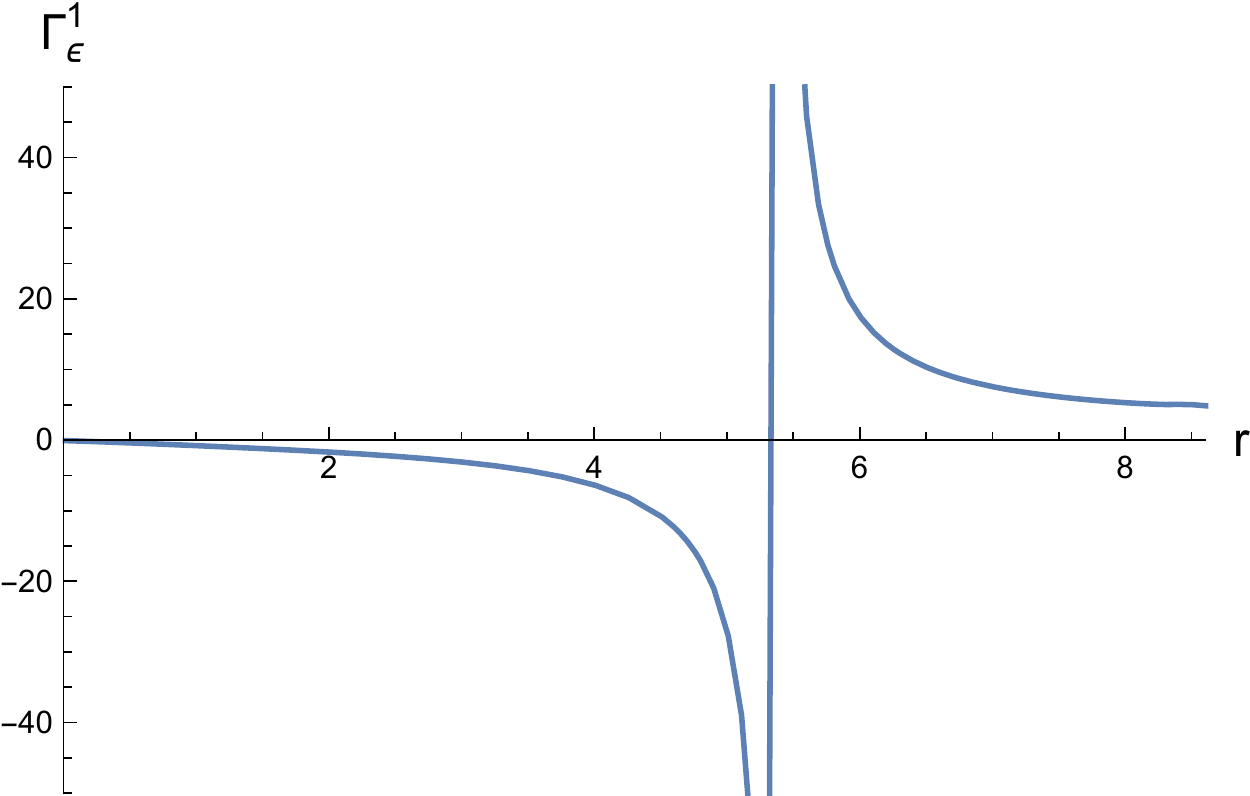}
\end{minipage}%
\begin{minipage}[b]{0.5\linewidth}
\centering
\includegraphics[scale=0.5]{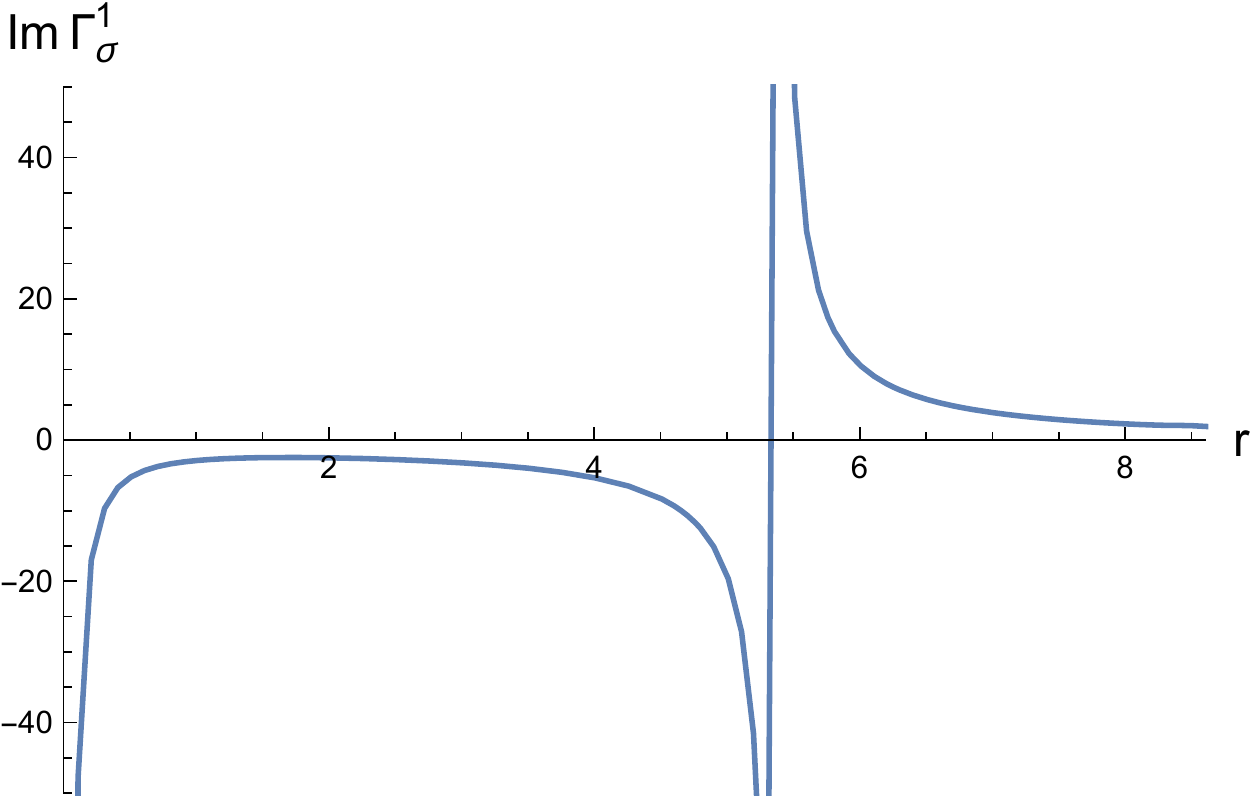}
\end{minipage}
\caption{The component ratios $\Gamma^{1}_{\epsilon,\sigma}$ at $y=y_{\rm cr}$, $n_{c}=12$.}
\end{figure}
\end{center}
\begin{center}
\begin{figure}[H]
\begin{minipage}[b]{0.5\linewidth}
\centering
\includegraphics[scale=0.5]{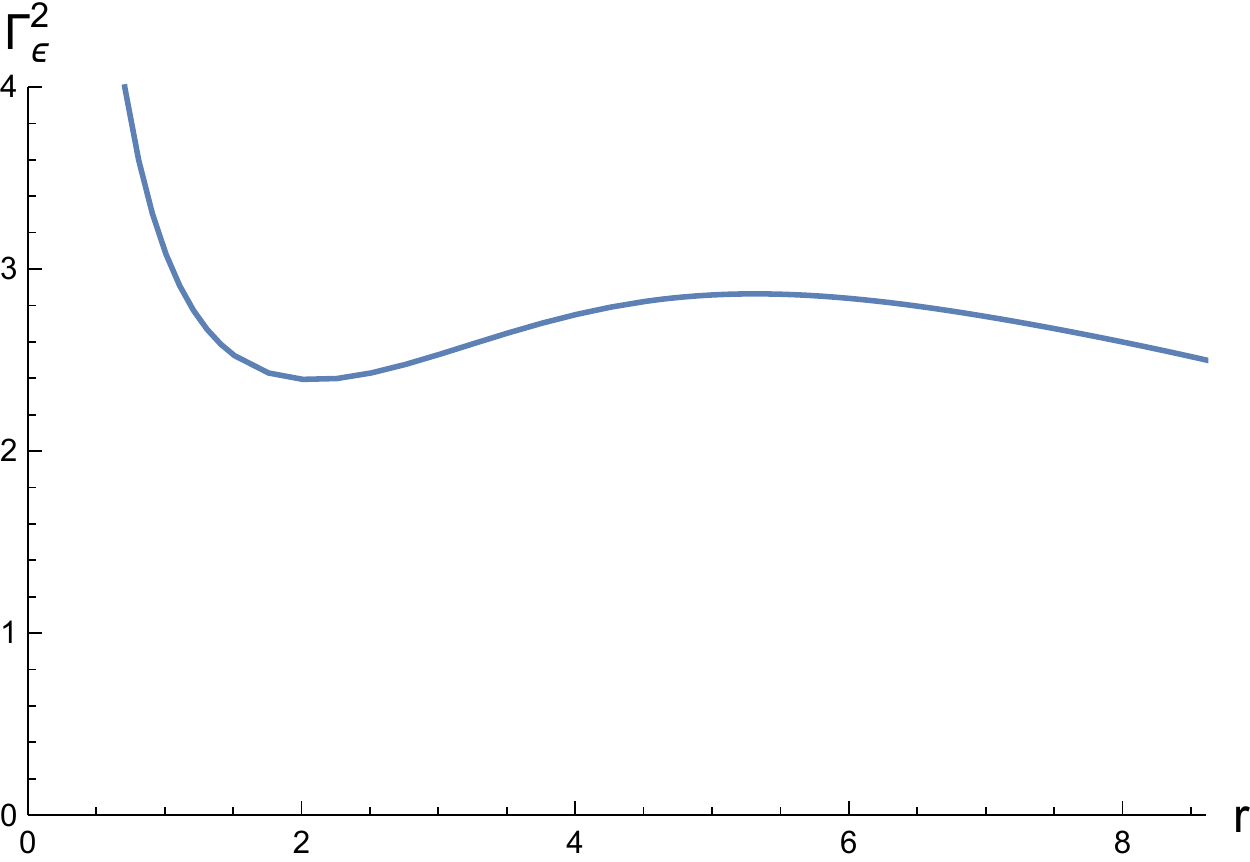}
\end{minipage}%
\begin{minipage}[b]{0.5\linewidth}
\centering
\includegraphics[scale=0.5]{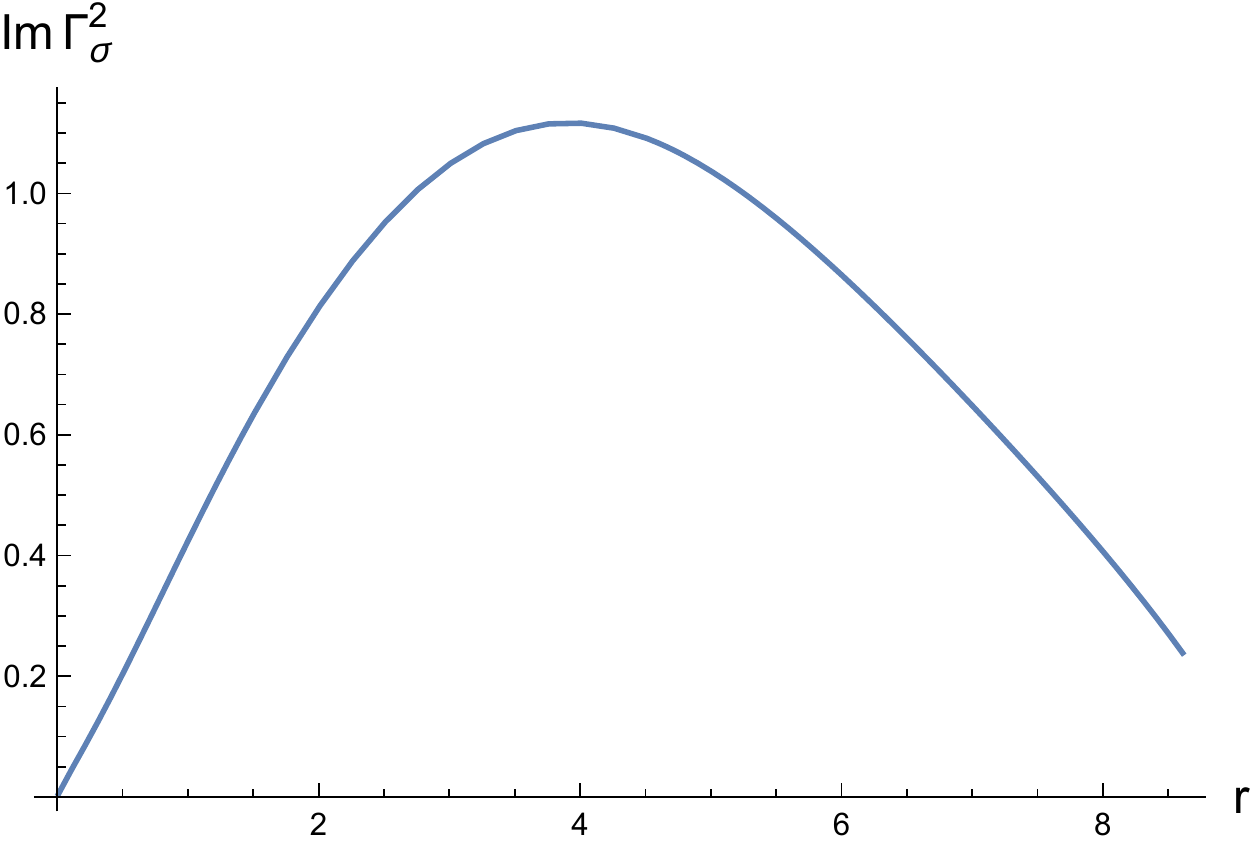}
\end{minipage}
\caption{The component ratios $\Gamma^{2}_{\epsilon,\sigma}$ at $y=y_{\rm cr}$, $n_{c}=12$.}
\end{figure}
\end{center}
We see that the ratios for the vacuum and the first excited level each behave qualitatively very similar 
to the vacuum ratios for the massive flows  discussed in section \ref{Im_massive}. The ratios 
$\Gamma^{0}_{\epsilon,\sigma}$ blow up near $r=6.4$ while $\Gamma^{1}_{\epsilon,\sigma}$ blow up near $r=5.3$. 
In each case the ratios $\Gamma^{0}_{\epsilon}/\Gamma^{0}_{\sigma}$, $\Gamma^{1}_{\epsilon}/\Gamma^{1}_{\sigma}$
remain finite and increasing functions past the blow up points. The ratios $\Gamma^{2}_{\epsilon,\sigma}$ while remaining 
finite do not show any tendency to level at a constant value. While we cannot say with certainty that the oscillations between the 
blow up points will continue everything points towards the non-convergent oscillatory scenario. 

In addition to this numerical indications we would like to remark that the alternative scenario in which we approach 
a conformal  interface symmetric under spin reversal would be very hard to envisage from a general point of view. 
Clearly at large distances the magnetic field perturbation dominates and it breaks this symmetry. 

\section{Open problems}
In this paper we 
showed how  pairings between states arising in  RG interfaces of  \cite{BR} can be calculated numerically using  truncated Hamiltonian  techniques. 
As illustrated by the exactly solvable case of the Ising model with zero magnetic case in order to read off the numerical values of the pairings (or rather their ratios) 
one needs to go to very large values of dimensionless couplings. At these values  one cannot a priory trust any  results.
Perturbative corrections are organised in ratios of couplings to the truncation energy 
and are very large in this region. Some aberrations in the low-lying spectrum that one can easily spot are the non-linearity of the vacuum energy 
dependence on scale and non-constance of the mass gap\footnote{The author thanks V. Rychkov for suggesting to discuss these quantities.}: 
$\tilde E_{1}-\tilde E_{0}$. These can be easily spotted in the TCSA sample of data 
presented on the plots below. 

\begin{center}
\begin{figure}[H]
\begin{minipage}[b]{0.5\linewidth}
\centering
\includegraphics[scale=0.8]{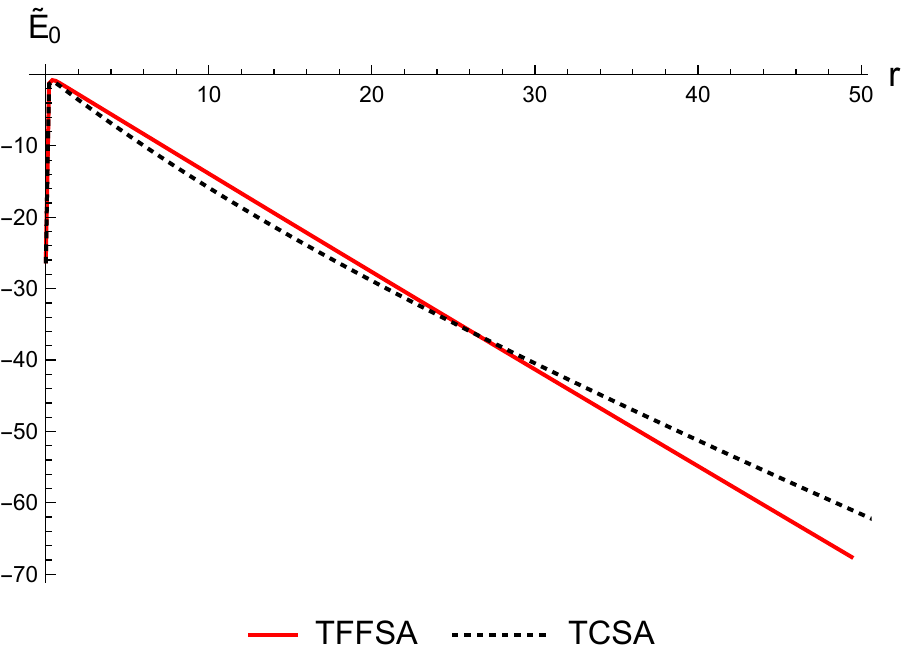}
\end{minipage}%
\begin{minipage}[b]{0.5\linewidth}
\centering
\includegraphics[scale=0.8]{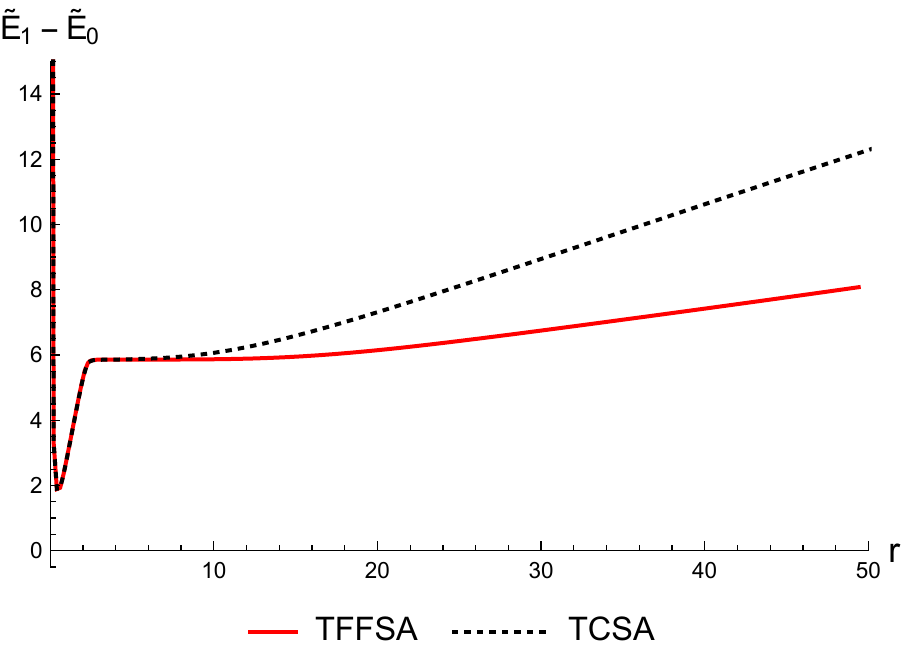}
\end{minipage}
\caption{The vacuum energy and the energy gap  at $y=1$, $n_{c}=12$.}
\end{figure}
\end{center}

The TFFSA vacuum energy behaves much better even at very large scales $r\sim 50$. The vacuum energy remains linear with a very good 
accuracy (regression variance is about 0.01). The mass gap calculated using TFFSA  deviates from constant at big enough $r$ but the 
deviation is significantly smaller than in the TCSA data. 

This means that in general we cannot trust the energy eigenvalues and eigenvectors for such large values of scale, and yet our numerical 
results (both TCSA and TFFSA) indicate that  ratios of low lying components of the vacuum eigenvector are well behaved at large scales and 
 the non-perturbative errors for them remain well bounded. This suggests that although we need to go to very large scales to read off the asymptotic 
 values,  these quantities are more robust against truncation errors than the energy eigenvalues and eigenvectors as a whole. 

At the moment we 
have no conceptual understanding of this fact which is crucial for the TCSA scheme to work for calculating the IR asymptotics of component ratios. 
Perturbative corrections also need to be understood  as well with a suitable quantitative method for their incorporation needs to be worked out.

 Another important issue that needs further understanding is renormalisation of wave functionals. 
 As discussed in the introduction in some cases additional boundary counter terms may be needed to renormalise 
  wave functionals in the interaction picture. This may occur when vector fields are present in the OPE of the perturbing operators. 
 In the bulk perturbation such terms do not need any counter terms due to rotational symmetry but from the point of view of the boundary 
 they are scalars and may lead to additional divergences for collisions at the boundary (where rotational symmetry is broken). 
 Such counter terms are local along the boundary but what is their manifestation in the Hamiltonian formalism (and TCSA) is not clear. 
 
 There is a good chance, in our opinion, that for superrenormalisable perturbations the above conceptual issues do not impede using  
 the raw TCSA data for  identifying RG interfaces and RG boundaries. It would be interesting to explore systematically other examples 
 such as the tricritical Ising model or Potts model. For the tricritical Ising model there are 4 relevant operators and 6 elementary 
 conformal boundary conditions. It would be interesting to find out  how the three-dimensional space of massive flows breaks up into regions 
 according to their RG boundaries. 
 Also a  conformal interface for the RG flow from the tricritical Ising to the critical Ising was put forward in 
 \cite{Gaiotto}. It will be interesting to test that proposal numerically using TCSA. We hope to report some answers to these questions in a near future 
 \cite{inprogress}.

\setcounter{equation}{0}

\vspace{1cm}

\begin{center}
{\bf \Large Acknowledgments }
\end{center}
The author is grateful to Adam Nahum,  Slava Rychkov, Cornelius Schmidt-Colinet,  Stefan Sint, G\' abor Tak\' acs, Balt van Rees, and Gerard Watts
   for stimulating discussions and 
comments on the manuscript. 
He also thanks the organisers of the workshop "Conformal Field Theories and Renormalisation Group 
Flows in Dimensions $d>2$" at Galileo Galilei Institute (GGI) in Florence and GGI staff for hospitality. 
This work was supported in part by STFC grant "Particle Theory at the Higgs Centre", ST/L000334/1.
All numerical results presented in the paper were obtained using Wolfram {\it Mathematica } package v. 10.2 .


\begin{thebibliography}{99}
 \bibitem{BR} I. Brunner and D. Roggenkamp, {\it Defects and bulk perturbations of boundary 
  Landau-Ginzburg orbifolds}, JHEP {\bf 0804} (2008) 00; arXiv:0712.0188.
  
  \bibitem{FQ} S. Fredenhagen and T. Quella, {\it 	
Generalised permutation branes}, JHEP {\bf 0511} (2005) 004;  arXiv:hep-th/0509153.
  
\bibitem{Gaiotto} D. Gaiotto, {\it Domain walls for two-dimensional renormalization group flows}, JHEP 1212 (2012) 103;
arXiv:1201.0767.


\bibitem{FZ} P. Fonseca and A. Zamolodchikov, {\it 	
Ising field theory in a magnetic field: Analytic properties of the free energy}, Journal of Statistical Physics,
 Vol. 110, Issue 3 (2003) pp. 527-590; arXiv:hep-th/0112167.

\bibitem{QRW} T. Quella, I. Runkel and G. Watts, {\it Reflection and Transmission for Conformal Defects},  
JHEP {\bf 0704} (2007) 095;  arXiv:hep-th/0611296.


\bibitem{YZ1} V. Yurov and Al. Zamolodchikov, {\it Truncated Conformal Space Approach To Scaling Lee-yang Model}, 
Int. J. Mod. Phys. {\bf A5} (1990) 3221-3246.  

\bibitem{YZ2} V. Yurov and Al. Zamolodchikov, {\it 	
Truncated fermionic space approach to the critical 2-D Ising model with magnetic field}, Int. J. Mod. Phys.  {\bf A6}  (1991) 4557-4578.

\bibitem{GW} P. Giokas and G. Watts, {\it The renormalisation group for the truncated conformal space approach on the cylinder},  arXiv:1106.2448.

\bibitem{HRvR} M. Hogervorst, S. Rychkov, and B. C. van Rees,{\it A Cheap Alternative to the Lattice?},  Phys. Rev. {\bf D91} (2015) 025005; arXiv:1409.1581.

\bibitem{RV} S. Rychkov, L. G. Vitale, {\it Hamiltonian Truncation Study of the $\Phi^4$ Theory in Two Dimensions}, Phys. Rev. {\bf D 91} (2015) 085011; 
 arXiv:1412.3460. 



\bibitem{LT_RG} M. Lencs\' es, G. Tak\' acs, {\it Excited state TBA and renormalized TCSA in the scaling Potts model},  
JHEP {\bf 09} (2014) 052; arXiv:1405.3157.  

\bibitem{LT_Potts} M. Lencs\' es, G. Tak\' acs, {\it Confinement in the q-state Potts model: an RG-TCSA study}, JHEP {\bf 09} (2015) 146;  arXiv:1506.06477.

\bibitem{Takacs_quenches} T. Rakovszky, M. Mesty\' an, M. Collura, M. Kormos, and G. Tak\' acs, 
{\it Hamiltonian truncation approach to quenches in the Ising field theory}, arXiv:1607.01068.

\bibitem{Cardy_etal} M. L\" assig, G. Mussardo, and John L. Cardy, {\it The scaling region of the tricritical Ising model in two-dimensions}, 
Nucl. Phys. {\bf B348} (1991) 591-618.  


\bibitem{Symanzik} K. Symanzik, {\it Schr\" odinger representation and Casimir effect in renormalizable quantum field theory}, 
Nucl. Phys. {\bf B190} (1981) 1-44.

\bibitem{Luscher} M. L\" usher, {\it Schr\" odinger representation in quantum field theory}, Nucl Phys. {\bf 254} (1985) 52. 

\bibitem{Minic_Nair} D. Minic and V.P. Nair, {\it Wave Functionals, Hamiltonians and Renormalization Group}, 
 Int. J. Mod. Phys. {\bf A11} (1996) 2749;  arXiv:hep-th/9406074.
 
 \bibitem{FLP} J. R. Fliss, R. G. Leigh, and O. Parrikar, {\it Unitary Networks from the Exact Renormalization of Wave Functionals}, 
  arXiv:1609.03493.
  
  \bibitem{H_theorem} L.~Klaczynski, {\it Haag's theorem in renormalised quantum field theories},  arXiv:1602.00662.


\bibitem{TWunpub} G.  Tak\' acs (2012), unpublished.

\bibitem{CEF} P. Calabrese, F. H. L. Essler, and M. Fagotti, {\it Quantum Quench in the Transverse Field Ising chain I: Time evolution of order parameter correlators},  J. Stat. Mech. (2012) P07016;  arXiv:1204.3911.


\bibitem{YL1} C.N. Yang and T. D. Lee, {\it Statistical theory of equations of state and phase transitions. I. Theory of condensation.}, 
Phys. Rev. {\bf 87} (1952) 404.

\bibitem{YL2} C.N. Yang and T. D. Lee, {\it Statistical theory of equations of state and phase transitions. II. Lattice gas and Ising model. }, 
Phys. Rev. {\bf 87} (1952) 410.

\bibitem{Cardy2} J. L. Cardy, {\it 	
Conformal Invariance and the Yang-lee Edge Singularity in Two-dimensions},  Phys. Rev. Lett. {\bf 54} (1985) 1354.



\bibitem{Berezin} F. A. Berezin, {\it The method of second quantization}, Academic Press, 1966. 

\bibitem{Cardy} J. Cardy, {\it 	
Boundary Conditions, Fusion Rules and the Verlinde Formula }, Nucl. Phys. {\bf B324} (1989) 581-596. 

\bibitem{BPPZ} R. E. Behrend, P.A. Pearce, V. B.  Petkova, and  J.-B. Zuber, {\it Boundary Conditions in Rational Conformal Field 
Theories}, Nucl. Phys. {\bf B579} (2000) 707-773; arXiv:hep-th/9908036.

\bibitem{TFT} J. Fuchs, I. Runkel, and C. Schweigert, {\it TFT construction of RCFT correlators I: Partition functions}, Nucl. Phys. {\bf B646} (2002) 
353-497;  arXiv:hep-th/0204148.

\bibitem{BBDO} C. Bachas, J. de Boer, R. Dijkgraaf, and H. Ooguri, {\it Permeable conformal walls and holography}, JHEP 06 (2002) 027; 
arXiv:0111210. 

\bibitem{BB} C. Bachas and I. Brunner, {\it Fusion of conformal interfaces}, JHEP 0802:085,2008; arXiv:0712.0076. 

\bibitem{Zam1} A. Zamolodchikov, {\it Ising Spectroscopy I:  Mesons at $T\! <\! T_c$}, arXiv:1310.4821.

\bibitem{Zam2} A. Zamolodchikov, {\it Ising Spectroscopy II: Particles and poles at $T\!>\!T_{c}$},   arXiv:1310.4821. 

\bibitem{CZ} R. Chatterjee and A. Zamolodchikov, {\it Local Magnetization in Critical Ising Model with Boundary Magnetic Field},  
Mod. Phys. Lett. {\bf A9} (1994) 2227; arXiv:hep-th/9311165.

\bibitem{Chat} Chatterjee, {Exact Partition Function and Boundary State of Critical Ising Model with Boundary Magnetic Field}, 
Mod. Phys. Lett. {\bf A10} (1995) 973;  arXiv:hep-th/9412169.

\bibitem{inprogress} A. Konechny, {\it RG boundaries and interfaces for tricritical Ising field theory}, work in progress.
\end{thebibliography}
\end{document}